\documentclass[twoside,12pt,a4paper]{article}
\usepackage{graphicx}
\usepackage{amsmath}
\usepackage{upgreek}
\usepackage{mcite}
\usepackage{feynmf}
\usepackage{umlaut}
\usepackage{xspace}

\newcommand{\elsa}{{\sc Elsa}\xspace}
\newcommand{\mami}{{\sc Mami}\xspace}
\newcommand{\gdhsr}{GDH Sum Rule\xspace}
\newcommand{\gdhex}{GDH-Experiment\xspace}
\newcommand{\gdhdet}{{GDH-Detector}\xspace}
\newcommand{\gdhcol}{{GDH-Collaboration}\xspace}
\newcommand{\daphne}{{\sc Daphne}\xspace}
\newcommand{\cer}{\v{Cerenkov}\xspace}
\newcommand{\mol}{M{\o}ller\xspace}
\newcommand{\molpol}{\mol polarimeter\xspace}
\newcommand{\maid}{{\sc Maid}\xspace}
\newcommand{\said}{{\sc Said}\xspace}

\def\graph(#1,#2)#3{
 \begin{minipage}{#1\unitlength}
  \begin{picture}(#1,#2)
   \begin{fmfgraph}(#1,#2)
    #3
   \end{fmfgraph}
  \end{picture}
 \end{minipage}
}
\def\graphs(#1,#2)#3{
 \begin{minipage}{#1\unitlength}
  \begin{picture}(#1,#2)
   \begin{fmfgraph*}(#1,#2)
    #3
   \end{fmfgraph*}
  \end{picture}
 \end{minipage}
}

\def\be{\par\nobreak\noindent\begin{equation}}
\def\ee{\end{equation}\par\nobreak\noindent}
\def\beqn{\par\nobreak\noindent\begin{eqnarray}}
\def\eeqn{\end{eqnarray}\par\nobreak\noindent}

\begin{document}

\title{ \vspace{1cm} The Gerasimov-Drell-Hearn Sum Rule}
\author{Klaus\ Helbing\\
\\
Physikalisches Institut, Universit\"at Erlangen-N\"urnberg, Erlangen, Germany$^1$}
\maketitle
\vspace{-15pt}
\begin{equation}\Large
\boxed{\boxed{
~\int\limits_0^\infty \frac{d\nu}{\nu} 
\left[ \sigma_{3/2}(\nu) - \sigma_{1/2}(\nu) \right] =
\frac{2\pi^2\alpha}{m^2}\kappa^2~
}}
\label{eqn:gdh}
\end{equation}

\begin{abstract} 
Sum rules measurements involving the spin structure of the nucleon
like those due to Bjorken, Ellis and Jaffe and the one due to
Gerasimov, Drell and Hearn allow to study the structure of strong
interactions. At long distance scales in the confinement regime the
Gerasimov-Drell-Hearn (GDH) Sum Rule (Eq.~(\ref{eqn:gdh})) connects
static properties of the nucleon - like the anomalous 
magnetic moment and the nucleon mass - with the difference of spin
dependent doubly polarized total absorption cross sections of real photons.
Hence, the full spin-dependent excitation spectrum of the nucleon is
being related to its static properties. The sum rule has not been
investigated experimentally until recently. Now, for the first time this
fundamental sum rule is verified by the \gdhcol with
circularly polarized real photons and longitudinally polarized nucleons
at the two accelerators \elsa and \mami. The investigation of the response
of the proton as well as of the neutron allows to perform an isospin
decomposition. Further investigations with real photons are
scheduled at \textsc{Slac}, \textsc{JLab}, \textsc{Spring-8},
\textsc{Legs} and \textsc{Graal}. 
The integral (sum) of the \gdhsr can be generalized to the case of
virtual photons. This allows to establish a $Q^2$ dependency and to study
the transition to the perturbative regime of QCD. Ultimately, the GDH
Sum Rule can be related to the Bjorken and the Ellis-Jaffe Sum Rule. 
This transition is the subject of several experiments e.g. at \textsc{JLab} for
the resonance region and of the \textsc{Hermes} experiment at \textsc{Desy} for higher $Q^2$.

This contribution covers the status of theory concerning the \gdhsr as
well as the experimental approaches and their results for the
absorption of real and virtual photons. We point out that the
so-called No-Subtraction hypothesis, often considered the weakest
part of the derivation of the \gdhsr, in fact follows from unitarity
and does not impair the fundamental character of the \gdhsr.
The experimental data verify the \gdhsr for the proton at the level of
8~\% including the systematic uncertainties from extrapolations to
unmeasured energy regions. For the \gdhsr on the neutron and the
isovector case we find unexpected contributions at photon energies
above 1~GeV. 
\end{abstract}

\setcounter{footnote}{1}
\footnotetext{Now at Bergische Universit\"at Wuppertal, Fachbereich C -- Physik, Gau{\ss}str. 20, D-42119 Wuppertal, Germany; helbing@uni-wuppertal.de }

\newpage

\tableofcontents
\newpage

\section{Introduction}
Understanding the spin structure of the nucleon is at the heart of
present nuclear and particle physics activities.
Of particular interest are sum rules which connect information from all
energies to fundamentals of our current view of nature's
laws. The Gerasimov-Drell-Hearn (GDH) sum rule is an excellent example
of a whole class of dispersive sum rules which are consequences of general
principles. They can be used to test these fundamental
principles and thus probe deep mysteries of nature. Also, these sum
rules provide a vehicle to access new experimental observables and to
study the physics of strongly interacting systems in refined detail.

The Gerasimov-Drell-Hearn sum rule was established in the second half
of the 1960ies. The two independent derivations presented by
Sergei~B.~Gerasimov~\cite{Gerasimov:1965et} and by Sidney~D.~Drell and
Anthony~C.~Hearn~\cite{Drell:1966jv} both appeared in 1966 in English
language, while Gerasimov's original
publication~\cite{Gerasimov:1965rus} was available in Russian language
already in 1965. We owe this historic detail today's naming sequence
for the sum rule: Gerasimov-Drell-Hearn sum rule.
In fact, initially the sum rule was called Drell-Hearn-Gerasimov sum
rule. The change in sequence was adopted first for the naming of the
\gdhcol --- the collaboration that finally took the challenge
to verify this fundamental sum rule. This sequence is widely accepted today.
One might even add two more names to the sum rule: Hosoda
and Yamamoto~\cite{Hosoda:1966} used the current algebra formalism for
their derivation also dated 1966. 

\begin{figure}
\begin{center}
\includegraphics[width=0.75\textwidth]{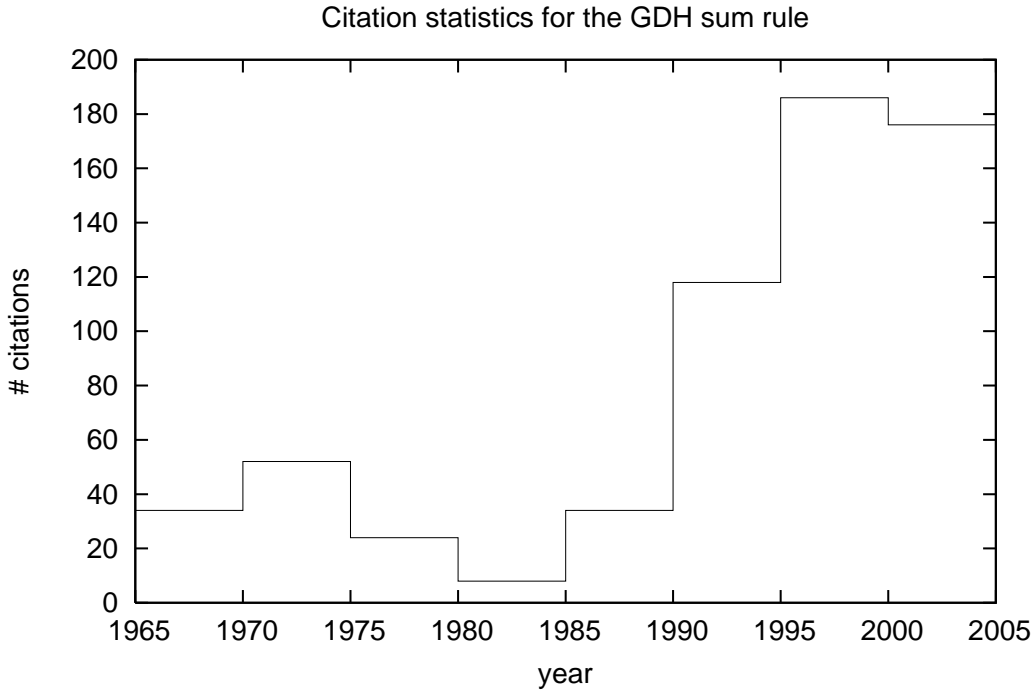}
\label{fig:PubsGDH}
\caption{Citation statistics for the two sum rule
derivations of Gerasimov~\protect\cite{Gerasimov:1965et} and Drell and
Hearn~\protect\cite{Drell:1966jv} according to \textsc{Spires}.} 
\end{center}
\end{figure}

Fig.~\ref{fig:PubsGDH} shows the citation statistics for the sum rule
derivations of Gerasimov~\cite{Gerasimov:1965et} and Drell and
Hearn~\cite{Drell:1966jv} according to \textsc{Spires} as a function of
time. One observes that   
until 1990 the interest in the \gdhsr is essentially
constant. This hesitant reaction was probably driven by the problem to
estimate the experimental feasibility of a verification at that time. 
For example: Gerasimov rated the sum rule mainly to be of academic
interest only, while Hosoda and Yamamoto were convinced that it would be
straightforward to test it experimentally. Drell and Hearn, however,
state that a test would be a formidable experimental challenge and
call for it.  

Consequently, the early discussion was centered around questions
connected to the validity of the \gdhsr. As an example the title
of Ref.~\cite{Barton:1967at} might serve: ``Drell-Hearn-Gerasimov
sum rule: Examples and Counterexamples''.
Experimentally, only multipole analyses of
unpolarized single pion photoproduction data were possible, from which
--- in an indirect fashion ---  estimates of the contribution from
low-lying resonances to the GDH integral were extracted. 

Albeit by then about a quarter of a century old, in the early 1990ies the GDH
sum rule still lacked a direct experimental check, since a doubly
polarized photoabsorption measurement is needed covering a wide energy
range. This challenge had never been taken up so far.
The \gdhcol was established to bring together polarized
sources and polarized targets as well as detectors suited to measure
total photoabsorption cross sections at two electron accelerators ---
\elsa (Bonn) and \mami (Mainz). The collaboration has
about 60 members coming from 16 institutions from all over Europe,
Japan and Russia. 

The renewed interest in the \gdhsr arose not only due to the
availability of experimental techniques but was also stimulated by the
apparent failure of the Ellis-Jaffe sum rule~\cite{Ellis:1973kp} as
reported in polarized deep inelastic scattering experiments at
\textsc{Slac}~\cite{Baum:1983ha} and \textsc{Cern}~\cite{Ashman:1989ig}. 
Although the Ellis-Jaffe sum rule cannot be regarded to be quite as
fundamental as the \gdhsr it gave rise to the so-called ``spin
crisis'' in the late 1980ies. It became clear that further precision tests
were needed to improve our understanding of the spin structure of the
nucleon. 

In 1998 the \gdhcol at \mami provided the first
direct information on the helicity structure as probed by real photons. The
\gdhsr for the proton was verified in 2001 together with the
data taken at \elsa by the
\gdhcol~\cite{Dutz:2003mm,Helbing:2002eg}. The data taking of the
\gdhcol is completed since 2003.

In the aftermath of the efforts to determine the helicity structure
with real photons also experiments at \textsc{JLab} in Hall~A and with
the \textsc{Clas} detector and at \textsc{Desy} with the
\textsc{Hermes} detector help to identify the spin structure of the
nucleon for intermediate photon virtualities. 
Besides the direct verification of the \gdhsr a wealth of new
data is now available to disentangle the involved structure
of the nucleon aided by polarization observables. For a recent reviews
on this subject see for example~\cite{Drechsel:2004ki,Bass:2004xa}.

This review in detail discusses the status of theoretical considerations
concerning the \gdhsr at the real photon point including a
survey of problems with the interpretation of the sum rule and of
possible modifications. Also, the generalizations of the \gdhsr
to finite photon virtuality and the connection to other sum rules are
outlined. 

On the experimental side experiments with 
a connection to the \gdhsr are addressed. 
Virtues of future experiments and their chances to improve the current
understanding of the \gdhsr are also discussed.
The experiments of the \gdhcol will receive the most
attention as these measurements are -- so far -- the only ones that have
lead to a direct test of the \gdhsr with real photons.

\newpage

\section{The \gdhsr for real photons}
As already mentioned in the introduction, there is more than one
method to derive the Gerasimov-Drell-Hearn sum rule:
Gerasimov~\cite{Gerasimov:1965et} as well as Drell and
Hearn~\cite{Drell:1966jv} used a dispersion theoretic approach. Hosoda
and Yamamoto~\cite{Hosoda:1966} on the other hand used the current
algebra formalism for their derivation of the very same sum rule. 
And there is yet another way: In 1972, Dicus and
Palmer~\cite{Dicus:1972vp} presented a derivation of the \gdhsr
from the algebra of currents on the light-cone.

In the following all three of these derivations will be discussed with
their main aspects. We present the
dispersion theoretic derivation in most detail as it is very
instructive. Also, the dispersion theoretic approach is the most
illustrative in identifying where the fundamental principles actually
enter the derivation.  

We will show that the dispersion theoretic derivation of the \gdhsr is
possible without the restriction to lowest non-vanishing order in
electromagnetic coupling. While the low energy theorems indeed were shown 
only in low orders of coupling all other steps of the dispersion theoretic
derivation hold without this limitation; especially the validity of the
No-Subtraction hypothesis is guaranteed only without this restriction.
Concerning the low energy theorems we argue that the magnetic moment
of the nucleon in low electromagnetic order may deviate only
insignificantly from the measured one, so that in view of the
experimental errors of the verification of the \gdhsr the limitation
to low orders can be neglected here. 
This way we overcome a discussion of the validity of the
No-Subtraction hypothesis which so far has resisted attempts of an
interpretation in terms of the internal dynamics of the nucleon.

\subsection{Dispersion theoretic derivation}
\label{sec:dispersion}
The dispersion theoretic derivation exclusively relies on the
following assumptions: 
\begin{itemize}
\item Lorentz invariance
\item Gauge invariance
\item Crossing symmetry
\item Rotational invariance
\item Causality and
\item Unitarity
\end{itemize}
All our modern relativistic quantum field
theories rely on these principles. Hence a verification of the GDH sum
rule provides a vital cross check of the foundations of modern
physics. 

By means of crossing symmetry, rotational invariance and gauge
invariance the Compton forward scattering amplitude takes a very
simple form with analytical behavior. Causality leads to the analytic
continuation of the Compton forward scattering to complex values of
the photon energy which leads to the Kramers-Kronig dispersion
relation. The Kramers-Kronig dispersion relation connects the static
limit with the integral of the elastic scattering amplitude of all
energies. Elastic scattering (here, Compton scattering) is connected
to the total cross section by unitarity, the optical theorem.
Finally, the elastic scattering is connected to static properties of
the nucleon by means of low-energy limits following from gauge and
Lorentz invariance and crossing symmetry.
In the following we will outline this derivation in detail.

\subsubsection{Spin dependent Compton forward scattering amplitude}
\label{sec:ComptonForward}
The elastic scattering of light on elementary particles has been
a key subject in the course of the formulation of
particle physics especially for the electromagnetic
force. Compton scattering is the cornerstone also of the derivation of
the \gdhsr.

To discuss the spin content of general Compton scattering off
spin-1/2 systems with real or
virtual photons one uses helicity
amplitudes~\cite{Jacob:1959at} by choosing appropriate
photon and nucleon polarization states. We denote the helicity
amplitudes by $M_{\lambda',\nu';\lambda,\nu}$ with $\lambda,\lambda' =
\pm 1$ and $\nu,\nu' = \pm 1/2$. These 16 amplitudes depend on the
Lorentz-invariant Mandelstam variables s and t. By parity invariance
$M_{-\lambda',-\nu;-\lambda,-\nu} = (-1)^{\lambda-\nu-\lambda'+\nu'}
M_{\lambda',\nu';\lambda,\nu}$ only 8 amplitudes are
independent. Time-reversal invariance
$M_{\lambda',\nu;\lambda,\nu} = (-1)^{\lambda-\nu-\lambda'+\nu'}
M_{\lambda',\nu';\lambda,\nu}$ reduces the number of amplitudes to
6~\cite{Rollnik:1976} for which one may take 
\par\nobreak\noindent
\begin{eqnarray}
\phi_1 = M_{1 \frac{1}{2},1 \frac{1}{2}} \quad & \quad \phi_2 = M_{-1
-\frac{1}{2},1 \frac{1}{2}} \quad & \quad \phi_3 = M_{-1 \frac{1}{2},1
\frac{1}{2}} \nonumber \\
\phi_4 = M_{1 -\frac{1}{2},1 \frac{1}{2}} \quad & \quad \phi_5 = M_{1 -\frac{1}{2},1 -\frac{1}{2}} \quad & \quad \phi_6 = M_{-1 \frac{1}{2},1 -\frac{1}{2}}
\label{eqn:phi}
\end{eqnarray}
By angular momentum conservation in forward direction only $\phi_1$ and
$\phi_5$ can contribute.

The requirement of C, P and T invariance is somewhat cumbersome and
indeed if we restrict the discussion to real photons we can get over
it and instead use assumptions that are essential to other parts of
the derivation as well:
We use the special gauge useful for real photons with the time component
$\mathcal{A}^0=0$ of the photon field $\mathcal{A}$. 
Compton scattering is symmetric under the exchange of the in- and
outgoing photons ($\vec{k_1} \leftrightarrow - \vec{k_2}$ and
$\vec{\epsilon_1} \leftrightarrow 
\vec{\epsilon_2}^*$). $\vec{\epsilon}_{1,2}$ and $\vec{k_{1,2}}$ label
the initial and the 
final polarization of the Compton scattered photon and their momenta
respectively. 
This symmetry is called crossing-symmetry and is exact for all orders
of electromagnetic coupling. Therefore we have the following crossing
properties for the Compton amplitude
\be F(\vec{k_1},\vec{\epsilon}_1,\vec{\epsilon}_2) =
F^*(-\vec{k_2},\vec{\epsilon_2}^*,\vec{\epsilon_1}^*) \ . 
\label{eqn:crossing}
\ee
Due to the
superposition principle $F$ has to be linear in $\vec{\epsilon_1}$ and
$\vec{\epsilon_2}$. We now restrict the discussion to forward
scattering where $\vec{k} \equiv \vec{k_1} = \vec{k_2}$. In the
nucleon rest frame the amplitude
$F$ can be written as a linear combination with scalar functions $f_i$:
\par\nobreak\noindent
\begin{eqnarray}
F & = & \left< \chi_2^\dag \left| 
	\sum_{i=1}^5 f_i \  K_i 
	\right| \chi_1 \right>  \nonumber\\ & & \nonumber\\
	& = & \left< \chi_2^\dag \right| f_1 \  \left( \vec{\epsilon_2}^* \cdot
	\vec{\epsilon_1} \right) \qquad+\nonumber\\
	&  & \qquad f_2 \  i \vec{\sigma}\left(\vec{\epsilon_2}^* \times \vec{\epsilon_1}\right) \qquad+\nonumber\\
	&  & \qquad f_3 \  \left(\vec{\epsilon_2}^*\cdot\vec{k}\right)\left(\vec{\epsilon_1}\cdot\vec{k}\right) \qquad+\nonumber\\
	&  & \qquad f_4 \  
	i\vec{\sigma}\left[\vec{\epsilon_2}^*\left(\vec{\epsilon_1}\cdot\vec{k}\right)
	- 	\vec{\epsilon_1}\left(\vec{\epsilon_2}^*\cdot\vec{k}\right)\right] \quad+\nonumber\\
	&  & \qquad \left. \left. f_5 \  
	i\vec{\sigma}\left[\left(\vec{\epsilon_2}^*\times\vec{k}\right)\left(\vec{\epsilon_1}\cdot\vec{k}\right) -
	\left(\vec{\epsilon_1}\times\vec{k}\right)\left(\vec{\epsilon_2}^*\cdot\vec{k}\right)\right]
	\qquad \right| \chi_1 \right>
\label{eqn:ForwardSuper}
\end{eqnarray}
$\vec{\sigma}$ is the vector of Pauli spin matrices and $\chi_{1,2}$
are the initial and final spinors of the nucleon. With the
theory of the rotation group one can show that no more linearly
independent combinations can be found. 
The transversality condition for real photons reads $\vec{\epsilon}
\cdot \vec{k} = 0$ and one observes that only the first two terms
contribute. Hence, we can write the forward scattering
amplitude $F(\theta=0,\nu)$ with the photon energy $\nu$:
\par\nobreak\noindent
\begin{equation}
F(\theta=0,\nu) = \left< \left. \left. \chi_2^\dag \right| \  f_1(\nu)
	 \  \vec{\epsilon_2}^* \cdot \vec{\epsilon_1} +  f_2(\nu) \  i\vec{\sigma} \left(\vec{\epsilon_2}^*\times \vec{\epsilon_1}
	 \right) \  \right| \chi_1 \right>
\label{eqn:CompForw}
\end{equation}
The polarization vectors for left and right handed photons are 
\par\nobreak\noindent
\begin{equation}
  \vec{\epsilon}_R = -\frac{1}{\sqrt{2}} ( \vec{\epsilon}_x + i \vec{\epsilon}_y ) \quad, \qquad
  \vec{\epsilon}_L = +\frac{1}{\sqrt{2}} ( \vec{\epsilon}_x - i
  \vec{\epsilon}_y )
\label{eqn:leftrightpol}
\end{equation}
with the z-axis being the direction of motion of the photon. For the
two terms of Eq.~(\ref{eqn:CompForw}) one obtains the following
combinations: 
\par\nobreak\noindent
\begin{equation}
  \vec{\epsilon}_2^{\ \ast} \cdot \vec{\epsilon}_1 = 
  \left\{ 
  \begin{array}{r@{\quad:\quad}l} 1 & \vec{\epsilon}_1 = \vec{\epsilon}_2 = \vec{\epsilon}_R \\
				  1 & \vec{\epsilon}_1 = \vec{\epsilon}_2 = \vec{\epsilon}_L \\
				  0 & \text{else}
  \end{array}
  \right. \quad , 
\end{equation}
\begin{equation}
  \vec{\epsilon}_2^{\ \ast} \times \vec{\epsilon}_1 =
  \left\{
  \begin{array}{r@{\quad:\quad}l} 
	-i \vec{\epsilon}_z & \vec{\epsilon}_1 = \vec{\epsilon}_2 = \vec{\epsilon}_L \\
	+i \vec{\epsilon}_z & \vec{\epsilon}_1 = \vec{\epsilon}_2 = \vec{\epsilon}_R \\
	0 & \text{else}
  \end{array}
  \right.
\end{equation}
We can now compute the Compton amplitude for all possible spin
configurations. As it turns out, we only need to distinguish the
orientation\footnote{We have again used the usual convention to
observe the spins in the nucleon rest frame.} of nucleon and photon
spins in parallel~($3/2$) and anti-parallel~($1/2$):
\par\nobreak\noindent
\begin{equation}
\label{eqn:fpol}
  f_{3/2}(\nu ) = f_1(\nu ) - f_2(\nu ) \quad, \qquad
  f_{1/2}(\nu ) = f_1(\nu ) + f_2(\nu )
\end{equation}
$f_{3/2}$ and $f_{1/2}$ can be associated with $\phi_1$ and $\phi_5$
in Eq.~(\ref{eqn:phi}). 

For the treatment of analyticity in Sec.~\ref{sec:KramersKronig}
however, it is important to work with a set of amplitudes
free of kinematic singularities and 
zeros~(KSZF). It is not clear a priori if $f_1$ and $f_2$ fulfill
this requirement. The KSZF invariant amplitudes can be obtained by a
rather tedious construction procedure~\cite{Kellett:1973ea}. For their
definition one writes the general Compton amplitude like
$T = {e^\mu}^{^*}_{(\lambda')}{e^\nu}_{(\lambda)} T_{\mu\nu}$. The
tensor $T_{\mu\nu}$ may be expanded with respect to a tensor basis
$I^i_{\mu\nu}$:
\par\nobreak\noindent
\begin{equation}
T_{\mu\nu} = \sum_{i=1}^6 A_i I^i_{\mu\nu}
\end{equation}
where the KSZF invariant amplitudes $A_i$ depend on the Lorentz invariant
variables s, t and u. The explicit construction leads to the following
relation of the $A_i$ to the $f_{1,2}$ of Eq.~(\ref{eqn:CompForw}):
\par\nobreak\noindent
\begin{eqnarray}
f_1 & = & \frac{\nu^2 m}{4\pi}\left[ m A_2 + 2 A_3 \right] \nonumber
\\
f_2 & = & \frac{\nu^2 m}{4\pi} A_4
\label{eqn:KSZF}
\end{eqnarray}
with the nucleon mass $m$. Due to the work of
Ref.~\cite{Kellett:1973ea} we see that also $f_1, f_2$ and even 
$f_2/\nu$ are free of kinematic zeros and singularities\footnote{we
will need $f_2/\nu$ (not $f_2$ alone) to 
derive the \gdhsr.}. This is important as we will need the
analyticity of these functions of the photon energy on the real axis in
Sec.~\ref{sec:KramersKronig}. \\
In conclusion, we have shown that the forward Compton scattering takes
the very simple form of Eq.~(\ref{eqn:CompForw}) with only two
amplitudes that have the mathematical properties we will need in the
following section. 

\subsubsection{Kramers-Kronig dispersion relation}
\label{sec:KramersKronig}
Causality applied to scattering implies that the scattered wave $\psi_\text{scatt}(z,t)$
at time $t$ can be influenced by the incoming wave $\psi_\text{inc}(z',t')$ only at times $t'$
prior to $t$ with $z=z'$. The scattered wave depends linearly on the incoming
wave\footnote{The linearity is essentially Huygens' principle or the
superposition principle we have already used 
in Sec.~\ref{sec:ComptonForward}.}: 
\par\nobreak\noindent
\begin{equation}
\psi_\text{scatt}(z,t) = \int\limits_{-\infty}^{\infty}
dt' \  K(z'-z,t-t') \  \psi_\text{inc}(z',t') \qquad \text{with }
K(\zeta,\tau) = 0 \text{ for } \tau < \zeta.
\end{equation}
In the context of wave packets we can assume without loss of
generality that $\psi_\text{inc}(z,t) = 0$ for $z>t$. Then, due to the
characteristics of $K(\zeta,\tau)$, the same also holds for
$\psi_\text{scatt}$. We can now obtain the scattering amplitude as a
function of the energy $\nu$ or frequency $\omega = \nu/2\pi$ by
performing a Fourier transform of $\psi_\text{scatt}(z,t)$:
\par\nobreak\noindent
\begin{equation}
f_\text{scatt}(\omega) = \frac{1}{\sqrt{2\pi} } \int\limits_{-\infty}^z dt
\  \psi_\text{scatt}(z,t) \  e^{-i\omega (t-z)}
\end{equation}
By taking $z=0$ it is easy to demonstrate that
$f_\text{scatt}$ can be extended analytically into the full upper half
of the complex plane:
\par\nobreak\noindent
\begin{equation}
f_\text{scatt}(\omega+i\left|\gamma\right|) =
\frac{1}{\sqrt{2\pi}} \int\limits_{-\infty}^0 dt \  \psi_\text{scatt}(0,t)
\exp(-i\omega t-\left|\gamma\right|\left|t\right|) 
\end{equation}
Also, we have to keep in mind that $\psi_\text{scatt}$ is bounded which is
guaranteed by unitarity. For wave packets we have $t \to
-\infty: \psi_\text{scatt}(z,t) \to 0$. This in turn tells us that
for a given $\omega$ one can find a local
neighborhood of $\omega$ where $f_\text{scatt}$ can be continued into
the lower half of the complex plane. It shall be noted that
poles that stem for nucleon resonances do {\em not}\/ lie {\em on}\/ the
real axis but rather along the real axis --- with our conventions here in the
lower half of the complex plane.\footnote{Poles on the real axis 
-- often alleged in the literature -- by virtue of the optical theorem
would lead to divergences of the cross section and would make the
world a different place. Also the pole of the free nucleon on the real
axis at vanishing photon energy is kinematically suppressed
giving rise to the low-energy theorems (see Sec.~\ref{sec:low}). This
is a unique feature of forward scattering. The distance of poles
arising from resonances from the real axis reflects the widths of the
resonances.}  

Admittedly, the above motivation of the analyticity of the Compton
forward scattering is based on ideas of classical electrodynamics. A
derivation of the same can also be done in terms of quantum
fields~\cite{Gell-Mann:1954db}. Here causality implies the
vanishing of the commutator of two field operators $\phi(x)$ and
$\phi(x')$ if $x-x'$ is space-like. Later Goldberger was able to
generalize the argument leading to the derivation of the
Kramers-Kronig dispersion relation without the use of perturbation
theory~\cite{Goldberger:1955}. 

We are now in the position to apply Cauchy's integral formula:
\par
\begin{figure}
\begin{center}
	\includegraphics[width=0.7\textwidth]{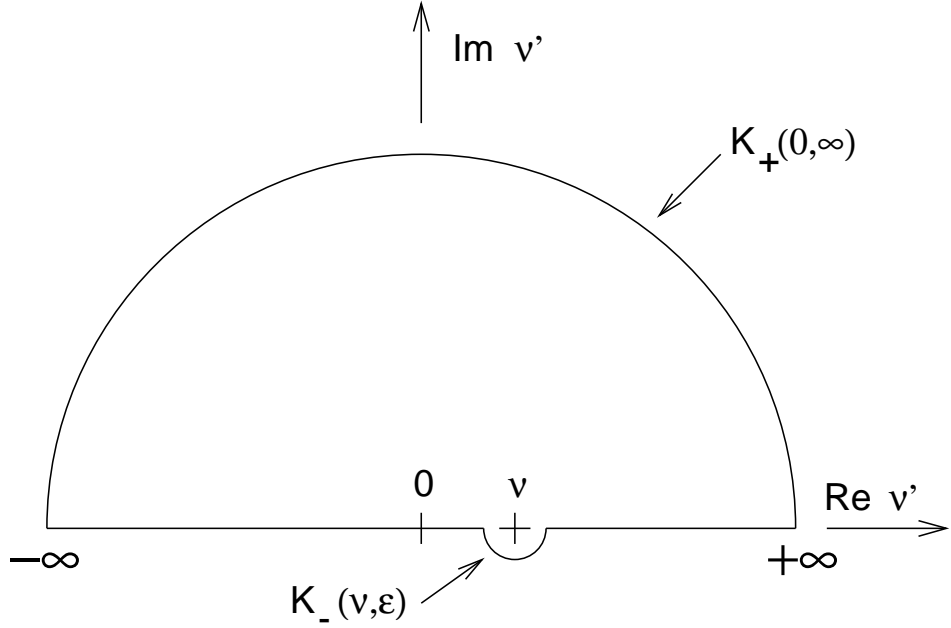}
	\caption{Path of integration for Cauchy's integral formula
	applied to the forward scattering amplitude}
	\label{fig:intweg}
\end{center}
\end{figure}

\nobreak\noindent
\begin{equation}
f(\nu) = \frac{1}{2\pi i}\oint\limits_{\cal C}d\nu'
\frac{f(\nu')}{\nu'-\nu}
\end{equation} 
We choose the path $\cal C$ as depicted in Fig.~\ref{fig:intweg}
where the integral is to be taken counter-clockwise. $K_+(0,\infty)$
is the half-circle at infinity in the upper half of the complex plane and
$K_-(\nu,\varepsilon)$ a small half circle around $\nu$ of radius
$\varepsilon$ in the lower half of the complex plane with the center
$\nu$ on the real axis.  

We can now evaluate the individual contributions from the segments of
the integration path:
\par\nobreak\noindent
\begin{eqnarray}
\label{eqn:intweg}
f(\nu) & = & \frac{1}{2\pi i}\; {\cal P}
\int\limits_{-\infty}^{\infty} d\nu' \frac{f(\nu')}{\nu'-\nu} \\
&& + \  \frac{1}{2\pi i}
\int\limits_{K_+(0,\infty)}d\nu'\frac{f(\nu')}{\nu'-\nu} \  + \
\lim_{\varepsilon\to 0} \  \frac{1}{2\pi i}\int\limits_{K_-(\nu,\varepsilon)} d\nu' 
\frac{f(\nu')}{\nu'-\nu}  \nonumber
\end{eqnarray}
with the Cauchy principle value
\par\nobreak\noindent
\begin{equation}
\;{\cal P} \int_{-\infty}^{\infty} \equiv \lim_{\varepsilon \to
0}\int_{-\infty}^{\nu-\varepsilon} + \int_{\nu+\varepsilon}^{\infty} \  .
\end{equation}
The integral for $K_-(\nu,\varepsilon)$ is simply half the residue we
would get for the full circle: $\frac{1}{2}f(\nu)$.  We {\em assume} 
that the integral along the path 
$K_+(0,\infty)$ vanishes which is called the No-Subtraction
hypothesis (see Sec.~\ref{sec:nosub}). One then obtains
\par\nobreak\noindent
\begin{equation}
f(\nu) = \frac{1}{\pi i} \  {\cal P} \int\limits_{-\infty}^{+\infty} d\nu' 
\frac{f(\nu')}{\nu'-\nu} \ .
\label{eqn:nosub}
\end{equation}
Recall the crossing properties of Compton scattering mentioned in
Sec.~\ref{sec:ComptonForward}. Applied to $F, f_1, f_2$ and $f_2/\nu$
we obtain 
\par\nobreak\noindent
\begin{equation}
F(\theta=0,-\nu) = F^*(\theta=0,\nu) \  , \quad 
f_1(-\nu) = f_1^*(\nu) \  , \quad 
f_2(-\nu) = -f_2^*(\nu) \  , \quad 
\frac{f_2(-\nu)}{-\nu} = \frac{f_2^*(\nu)}{\nu}
\end{equation}
As we will use Eq.~(\ref{eqn:nosub}) only for $f_1$ and $f_2/\nu$
and not for $f_2$ we can continue with the crossing relation for $f:
f(-\nu) = f^*(\nu)$. We can now write Eq.~(\ref{eqn:nosub}) as:
\par\nobreak\noindent
\begin{equation}
f(\nu) = \frac{1}{\pi i} \  {\cal P} \int\limits_0^{+\infty}d\nu'
\left( \frac{f(\nu')}{\nu'-\nu} + \frac{f^*(\nu')}{-\nu'-\nu} \right)
\label{eqn:positive}
\end{equation}
Considering the real part only, this further simplifies to the famous
Kramers-Kronig dispersion relation:
\par\nobreak\noindent
\begin{equation}
\label{eqn:dispersion}
\text{Re} \, f(\nu) = \frac{2}{\pi} \  {\cal P} \int\limits_{0}^{\infty}\,
d\nu' \, \nu' \  \frac{\text{Im}\; f(\nu')}{\nu'^2 - \nu^2}
\end{equation}
\subsubsection{Optical theorem}
\label{sec:OptTheo}
The optical theorem can be derived from probability current
conservation which is also called unitarity. It connects the elastic
forward amplitude to the 
total cross section:
\par\nobreak\noindent
\begin{equation}
\label{eqn:optical}
\text{Im}\; f(\nu) = \frac{\nu}{4\pi}\sigma
\end{equation}
For the amplitudes $f_{1,2}$ as defined in
Eq.~(\ref{eqn:CompForw}) and~(\ref{eqn:KSZF}) one
obtains\footnote{remember that we have derived the Kramers-Kronig
dispersion relation for $f_1$ and $f_2/\nu$}
\par\nobreak\noindent
\begin{eqnarray}
\label{eqn:Opticf1}
\text{Im } f_1(\nu) & = & \frac{\nu}{8\pi} \left[ \sigma_{1/2}(\nu) +
\sigma_{3/2}(\nu)\right] =  \frac{\nu}{4\pi}\sigma_T(\nu) \  , \\
\text{Im } \frac{f_2(\nu)}{\nu} & = & \frac{1}{8\pi} \left[
\sigma_{1/2}(\nu) - \sigma_{3/2}(\nu)\right] =
\frac{1}{4\pi}\sigma_{TT}(\nu) \  .
\label{eqn:Opticf2}
\end{eqnarray}
For the transverse polarization of the photon the subscripts of the
total cross sections $\sigma_{3/2}$ and $\sigma_{1/2}$ denote the
total helicity of the photon-nucleon system in the nucleon rest frame
with respect to the center of mass momentum. 
The symbols $\sigma_{TT}$ and $\sigma_T$ are commonly used in electron
scattering or better to say virtual Compton scattering (see Sec.~\ref{sec:PolAbs}).

Here, it is important to understand that the optical theorem is a
statement for the total cross section including both elastic and inelastic
contributions: $\sigma = \sigma_\text{elast} +
\sigma_\text{inel}$. Actually a sizable fraction of text books get
it wrong and claim the theorem relates the elastic forward scattering
to the inelastic part of the cross section only. 
For the special case of photon scattering the optical theorem relates
the imaginary part of the Compton forward scattering amplitude to the
total cross section for photon scattering --- that is photoabsorption
and Compton scattering. 

However, if one considers the Compton forward scattering amplitude only
up to the lowest non-trivial order in
electromagnetic coupling
\par\nobreak\noindent
\begin{equation} \label{eqn:fBorn}
f_\text{low} = i \  \frac{e^2}{16\pi M}\,\epsilon_\mu^*\epsilon_\nu
\int\! d^4x\, e^{ik\cdot x}
\sum\limits_{X_\text{had}}  \left< p_2,s_2 \left| {J^\mu(x) \left|X_\text{had}\right>
\left< X_\text{had} \right| J^\nu(0)} \right| p_1,s_1 \right>
\end{equation}
with the sum over purely hadronic intermediate states $X_\text{had}$
only the optical theorem instead reads 
\par\nobreak\noindent
\begin{equation}
\text{Im}\; f_\text{low}(\nu) =
\frac{\nu}{4\pi}\sigma_{\gamma\text{-abs}} 
\label{eqn:AmpliBorn}
\end{equation}
with $\sigma_{\gamma\text{-abs}}$ denoting the photoabsorption cross
section only.
$J^\mu$ is the electromagnetic current of the photon field, $p_{1,2}$
are the initial and final nucleon four-momenta, $s_{1,2}$ the spins.
So, to lowest non-vanishing order the optical theorem relates the
elastic forward Compton scattering amplitude to the photoabsorption
cross section only. It can be proven rigorously that the
optical theorem even holds for all orders of coupling individually.

In the language of Feynman diagrams the optical theorem and the
question of inclusion of elastic processes in the cross section can be
understood graphically in terms of cutting through the diagram for the
amplitude of the elastic process: 
~\vspace{3mm}\par\nobreak\noindent
\begin{fmffile}{fd}
\begin{equation}
\label{eqn:OpticBorn}
  {\Large\text{Im}_{\theta=0}} \left( \quad\ \  
  \graphs(100,62){
   \fmfleft{i1,i2} \fmfright{o1,o2}  \fmfbottom{b} \fmftop{t}
   \fmf{dashes}{t,b}
   \fmflabel{$\gamma$}{i2} \fmflabel{$\gamma$}{o2}
   \fmflabel{n,p}{i1} \fmflabel{n,p}{o1}
   \fmf{fermion,width=thick,tension=3}{i1,v1} 
   \fmf{phantom,tension=3.5}{b,h1} \fmf{phantom,tension=2}{h1,t}
   \fmf{phantom,tension=2}{b,h2} \fmf{phantom,tension=3.5}{h2,t}
   \fmf{phantom}{v1,v2}
   \fmf{fermion,width=thick,tension=3}{v2,o1}
   \fmf{photon,tension=3}{i2,v1}  \fmf{photon,tension=3}{v2,o2}
   \fmfdot{v1,v2}
   \fmflabel{\scriptsize cut}{t}
   \fmfpoly{square,smooth,pull=1.5,tension=0.5,filled=10,label={\Large\bf X}}{v1,h1,v2,h2} 
  } \quad\ \  \right)
	\quad  = \quad 
  \left| \quad
  \graphs(100,62){
  \fmfleft{i1,i2} \fmfright{o} \fmfbottom{b} \fmftop{t}
  \fmflabel{$\gamma$}{i2}
  \fmflabel{n,p}{i1}
  \fmf{fermion,width=thick}{i1,v}
  \fmf{boson}{i2,v}
  \fmf{phantom,tension=1.5}{v,o}
  \fmfblob{0.15w}{v}
  \fmf{phantom}{t,a1} \fmf{plain,tension=6}{a1,a2} \fmf{plain,tension=2}{a2,o}
  \fmf{phantom}{b,a3} \fmf{plain,tension=6}{a3,a4} \fmf{plain,tension=2}{a4,o}
  \fmfdot{v}
  \fmffreeze
  \fmf{plain}{a2,v,a4}
  \fmflabel{\Large\bf X}{o}
  } \quad \quad \right|^{\mathbf{2}}
\end{equation}
\end{fmffile}
~\vspace{3mm}\par\nobreak\noindent
The left hand side of Eq.~(\ref{eqn:OpticBorn}) represents only the
contribution from the lowest order in electromagnetic coupling like in
Eq.~(\ref{eqn:AmpliBorn}). Since there are 
no photons allowed 
in the intermediate state for the Compton amplitude the corresponding
cross section has no elastic photon in the final state. X symbolizes
states originating from strong interactions like resonance excitation.
However,
radiative corrections in higher orders change this
picture. Eq.~(\ref{eqn:OpticRadi}) shows a typical example.
~\vspace{3mm}\par\nobreak\noindent
\begin{fmffile}{ff}
\begin{eqnarray}
  {\Large\text{Im}_{\theta=0}} \left( \quad\ \  
  \graphs(100,62){
   \fmfleft{i1,i2} \fmfright{o1,o2}  \fmfbottom{b} \fmftop{t}
   \fmf{dashes}{t,b}
   \fmflabel{\scriptsize cut}{t}
   \fmflabel{$\gamma$}{i2} \fmflabel{$\gamma$}{o2}
   \fmflabel{n,p}{i1} \fmflabel{n,p}{o1}
   \fmf{fermion,width=thick}{i1,v1}
   \fmf{plain,width=thick,tension=2}{v1,v3}
   \fmf{plain,width=thick}{v3,v4}
   \fmf{plain,width=thick,tension=2}{v4,v2}
   \fmf{fermion,width=thick}{v2,o1} 
   \fmf{photon,right,tension=0}{v4,v3}
   \fmf{photon}{i2,v1}  \fmf{photon}{v2,o2}
   \fmfdot{v1,v2,v3,v4}
  } \quad\ \  \right)
	\quad  = \quad 
  \left| \quad
  \graphs(100,62){
   \fmfleft{i1,i2} \fmfright{o1,o2}  \fmfbottom{b} \fmftop{t}
   \fmflabel{$\gamma$}{i2} \fmflabel{$\gamma$}{o2}
   \fmflabel{n,p}{i1} \fmflabel{n,p}{o1}
   \fmf{fermion,width=thick}{i1,v1}
   \fmf{plain,width=thick}{v1,v2}
   \fmf{fermion,width=thick}{v2,o1} 
   \fmf{photon}{i2,v1}  \fmf{photon}{v2,o2}
   \fmfdot{v1,v2}
  } \quad \right|^{\mathbf{2}}
\label{eqn:OpticRadi}
\end{eqnarray}
\end{fmffile}
~\vspace{3mm}\par\nobreak\noindent
In conclusion, we have two options here to choose from. While Compton
scattering to lowest non-trivial order in principle is not an
experimental observable it is related to the total photoabsorption
cross section which indeed is an observable. Actually, 
experimentally it is easier to measure the photoabsorption
cross section only, excluding the elastic part. This is indeed what
constitutes the experimental data of Sec.~\ref{sec:results}. We will
come back to this feature when we discuss the Low-energy theorem and
when we evaluate the credibility of the No-Subtraction hypothesis.

\subsubsection{Low-energy theorem}
\label{sec:low}
Within modern particle physics among the first important achievements
was the proof of the Low-energy theorem in Compton scattering by
Thirring in 1950~\cite{Thirring:1950}. According to this theorem the
Thomson formula 
\par\nobreak\noindent
\begin{equation}
\sigma_\text{Thomson} = \frac{8\pi}{3} \left( \frac{e^2}{m_e} \right)^2
\end{equation}
is exactly valid at threshold --- to any order in the electromagnetic coupling
--- if $e$ and $m_e$ are interpreted as the renormalized charge and
electron mass. 

A generalization of this result was obtained in 1954 by
Low~\cite{Low:1954kd} and Gell-Mann
and Goldberger~\cite{Gell-Mann:1954kc}.
Both papers appeared face to face in the respective Physical
Review volume.
The derivation is done for scattering off spin-1/2 systems without
specific assumptions about a possible substructure of the system.
The proof is based on Lorentz invariance, gauge invariance and
crossing symmetry only. It can be used directly also for the
Compton scattering off strongly interacting particles, especially off
protons and neutrons with the appropriate values
for the mass, charge and the anomalous moment $\kappa$.
Strong interactions and the substructure modify the response of the
nucleon in photon scattering with respect to the expected behavior
of a point-like spin-1/2. Manifestations are the values of nucleon
magnetic moments: The proton and neutron moments ($\mu_p = 2.79\,\mu_N$,
$\mu_n = -1.91\,\mu_N$ with $\mu_N = e\hbar/2m_N$) are not at all close to the
expectations ($1.0\,\mu_N$, $0.0\,\mu_N$) based on the Dirac equation.

We can sketch the derivation of the Low-energy theorem based on the
principle of minimal coupling including the anomalous
moment~\cite{Gell-Mann:1956}. The modified Dirac equation for the 
nucleon then reads
\par\nobreak\noindent
\begin{equation}
\left( \gamma^\mu \left(i \partial_\mu - eA_\mu \right) -
\frac{\kappa\mu_N}{2} \sigma_{\mu\nu}F^{\mu\nu} - m_N \right) \psi =
0 \quad ,
\label{eqn:anomaldirac}
\end{equation}
with $\kappa_p = 1.79$ and $\kappa_n = -1.91$ for the anomalous
magnetic moments.
The Compton scattering amplitude up to lowest non-trivial order in
electromagnetic coupling is  
\par\nobreak\noindent
\begin{eqnarray}
\label{eqn:lowestorder}
f&=&\alpha \frac{|k_2|}{|k_1|} \bar{u}(p_f,s_f) \left[ \left( \not\!\epsilon_2+
\frac{i\kappa\mu_N}{2} \sigma_{\mu\nu} ( \epsilon_2^\nu k_2^\mu -
\epsilon_2^\mu k_2^\nu ) \right) \right. \\
&& \cdot\frac{1}{\not\!p_i + \not\!k - m_N} \left. \cdot \left(\not\!\epsilon_1+
\frac{i\kappa\mu_N}{2} \sigma_{\mu\nu} (\epsilon_1^\nu k_1^\mu -
\epsilon_1^\mu k_1^\nu ) \right) + \mbox{crossed} \right] u(p_i,s_i) 
\quad . \nonumber
\end{eqnarray}
with the polarization and momentum 4-vectors for initial and final state
$\epsilon_1,\epsilon_2$ and $k_1,k_2$.
Up to expressions linear in $k$ we get
\par\nobreak\noindent
\begin{eqnarray}
\label{eqn:compton} 
f & = & -\frac{e^2}{m_N} (\vec{\epsilon}_2\cdot{\vec{\epsilon}_1}) -
2i\mu^2 \left| \vec{k}_1 \right| \vec{\sigma} \cdot \left[
(\vec{n}_2\times\vec{\epsilon}_2) \times 
(\vec{n}_1\times\vec{\epsilon}_1) \right] \nonumber\\
&& - \frac{ie\mu}{m_N} \left| \vec{k}_1 \right| \left[ \vec{\sigma}\cdot\left\{
\frac{\vec{n}_1(\vec{n}_1 \times \vec{\epsilon}_1) + (\vec{n}_1\times\vec{\epsilon}_1)\vec{n}_1}{2} \right\}
\cdot \vec{\epsilon_2} +\mbox{crossed} \right] \\
&& + \frac{ie\mu_A}{m_n} \left| \vec{k}_1 \right| \ \vec{\sigma}\cdot(\vec{\epsilon}_2 \times
\vec{\epsilon}_1) \quad , \nonumber
\end{eqnarray}
with $\vec{n}_{1,2} = \vec{k}_{1,2}/\left|\vec{k}_{1,2}\right|$.
The first term is the classical Thomson scattering. In forward direction
($\vec{k}_1 = \vec{k}_2$) the expression simplifies considerably:
\par\nobreak\noindent
\begin{equation}
f(\nu) = -\frac{\alpha}{m}\vec{\epsilon^*_2}\cdot\vec{\epsilon_1} -
\frac{\alpha}{2m^2}\kappa^2\nu
i\:\vec{\sigma}\cdot(\vec{\epsilon^*_2}\times\vec{\epsilon_1}) 
\label{eqn:low}
\end{equation}
While the extension of the Dirac Eq.~(\ref{eqn:anomaldirac})
is a rather heuristic approach F.E.~Low~\cite{Low:1954kd} and
M.~Gell-Mann and M.L.Goldberger~\cite{Gell-Mann:1954kc} present a
rigorous proof in field theory. Both works, however, are limited to
lowest-order in electromagnetic coupling like the derivation above. 
The result is a low energy expansion in the photon energy $\nu$:
\beqn
f_1(\nu) & = & -\frac{\alpha}{m} + (\alpha_E+\beta_M)
\,\nu^2+ {\mathcal{O}}(\nu^4) \ , \label{eqn:Lowf1} \\
f_2(\nu)/\nu & = & -\frac{\alpha}{2m^2}\kappa^2_N + \gamma_0\nu^2 + {\mathcal{O}}(\nu^4)
\ . \label{eqn:Lowf2}
\eeqn
Observe that due to the crossing relation Eq.~(\ref{eqn:crossing}) $f_1$
is an even and $f_2$ is an odd function of $\nu$.
The leading term of the spin-independent amplitude, $f_1(0)$, is the
Thomson term. All odd terms vanish because of
crossing symmetry. The term ${\mathcal{O}}(\nu^2)$ 
describes Rayleigh scattering and reveal information on the internal nucleon
structure through the electric ($\alpha_E$) and magnetic ($\beta_M$) dipole
polarizabilities. In the case of the
spin-flip amplitude $f_2/\nu$, the leading term is determined by the
anomalous magnetic moment. The term quadratic in the photon energy
$\nu$ is connected to the spin structure through
the forward spin polarizability $\gamma_0$. 

At this point it is important to understand that these low energy
theorems do {\em not}\/ rest upon the assumption that the nucleon can be
treated like a fundamental point like particle without
substructure. Gell-Mann and Goldberger present three alternative
derivations of the low energy theorem~\cite{Gell-Mann:1956}. The derivation
they call the ``Classical calculation'' is most explicit in this
respect. The anomalous magnetic moment accounts for the 
$E1$ scattering while the total magnetic moment is responsible
for $M1$ scattering as well as the absorption of $E1$ radiation and emission
of $M2$ radiation and the reverse process.
The magnetic moment of the nucleon interacts with the
gradient of the magnetic field of the photon and the finite size of the
nucleon is irrelevant since terms quadratic in the field strengths
are dropped. The restriction to terms linear in the field strengths is
legitimate because we are considering the limit of vanishing 
energies of the fields. 

\subsubsection{Synthesis}
We can now connect the static properties of
Eqs.~(\ref{eqn:Lowf1}) and (\ref{eqn:Lowf2}) via the dispersion
relation~(\ref{eqn:dispersion}) with the cross sections of
Eqs.~(\ref{eqn:Opticf1}) and (\ref{eqn:Opticf2}).  
To compare the Kramers-Kronig relation with the low energy expansion
we write it as Taylor series -- here applied to the two relevant
amplitudes $f_1(\nu)$ and $f_2(\nu)/\nu$ with the optical theorem
already incorporated:
\par\nobreak\noindent
\begin{eqnarray}
{\mbox{Re}}\ f_1(\nu) & = & 
\frac{1}{2\pi^2} \sum_{n=0}^\infty
\left[ \int\limits_{0}^{\infty}d\nu'\,
\sigma_T(\nu')\ \left( \frac{\nu}{\nu'} \right)^{2n} \right]
\label{eqn:KKf1} \\
{\mbox{Re}}\ \frac{f_2(\nu)}{\nu} & = &
 \frac{1}{4\pi^2} \sum_{n=0}^{\infty}
\left[\int\limits_{0}^{\infty} \frac{d\nu'}{\nu'}
\left(\sigma_{1/2}(\nu')-\sigma_{3/2}(\nu')\right)
\left( \frac{\nu}{\nu'}\right)^{2n}
 \right ]
\ . \label{eqn:KKf2}
\end{eqnarray}
Due to the crossing relation (Eq.~(\ref{eqn:crossing})) $f_1$ and
$f_2/\nu$ are even functions of $\nu$. So, only the even terms in
the Taylor expansions are accounted for in Eqs.~(\ref{eqn:KKf1},\ref{eqn:KKf2}).  \\
In particular for the leading term for $f_1$
we obtain Baldin's sum rule~\cite{Bal60} for the electric and magnetic
polarizabilities $\alpha_E + \beta_M$,
\par\nobreak\noindent
\begin{equation}
\alpha_E + \beta_M = \frac{1}{2\pi^2}\,
\int\limits_{0}^{\infty}\,\frac{\sigma_T(\nu')}{\nu'^2} \,d\nu'\ ,
\end{equation}
the \gdhsr,
\be
\label{eqn:GDHsr} \boxed{\boldsymbol{ 
\frac{\alpha\kappa^2_N}{2m^2} = \frac{1}{4\pi^2}
\int\limits_{0}^{\infty}\,\frac{\sigma_{3/2}(\nu') -\sigma_{1/2}(\nu')}{\nu'}\,d\nu' \,} ,}
\ee
and the forward spin polarizability $\gamma_0$~\cite{Gell-Mann:1954kc,Gell-Mann:1954db},
\par\nobreak\noindent
\begin{equation} \gamma_0=
\,-\,\frac{1}{4\pi^2}\,\int\limits_{0}^{\infty}\, 
\frac{\sigma_{3/2}(\nu')-\sigma_{1/2}(\nu')} {\nu'^3}\,d\nu'\ . 
\end{equation}
With Eq.~(\ref{eqn:GDHsr}) we have finally arrived at the GDH sum
rule. For the derivation we have used exclusively Lorentz invariance, gauge
invariance, 
causality and unitarity. 

The \gdhsr can even be established for
the deuteron with the appropriate anomalous magnetic moment as a
generalization of the derivation to spin-1 systems. The compositeness
of the deuteron complicates the measurement as photo-disintegration
has to be taken into account. Still, the \gdhsr for the deuteron is of
fundamental character as the finite size and compositeness of the
deuteron do not impair the validity of the Low-energy theorem. An
experimental verification, however, cannot reach the precision
achievable for the proton. 

\subsection{Equal-times and light-cone current algebra derivations}
\label{sec:EqualTimes}
In this section we will only sketch the derivations based on current algebra and
outline the assumptions and some intermediate steps needed to discuss the virtues of
these alternative approaches. 

\subsubsection{Equal-times current algebra}
\label{sec:EqTsub}
The general idea for this derivation can be found in Ref.~\cite{Hosoda:1966}
and a more detailed calculation in Ref.~\cite{Pantforder:1998nb}. 
The current density originating from a Dirac field $\psi(x)$ has the form
\par\nobreak\noindent
\begin{equation} \label{curr-def1}
 J^\mu(x) = \bar\psi(x) \gamma^\mu \psi(x)
 = \psi^\dagger(x) \gamma^0\gamma^\mu \psi(x).
\end{equation}
The central assumption of this type of algebra is that at equal
times, the charge density commutes with each component of the current
density:
\par\nobreak\noindent
\begin{equation} \label{eqn:eqt-comm}
\left[J^0(x),J^\mu(y)\right] = 0 \quad \text{for} \quad x^0=y^0.
\end{equation}
The commutator of electric dipole moments $D(0)$ then also vanishes:
\par\nobreak\noindent
\begin{equation} \label{eqn:DipolComm}
\left[ D^i(0), D^j(0) \right] = 0 \quad \text{with the definition}
\quad D^i(x^0) = e \!\int d^3\boldsymbol{x} \ x^i J^0(x).
\end{equation}
In analogy to Eq.~(\ref{eqn:leftrightpol}) one can define dipole
operators corresponding to left and right handed circularly polarized
photons:
\par\nobreak\noindent
\begin{equation} \label{LR-def}
 D^{\text{R},\text{L}}(0) = \frac{1}{\sqrt2}\, \left( D^1(0) \pm
 iD^2(0) \right) .
\end{equation}
One can now apply the commutator of the dipole operator to the nucleon
with equal initial and final spin and momentum $p^\mu = \left( E^0, 0,
0, \sqrt{(E^0)^2 - M^2} \right)$: 
\par\nobreak\noindent
\begin{equation} \label{naive-comm-me}
0 = \left< p', \frac{1}{2} \right| \left[ D^\text{L}(0), D^\text{R}(0)
\right] \left|
p, \frac{1}{2} \right> = 0.
\end{equation}
Like in Eq.~(\ref{eqn:fBorn}) a (not quite) complete set of
intermediate states with all purely hadronic intermediate states is
inserted. Again radiative corrections are disregarded. The result for
the one-nucleon (1-N) state reads 

\begin{align} \label{comm-1N}
 \left<p', \frac{1}{2} \right| \left[ D^\text{L}(0), D^\text{R}(0)
 \right]  \left|p, \frac{1}{2}\right>_{\text{1-N}}
 = (2\pi)^3\, 2p^0\, \delta(\boldsymbol p' - \boldsymbol p)
     \left( \frac{2\pi\alpha\kappa^2}{M^2} -
     \frac{2\pi\alpha(1+\kappa)^2}{(p^0)^2}\right).
\end{align}

while the contribution from all other hadronic intermediate states
(hadr) -- also called continuum contribution -- is
\par\nobreak\noindent
\begin{equation} \label{comm-cont}
 \left<p', \frac{1}{2} \right| \left[ D^\text{L}(0), D^\text{R}(0)
 \right]  \left|p, \frac{1}{2}\right>_{\text{hadr}} =
 (2\pi)^3\, 2p^0\, \delta(\boldsymbol{p' - p}) \,
 8\!\!\int\limits_{\nu_{\text{thr}}}^\infty\! \frac{d \nu}{\nu}\,
 \text{Im} f_2\left( \nu,\frac{M^2\nu^2}{(p^0)^2} \right). 
\end{equation}
with the photon virtuality $q^2 = M^2\nu^2/(p^0)^2$.
To obtain the usual form of the sum rule one takes the limit $q^0 \to
\infty$. Then one has to interchange taking the limit with the integration over
$\nu$ which is the second crucial main assumption. Of course, the
optical theorem applies once again and one obtains the GDH sum
rule. 

Both assumptions for this derivation -- the vanishing of the
equal-times commutator and the legitimacy of taking the infinite
momentum limit by interchanging it with the integration -- have been
questioned in the  literature. Electric charge density commuting at
equal times has been challenged by
Refs.~\cite{Kawarabayashi:1966,Khare:1974gr,Chang:1994td}.
Kawarabayashi and Suzuki as well as Chang, Liang, and Workman explicitly
bring up the question whether an anomaly of this commutator gives rise
to a 
modification of the \gdhsr. However, Pantförder, Rollnik and
Pfeil~\cite{Pantforder:1997ii} have shown that, at least in the
Weinberg-Salam model for photon scattering off electrons, up to order 
$\alpha^2$ the very same graphs that give rise to the anomaly of the
electric charge density commutator also prevent dragging the naive infinite
momentum limit as described above. Actually, both modifications cancel
exactly. 

\subsubsection{Light-cone current algebra}
In 1972, a few years after the derivations using dispersion theory and
equal-times commutator algebra, Dicus and Palmer~\cite{Dicus:1972vp}
used the algebra of currents on the light-cone for an alternative proof of
the \gdhsr. Here we recall the principle idea.

Light-cone coordinates $r^\pm, \boldsymbol{r}_\perp$ are defined as 
\par\nobreak\noindent
\begin{equation}
 r^\pm = \frac1{\sqrt{2}}\, (r^0 \pm r^3), \quad  \boldsymbol{r}_\perp = (r^1,r^2).
\end{equation}
Similar to the derivation with the equal-times commutator one assumes
a vanishing commutator of charge densities:
\par\nobreak\noindent
\begin{equation}
 \left[ J^+(x), J^+(y) \right] = 0 \quad \text{with} \quad x^+=y^+
\end{equation}
\par\nobreak\noindent
For a proof of this commutator relation on the light-cone see
Ref.~\cite{Harindranath:1994xm}. 
Now, one again defines a first moment, this time of $J^+(x)$ and
sandwiches the commutator of the left and right-handed dipole moments
with a complete set of hadronic intermediate states. With the
separation of the one-nucleon state one obtains
\par\nobreak\noindent
\begin{equation}
 (2\pi)^3\, 2p^+\, \delta(p'^+ - p^+)\,
   \delta(\boldsymbol{p}'_\perp)\, \frac{2\pi\alpha\kappa^2}{M^2} = 
 (2\pi)^3\, 2p^+\, \delta(p'^+ - p^+)\,
   \delta(\boldsymbol{p}'_\perp) \  8\!\!\int\limits_{q^-_\text{thr}}^\infty\!
   \frac{dq^-}{q^-}\, \text{Im} f_2(\nu,0).
\end{equation}
The one-nucleon state determines the left hand side and all
other hadronic intermediate states the right hand side. With $\nu = p\cdot
q/M = p^+q^-/M$ and the optical theorem for $\text{Im} f_2(x)$ the GDH
sum rule follows. \\

The derivation using light-cone algebra is more straight forward than
the one based on equal-times algebra as we don't have to deal with the
infinite-momentum limit.
The weakest point of both approaches is the assumption of a vanishing
commutator. Both original authors reflect this circumstance. Hosoda
and Yamamoto in Ref.~\cite{Hosoda:1966} write 
``We do not know of a general proof of Eq.~(1)\footnote{corresponds to our
Eq.~(\ref{eqn:DipolComm})}. But we also do not know of a counter
example for Eq.~(1).'' Dicus and Palmer~\cite{Dicus:1972vp} even
suggest a non-vanishing form of the commutator for the light-cone algebra.

\subsection{Analogies of current algebra and dispersion
theoretic proofs}
In both algebra derivations the vanishing of a specific current density
commutator is the starting point. It ultimately allows to connect the
static properties of the nucleon calculated from the one-nucleon
contribution to the integral of the total photoabsorption cross
section. The vanishing of the commutator is motivated by causality
arguments and by employing canonical anticommutator
relations~\cite{Harindranath:1994xm}.\\ 
The origin of the Kramers-Kronig dispersion relation is similar to the
vanishing commutator of charge densities and it
has the same virtues, namely connecting the static properties to the
dynamic observables. A potential failure of the No-Subtraction
hypothesis would probably be related to anomalies of the above
commutators.

The ``one-nucleon'' contributions calculated for the current density
algebra derivations are similar to the Low-energy theorem -- even to
the extent that the proofs do not address radiative corrections (see
Sec.~\ref{sec:LowTheorem}).  
Finally, the optical theorem is used in the same way for all 3 derivations.

\subsection{Potential challenges of the \gdhsr}
\subsubsection{Low-energy theorem and its validity at higher orders of
coupling}
\label{sec:LowTheorem}
The derivation of the Low-energy theorem by F.~E.~Low~\cite{Low:1954kd}
and also by M.~Gell-Mann and M.~L.~Goldberger~\cite{Gell-Mann:1954kc}
is done only up to lowest none-trivial order in electromagnetic
coupling. Another derivation is provided by H.~D.~I.~Abarbanel and
M.~L.~Goldberger~\cite{Abarbanel:1967wk} which clarifies the
assumptions but is also limited to the same order of
electromagnetic coupling while strong interactions are included with
all orders. The background of this limitation is that the derivations
rest on the crucial assumption that the single-particle intermediate
state is separated from the multiparticle states (continuum) by a
finite energy gap. Thus the presence of intermediate soft photons
would invalidate this assumption. However, Roy and
Singh~\cite{Roy:1968} and later also T.~P.~Cheng~\cite{Cheng:1970cy}
were able to overcome this limitation and established the low theorem
up to the order $\alpha^2$. Consequently, the anomalous magnetic
moment $\mu_N$ in Eq.~(\ref{eqn:Lowf2}) is not the observed anomalous
magnetic moment of the nucleon but rather a theoretical one limited in
the electromagnetic coupling. Briefly, we want to discuss if this
may be a relevant limitation in the present context.

In order to explain the spectra of atoms in magnetic fields, Uhlenbeck
and Goudsmit~\cite{Uhlenbeck:1926} postulated that the electron has an intrinsic (spin)
angular momentum $\hbar/2$ and a magnetic dipole moment $e\hbar/2mc$, the Bohr
magneton. Later, Dirac showed that both properties of the electron are
the consequences of relativistically invariant quantum
mechanics~\cite{Dirac:1928}. The magnitude of the electron magnetic
dipole moment is $e\hbar/m$, that is, the Lande
g-factor for electrons is 2. 
As in the
case of the Lamb shift, radiative corrections give a small departure
from this prediction. Schwinger calculated the anomaly $(g-2)/2$ as
$\alpha/2\pi \simeq 0.00116$~\cite{Schwinger:1951nm}. 
Like the Lamb shift, the anomalous magnetic moment of the electron also
provides one of the most sensitive tests of QED. The accurately
measured value of the anomalous magnetic moment today is
$0.001159652187(4)$~\cite{Eidelman:2004wy} which is an excellent proof
of QED.

Quantitatively, the anomaly of the magnetic moment of the
electron due to radiative corrections is about $0.1~\%$ compared to the
the total magnetic moment. It is suggestive to assume a similar
approximative equality of lowest-order (or next-to-lowest-order) to
the all-orders anomalous magnetic moment also for the
nucleon. Especially since the effect of radiative corrections 
is to be compared to the experimental accuracy to measure the right
hand side of the \gdhsr (Eq.~(\ref{eqn:GDHsr})). This
accuracy is of the order of several percent only. It is also
suggestive to assume that the Low-energy theorem is true to all orders
as is the case for the Thomson limit. Hence for the time being, it
appears safe to ignore this issue within the experimental context as
it presents a negligible systematic uncertainty.

However, it is important to keep this restriction of some derivations
in mind as the discussion in Sec.~\ref{sec:EqualTimes} has shown that
this limitation may entail unphysical anomalies that lead to further
complications. And indeed we will come back to this issue of all
orders versus low orders in our discussion of the No-Subtraction
hypothesis in Sec.~\ref{sec:nosub}.  

\subsubsection{Convergence}
\label{sec:Convergence}
There are two issues with the convergence of the GDH integral i.e. the
right hand side of Eq.~(\ref{eqn:GDHsr}): The saturation for the part
going to infinity but also the part below the pion production
threshold down to zero due to the $1/\nu$ weighting.

The \gdhsr is a ``superconvergence'' relation~\cite{Trueman:1966}.
The solution to the question of the saturation of the high energy part
has been provided already in 1967 by A.~H.~Mueller and
T.~L.~Trueman~\cite{Mueller:1967}. They present a 
proof based
on Regge theory and  show that --- unlike in the
unpolarized case --- the Pomeranchuk trajectory is not relevant here and the
polarized cross section difference drops off at least like
\par\nobreak\noindent
\begin{equation}
\lim_{\nu\to\infty}\sigma_{3/2} - \sigma_{1/2} \le 
C \cdot \lim_{\nu\to\infty} 1/\left(\ln^2 \nu\right).
\end{equation}
This ensures the convergence of the GDH integral whatever sign the
constant $C$ might have. De Alfaro, Fubini, Rossetti and
Furlan~\cite{Alfaro:1966} earlier had presented an illustration why
the famous Froissart bound~\cite{Froissart:1961ux} for scalar
particles (with cross sections rising at most with $\ln^2\nu$) is
replaced with much stronger bounds for particles with spin. This later
argument is based explicitly on unitarity which is a built-in feature
of Regge theory anyway. \\

The part below pion threshold is treated differently by different
authors. Some authors do not include the part of the
integration down to vanishing photon energy in their notation of the
\gdhsr. Most prominently, even the original publications from Drell and
Hearn~\cite{Drell:1966jv} and Gerasimov~\cite{Gerasimov:1965et} do
not agree on this. Gerasimov starts the integral at some 
$\nu_\text{thr}$ --- presumably the pion threshold for the photon
scattering off nucleons or the photo-disintegration threshold for the
deuteron --- while Drell and Hearn take the full integral. Apparently,
Gerasimov considers the Compton amplitude 
to lowest order only. As discussed in Sec.~\ref{sec:OptTheo} this
implies that the cross section under consideration reduces from the full
total cross section to photoabsorption only without the elastic
contribution. Hence, only contributions from above the pion threshold
or the photo-disintegration exist. 

In view of our discussion in
Sec.~\ref{sec:LowTheorem} this differentiation appears
insignificant: The ``missing'' contribution from the elastic photon
scattering (i.e. Compton scattering) must be the difference between
the anomalous magnetic moment of the nucleon up to only low
orders and the real physical one. As pointed out in
Sec.~\ref{sec:LowTheorem} this appears minuscule~\footnote{On top of
that, the Low-energy theorem ensures that this contribution is
bounded as $f_2/\nu$ has no imaginary part at threshold.}.

In conclusion, both the low energy part as well as the high energy
part are well under control and the convergence of the \gdhsr is
guaranteed. 

\subsubsection{No-Subtraction hypothesis}
\label{sec:nosub}
Reconsidering the dispersion theoretic derivation of the \gdhsr
presented in Sec.~\ref{sec:dispersion} all steps but the
No-Subtraction hypothesis rely on fundamental assumptions like
gauge invariance, Lorentz invariance, causality and the like. In
contrast, so far, we have given no reason why the No-Subtraction
hypothesis should hold. The failure of the equivalent assumption for
the spin independent amplitude $f_1$ is even more
irritating. Eq.~(\ref{eqn:KKf1}) to lowest order in $\nu$ leads to 
the ridiculous
prediction of a negative total cross section: $\int \sigma_T \, d\nu'
= -\alpha/m$. 
The integral over the total cross section is likely even divergent as
the cross section is still rising up to energies accessible with
today's accelerators and also the Froissart theorem (see
Sec.~\ref{sec:Convergence}) does not provide a helpful bound in this
respect. 
The failure of the No-Subtraction hypothesis for $f_1$ is a
consequence of the relevance of the 
Pomeranchuk trajectory in the case of unpolarized scattering.

However, the Pomeranchuk trajectory does not contribute in the
polarized cross section difference (see Sec.~\ref{sec:Convergence}). 
In the diffractive picture of high energy scattering the amplitude is
spin independent. This together with the weighting with the inverse of
the photon energy provides a clue why indeed the \gdhsr needs no
subtraction. 

To reduce the dispersion
relation for the spin-flip Compton forward amplitude $f_2(\nu)/\nu$ from a contour
integral in the complex plane to an integration along the real axis
-- i.e. to the Kramers-Kronig dispersion relation -- one has to
assume that $f_2(\nu)/\nu$ vanishes sufficiently fast for $\nu \to
\infty$ such that
\par\nobreak\noindent
\begin{equation}
\frac{1}{2\pi i} \int\limits_{K_+(0,\infty)} d\nu'
\frac{f_2(\nu')/\nu'}{\nu'-\nu} = 0 \quad .
\end{equation}
with the definitions as given in Sec.~\ref{sec:KramersKronig}. 
A violation of this hypothesis would imply that $f_2$ rises at least
linearly with the photon energy $\nu$. This in itself is not in
conflict with the convergence of the integral as described in
Sec.~\ref{sec:Convergence} as long as the imaginary part remains well
behaved. However, the violation of the \gdhsr would lead to a
weird behavior of the corresponding differential cross sections. Since 
\par\nobreak\noindent
\begin{equation}
\left. \frac{d\sigma_{1/2}-d\sigma_{3/2}}{d\Omega}(\nu) \right|_{\theta = 0} = 
\left| f_2(\nu) \right|^2
\end{equation}
the divergence of $f_2(\nu)$ translates into a divergence of the
differential forward cross section
\par\nobreak\noindent
\begin{equation}
\lim_{\nu \to \infty}
\frac{1}{d\Omega}\left.\left(d\sigma_{3/2}-d\sigma_{1/2}
\right)\right|_{\theta=0} = \infty
\end{equation}
On the other hand for the total cross section we know from the
arguments presented in Sec.~\ref{sec:Convergence} 
\par\nobreak\noindent
\begin{equation}
\lim_{\nu \to \infty} \ \left( \sigma_{3/2}^\text{tot} -
\sigma_{1/2}^\text{tot} \right) = 0
\end{equation}

\begin{figure}
\begin{center}
        \includegraphics[width=0.9\textwidth]{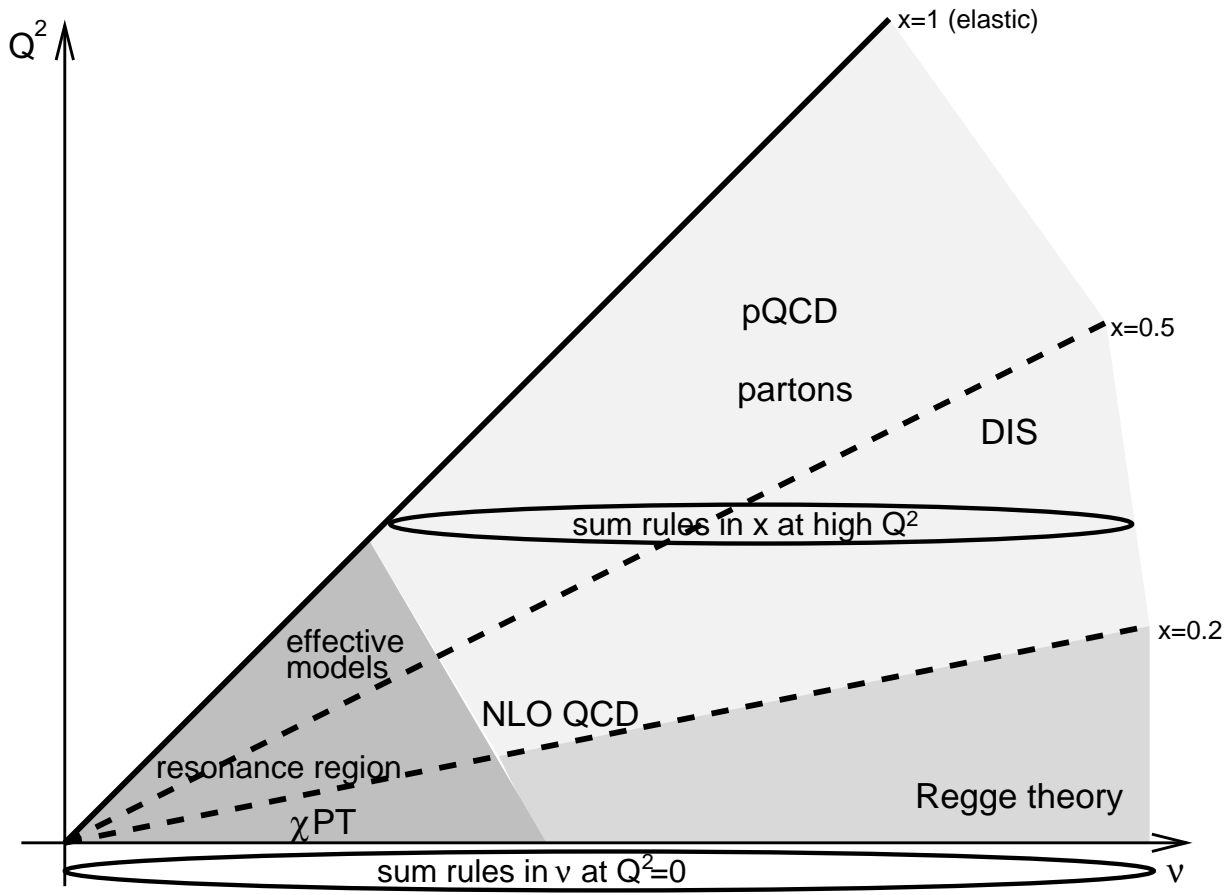}
        \caption{Kinematic regions and available descriptions of the
        strong interaction today}
        \label{fig:kinematics}
\end{center}
\end{figure}

Mathematically this is still quite possible. For example
$\lim\limits_{\nu\to\infty}d\sigma/d\Omega = \nu^{-1/2} \, \exp
(-\theta^2/\nu^2)$ would indeed show this characteristics.
However in terms of the internal dynamics of the nucleon it is
currently impossible to establish an understanding of such a
behavior. Fig.~\ref{fig:kinematics} sketches the various kinematic
domains of inelastic lepton scattering and coarsely attributes the
most prominent model descriptions to them.
In the kinematic domain under consideration here -- at high energies $\nu$
with vanishing $Q^2$ -- QCD and QCD-inspired models are not
applicable with present day techniques. Today, we are still left with Regge
theory which does not provide a picture of the internal
dynamics. 

Some evidence for the ``No-subtraction'' hypothesis of the
\gdhsr to be true may be assumed as calculations within several 
perturbative models have verified the \gdhsr. Moreover,
explicitly also the high energy limit of the spin flip
amplitude has been found to satisfy the ``No-subtraction''
hypothesis: Altarelli, Cabibbob and Maiani have verified that the
Compton amplitude to  fourth-order for the scattering of a photon off
a charged lepton is finite in Weinberg's 
model of weak and electro-magnetic interactions, and obeys the
Gerasimov-Drell-Hearn sum rule~\cite{Altarelli:1972nc}. Gerasimov and
Moulin~\cite{Gerasimov:1973ja} have successfully tested the GDH sum
rule in the pseudoscalar pion-nucleon model.
Brodsky and Schmidt~\cite{Brodsky:1995fj} have generalized the result
of Ref.~\cite{Altarelli:1972nc} to $2 \to 2$ Standard Model and
supersymmetric processes $\gamma a \to b c$ in the tree-graph
approximation.  

However, already in 1968, right
after the discovery of the \gdhsr,  Abarbanel and
Goldberger~\cite{Abarbanel:1967wk} considered a $J_0=1$ Regge
fixed pole being a possible source for the failure of the
No-Subtraction hypothesis.
Such a $J_0=1$ fixed pole would allow the imaginary part of the
spin-flip amplitude $f_2$ to vanish while $\lim\limits_{\nu\to\infty}
\text{Re} \left[ f_2(\nu)/\nu \right] \neq 0$.
In the case of real Compton scattering this fixed pole also does not
violate the Landau-Yang theorem: The Landau-Yang theorem forbids
two photons to have a total angular momentum of
$J_0=1$~\cite{Landau:1948,Yang:1950} in the center of mass
system\footnote{The Landau-Yang theorem is based on the Bose
statistics of the photons, transversality of real photons and
rotational symmetry.}. This is to be related to the t-channel process
with two external photon lines in the final state.
For the s-channel Compton forward ($t=0$) amplitude $f_2$ both photons
have the same helicity. Crossing relations (see
Ref.~\cite{Drechsler:1970}) then lead to a total helicity of 2 in the
t-channel~\cite{Rollnik:1976}. Hence, the total angular momentum has
to be at least 2 and the Landau-Yang theorem does not apply. Also, the
partial wave expansion includes only the values $J=2,3,\ldots$. This
actually allows to consider this fixed pole at all.

Such a fixed pole is forbidden for purely hadronic processes but it
cannot be ruled out a priori for electro-weak processes considered to
low-order coupling only.  
Fairly recently, Bass~\cite{Bass:2003vp} has revisited the
possibility of such a fixed pole in view of possible gluonic and sea
contributions. An observable effect of this would kick in only at very
high energies. A connection of the fixed pole to the gluon topology is
established. Bass conjectures a correction of up to 10~\% to the
\gdhsr due to the fixed pole~\cite{basspriv,Bass:2003vp}.

However, such a fixed pole has never been observed explicitly so far and
nature, of course, is not restricted in the electromagnetic
coupling. Abarbanel, Low, Muzinich, Nussinov and Schwarz were the
first to point out that bilinear unitarity in the $t$ channel also
forbids a $J=1$ fixed pole~\cite{Abarbanel:1967}. For a general
discussion of bilinear versus linear unitarity see for
example~\cite{Cutkosky:1960sp,Blankenbecler:1960,Cook:1962,Baker:1971sf}.
The principle idea is the following~\cite{RollnikPriv:2005}: Consider
the partial waves in the t-channel $a(t,J)$ of Compton
scattering. With the fixed pole one would have 
\be 
a(t,J) = {\beta \over J - J_0} \quad \text{in the proximity
of the pole } J_0 = 1\ .
\label{eqn:FixedPole}
\ee 
Now we reconsider the optical theorem in terms of
the $S$ and the $T$ matrix: $S = 1 + iT$. Unitarity gives $S^\dag S =
1$ or \be \text{Im } T = \frac{1}{2}\; T^\dag T\quad .\ee 
Here, the pole from Eq.~(\ref{eqn:FixedPole}) would enter
quadratically on the left hand side and only linearly on the right
hand side and we end up having a pole of second order on the left hand
side and a pole of first order only on the right hand side.
Due to this contradiction such a pole cannot contribute in nature with
all orders of electromagnetic coupling~\footnote{For a description why the
usual moving poles (also called trajectories) are not affected by this argument
see for example Ref.~\cite{Collins:1968}.} However, if we expand $T$ in
orders of the electromagnetic coupling $T = T_0 + T_1 + T_2 + \ldots$
where $T_0$ represents strong interactions only, one obtains separate
equations for each order in electromagnetic coupling:
\beqn
\text{Im } T_0 & = & \frac{1}{2}\,T_0^\dag T_0\\
\text{Im } T_1 & = & \text{Re } T_0^\dag T_1 \label{eqn:UniFirstOrder}\\
\text{Im } T_2 & = & \text{Re } T_2^\dag T_0 + \frac{1}{2}\,T_1^\dag T_1
\eeqn
When we consider lowest order coupling
Eq.~(\ref{eqn:UniFirstOrder}) is relevant only. Here the pole from
Eq.~(\ref{eqn:FixedPole}) appears to first order on both sides and it
may be relevant.\\

In conclusion, the fixed $J_0=1$ Regge pole may exist as an artifact of
the limitation of the calculation including only low-order coupling. In
nature however it is forbidden by full (bilinear or quadratic)
unitarity and a failure of the ``No-subtraction'' hypothesis also
would violate fundamental ingredients of today's field theories like
all the other steps of the derivation of the \gdhsr. 


\newpage
\section{Polarized virtual photoabsorption}
\label{sec:PolAbs}
Fig.~\ref{fig:Aparperp} depicts the kinematics and useful spin
orientations of lepton beam and nucleon target.
We have the following common invariant kinematic quantities:
$Q^2 = -q^2 = (k - k')^2 = 4EE'\sin^2\theta/2$ is the four-momentum
transfer to the nucleon or the virtuality of the photon with the
scattering angle $\theta$ of the lepton. $\nu = q \cdot P / m = E -
E'$ is the 
lepton's energy loss in the nucleon rest frame and the photon
energy. Bjorken-$x$ is defined as usual as $x = Q^2/2m\nu$ which, in the
parton model, is the fraction of the nucleon's momentum carried by the
struck quark. $y = q \cdot P / k \cdot P = \nu / E$ is the fraction of
the lepton's energy lost in the nucleon rest frame and $W^2 = (P+q)^2
= m^2 + 2m\nu - Q^2$ is the mass squared of the recoiling system
against the scattered lepton and $\gamma = \sqrt{Q^2} / \nu$. 

\begin{fmffile}{fe}
\begin{figure}
~\vspace{6mm}\par\nobreak\noindent
\begin{displaymath}
\label{eqn:Aparallel}
  A_\parallel \ = \ 
  \frac{ \sigma^{\uparrow\downarrow} - \sigma^{\uparrow\uparrow} }{
  \sigma^{\uparrow\downarrow} +  \sigma^{\uparrow\uparrow} }
  \  =  \  
  \left(\hspace{13mm}
  \graphs(130,81){
   \fmfleft{i3,i2,i1,i0} \fmfright{o5,o4,o3,o2,o1,o0}  \fmfbottom{b} \fmftop{t}
   \fmf{dashes,tension=0.1}{i1,o1}
   \fmf{fermion,tension=3}{i0,v1}
   \fmf{fermion}{v1,o0} 
   \fmf{fermion,width=thick,tension=1.2}{i3,v2} 
   \fmf{photon,label=$q$}{v1,v2}
   \fmf{fermion,width=thick,tension=0.75}{v2,o3} 
   \fmf{fermion,width=thick,tension=0.75}{v2,o4} 
   \fmf{fermion,width=thick,tension=0.75}{v2,o5} 
   \fmfblob{0.8ex}{v2}
   \fmflabel{$k,\Leftarrow$}{i0} 
   \fmflabel{$P,m,\Rightarrow$}{i3}  
   \fmflabel{$k'$}{o0}
   \fmflabel{$W$}{o4}
  } \ \   
  \ - \ \  
  \graphs(130,81){
   \fmfleft{i3,i2,i1,i0} \fmfright{o5,o4,o3,o2,o1,o0}  \fmfbottom{b} \fmftop{t}
   \fmf{dashes,tension=0.1}{i1,o1}
   \fmf{fermion,tension=3}{i0,v1}
   \fmf{fermion}{v1,o0} 
   \fmf{fermion,width=thick,tension=1.2}{i3,v2} 
   \fmf{photon,label=$q$}{v1,v2}
   \fmf{fermion,width=thick,tension=0.75}{v2,o3} 
   \fmf{fermion,width=thick,tension=0.75}{v2,o4} 
   \fmf{fermion,width=thick,tension=0.75}{v2,o5} 
   \fmfblob{0.8ex}{v2}
   \fmflabel{$\Rightarrow$}{i0} 
   \fmflabel{$\Rightarrow$}{i3}  
  } \right) / \sum
\end{displaymath}
~\vspace{15mm}\par\nobreak\noindent
\begin{displaymath}
\label{eqn:Aperp}
  A_\perp \ = \ 
  \frac{ \sigma^{\downarrow\leftarrow} - \sigma^{\uparrow\leftarrow} }{
  \sigma^{\downarrow\leftarrow} + \sigma^{\uparrow\leftarrow} }
  \  =  \  
  \left(\hspace{3mm}
  \graphs(130,81){
   \fmfleft{i3,i2,i1,i0} \fmfright{o5,o4,o3,o2,o1,o0}  \fmfbottom{b} \fmftop{t}
   \fmf{dashes,tension=0.1}{i1,o1}
   \fmf{fermion,tension=3}{i0,v1}
   \fmf{fermion}{v1,o0} 
   \fmf{fermion,width=thick,tension=1.2}{i3,v2} 
   \fmf{photon,label=$q$}{v1,v2}
   \fmf{fermion,width=thick,tension=0.75}{v2,o3} 
   \fmf{fermion,width=thick,tension=0.75}{v2,o4} 
   \fmf{fermion,width=thick,tension=0.75}{v2,o5} 
   \fmfblob{0.8ex}{v2}
   \fmflabel{$\Leftarrow$}{i0} 
   \fmflabel{$\Uparrow$}{i3}  
  } \ \  
  \ - \ \ \  
  \graphs(130,81){
   \fmfleft{i3,i2,i1,i0} \fmfright{o5,o4,o3,o2,o1,o0}  \fmfbottom{b} \fmftop{t}
   \fmf{dashes,tension=0.1}{i1,o1}
   \fmf{fermion,tension=3}{i0,v1}
   \fmf{fermion}{v1,o0} 
   \fmf{fermion,width=thick,tension=1.2}{i3,v2} 
   \fmf{photon,label=$q$}{v1,v2}
   \fmf{fermion,width=thick,tension=0.75}{v2,o3} 
   \fmf{fermion,width=thick,tension=0.75}{v2,o4} 
   \fmf{fermion,width=thick,tension=0.75}{v2,o5} 
   \fmfblob{0.8ex}{v2}
   \fmflabel{$\Rightarrow$}{i0} 
   \fmflabel{$\Uparrow$}{i3}  
  } \right) / \sum
\end{displaymath}
~\vspace{3mm}\par\nobreak\noindent
\caption{Kinematics of inelastic charged lepton scattering: $k, k'$
and $P$ are the four-momenta of the incoming and outgoing leptons and
of the incoming nucleon, $m$ is the mass of the nucleon and $W$ is the
invariant mass of the
recoiling system $X$. For the energy range under consideration the
exchanged particle can only be a photon. The exchanged photon has the
four-momentum  $q$. $\Rightarrow$ and $\Leftarrow$ denote spins in
parallel and antiparallel to the lepton beam momentum in the nucleon rest
frame. $\Uparrow$ indicates a nucleon spin perpendicular to the beam
momentum and $\sum$ stands for the sum of cross sections of the two relative spin
orientations. $A_\parallel$ and $A_\perp$ are the two
experimentally accessible asymmetries. The parts below the dashed
lines indicate the portion of the scattering process that correspond
to polarized virtual photoabsorption.}
\label{fig:Aparperp}
\end{figure}
\end{fmffile}

Fig.~\ref{fig:Aparperp} also provides the
definitions for the experimental asymmetries $A_\parallel, A_\perp$. We are,
however, interested in the photoabsorption subprocess which is the
lower part of the diagrams below the dashed lines and the asymmetries
of this subprocess.
The spin structure of a virtual photon is more involved than that of
real photon due to the longitudinal polarization component.
The polarized virtual photoabsorption cross section in the nucleon
rest frame may be written as
\beqn 
\frac{d\sigma}{d\Omega\ dE'} = \Gamma_V \left[ 
\sigma_T+\epsilon\sigma_L-P_eP_x\sqrt{2\epsilon(1-\epsilon)}\
         \sigma_{LT}-P_eP_z\sqrt{1-\epsilon^2}\ \sigma_{TT} \right]
\eeqn
with the photon polarization $\epsilon$ relative to the lepton
polarization $P_e$, the virtual photon flux factor $\Gamma_V$ and the
equivalent photon energy $K$
\beqn \epsilon = \frac{1}{1+2 (1+ \nu^2/Q^2) \tan^2 \theta/2}\ ,\quad
\Gamma_V = \frac{\alpha}{2\pi^2}\ \frac{E'}{E}\ \frac{K}{Q^2}\
\frac{1}{1-\epsilon}\ , \quad K = \nu(1-x) = \frac{W^2-m^2}{2m}\ .
\eeqn
$P_z$ and $P_x$ denote the components of the target polarization in the
direction of the virtual photon momentum ${\vec{q}}$ and perpendicular to that
direction in the scattering plane of the electron. 
In addition to the transverse cross sections $\sigma_{T}$ and
$\sigma_{TT}$, the virtuality of the photon gives rise to the
longitudinal 
$\sigma_{L}$ and the longitudinal-transverse $\sigma_{LT}$ cross sections.

With these definitions we have the photoabsorption 
asymmetries $A_1$ and $A_2$:
\par\nobreak\noindent
\begin{equation}
A_1 = \frac{\sigma_{1/2} - \sigma_{3/2}}{\sigma_{1/2} + \sigma_{3/2}}
 = \frac{\sigma_{TT}}{\sigma_{T}}\  , \quad A_2 =
\frac{2\, \sigma_{TL}}{ \sigma_{T} } 
\end{equation}
\par\nobreak\noindent
For transverse polarization of the photon the subscripts $3/2$ and
$1/2$ denote the total helicity of the photon-nucleon system in the
nucleon rest frame with respect to the center of mass
momentum. $\sigma_{TL}$ is the longitudinal-transverse virtual
photoabsorption cross section.  The experimentally
directly observable asymmetries $A_\parallel, A_\perp$ of the lepton
scattering process are related to the asymmetries
$A_1, A_2$ in the photoabsorption process by kinematic factors:
\par\nobreak\noindent
\begin{equation}
             A_{\parallel} = D (A_1 + \eta A_2)\  ,  
   \quad 
      A_{\perp}     =  d (A_2 -\xi A_1),
\end{equation}
\par\nobreak\noindent
The kinematic factors read
\par\nobreak\noindent
\begin{equation}
D = \frac{1-\epsilon E'/E}{1+\epsilon R} \, \quad
             d    = \frac{\sqrt{1-y-\gamma^2 y^2/4}}{1 -y/2}\,D \  , \quad
             \eta = \frac{\gamma\,(1 - y - \gamma^2 y^2/4)}
                                {(1 - y/2)(1 + \gamma^2 y/2)}\  , \quad
             \xi  = \frac{\gamma (1 - y/2)}{1 + \gamma^2 y/2}~\  .
\end{equation}
where $R$ is the ratio of longitudinal to transverse virtual
photoabsorption cross sections.

\subsection{Deep inelastic scattering}
The process in Fig.~\ref{fig:Aparperp} is called deep inelastic
scattering if $Q^2 >> m^2, W^2 >> m^2$. One can then neglect the
mass of the scattered lepton. In lowest order perturbation theory the
cross section for the scattering factorizes into a leptonic
$L^{\mu\nu}$ and a hadronic tensor $W^{\mu\nu}$:
\par\nobreak\noindent
\begin{equation}
\frac{d^2\sigma}{dxdy} = \frac{2\pi y \alpha^2}{Q^4} L^{\mu\nu} W_{\mu\nu}
\end{equation}
\par\nobreak\noindent
The lepton tensor associated with the exchange of a
photon\footnote{Recall that we have restricted the discussion to
energies where the exchange of a $Z^0$ boson is irrelevant.} reads
\par\nobreak\noindent
\begin{equation}
L_{\mu\nu} = 2 \left( 
k_mu k'_\nu + k'_\mu k_\nu - k \cdot k' g_{\mu\nu} - i \lambda
\epsilon_{\mu\nu\alpha\beta} k^\alpha k'^\beta
\right)
\end{equation}
\par\nobreak\noindent
with the helicity $\lambda = \pm 1$ of the incoming lepton. The
hadronic tensor describes the interaction of the virtual photon with
the target nucleon and this is where the internal structure of the
nucleon is manifest. Since this structure cannot (yet) be obtained directly by
application of QCD for all kinematic regions this hadron
tensor is parameterized by eight structure functions~\cite{Blumlein:1996vs}. 
For deep inelastic scattering where the momentum transfer is small
compared to the mass of the $Z^0$ boson, contributions from weak
interactions can be neglected and we have to consider
only four independent structure functions. 
$W^{\mu\nu}$ can 
be split into a symmetric and an anti-symmetric part:
\par\nobreak\noindent
\begin{equation}
        W^{\mu\nu}=W^{\{\mu\nu\}}+W^{[\mu\nu]},
\end{equation}
\par\nobreak\noindent
with
\par\nobreak\noindent
\begin{eqnarray}
\label{eqn:AntiSymmHadron}
        W^{\{\mu \nu\}} & = 
        & \left( -g^{\mu \nu}+{q^{\mu}q^{\nu}\over q^2}\right) 
        F_1 +\left[ \left( P^{\mu}-{\nu \over q^2}q^{\mu} \right) 
        \left( P^{\nu}-{\nu \over q^2}q^{\nu} \right) \right] {F_2 \over \nu}, 
        \label{eq:WS}\\
        W^{[\mu \nu]} & = & -i\varepsilon ^{\mu \nu \lambda \sigma} q_{\lambda}
        \left( {s_\sigma \over \nu} \left( g_1 + g_2 \right)
          -{q\cdot s \  P_\sigma \over \nu^2}g_2 \right), \label{eq:WA}
\end{eqnarray}
\par\nobreak\noindent
where $s$ is the nucleon covariant spin vector $(s^2=-m^2)$,
$P\cdot s=0$ and $\varepsilon ^{\mu \nu \lambda \sigma}$ is the
totally antisymmetric Levi-Civita tensor.
$g_{1,2}$ and $F_{1,2}$ are scalar dimensionless functions.
These structure functions are related to others in common use by:
\par\nobreak\noindent
\begin{equation}
        W_1 = F_1, \quad
        W_2 = {m^2 \over \nu} F_2, \quad
        G_1 = {m \over \nu}\, g_1, \quad
        G_2 = {m^2 \over \nu ^2}\, g_2.
\end{equation}
\par\nobreak\noindent
For the absorption of transversely polarized virtual photons by
longitudinally polarized nucleons with total spin 
$3/2 \  \mbox{and} \ 1/2$  the
result of this tensor product reads:
\par\nobreak\noindent
\begin{eqnarray}
\sigma_{3/2} & =  & \frac{8\pi^2\alpha}{2 m \nu - Q^2} \left( F_1(x,Q^2) -
g_1(x,Q^2) + \frac{Q^2}{\nu^2}g_2(x,Q^2) \right) \\
\sigma_{1/2} & =  & \frac{8\pi^2\alpha}{2 m \nu - Q^2} \left( F_1(x,Q^2) +
g_1(x,Q^2) - \frac{Q^2}{\nu^2}g_2(x,Q^2) \right)
\end{eqnarray}
\par\nobreak\noindent
We can now express the virtual photoabsorption asymmetry $A_1$ in
terms of these structure functions:
\be
A_{1} = \frac{g_1-\gamma^2\,g_2}{F_1} \simeq \frac{g_1}{F_1}
\ee

In the quark-parton model the quark densities depend only on the
momentum fraction~$x$ carried by the quark. In the infinite momentum
frame, due to angular momentum conservation, a virtual photon with
helicity $+1$ or $-1$ can only be absorbed by a quark with a spin
projection of $-\frac{1}{2}$ or $+\frac{1}{2}$, respectively. 
$g_1(x)$ is
then given by 
\par\nobreak\noindent
\begin{equation}
\label{eqn:g1_naive}
 g_{1}(x) = \frac{1}{2} \sum_{i=1}^{n_f} e_i^2 \Delta q_i(x),
\end{equation}
\par\nobreak\noindent
where 
\par\nobreak\noindent
\begin{equation}
  \Delta q_{i}(x) = q_i^+(x) - q_i^-(x) + \bar q_i^{~+}(x) - \bar q_i^{~-}(x),
\end{equation}
\par\nobreak\noindent
$q_i^+(\bar q_i^{~+})$ and  $q_i^-(\bar q_i^{~-})$ are  
the distribution functions of quarks~(antiquarks) with spin
parallel and antiparallel to the nucleon spin,
respectively, $e_i$  is the electric charge of the quarks of flavor $i$
and $n_f$ is the number of quark flavors involved. 

\subsubsection{Bjorken sum rule}

The Bjorken sum rule~\cite{Bjorken:1966jh} and the Ellis-Jaffe sum
rule~\cite{Ellis:1973kp} are the counterpieces of the GDH sum
rule. While the \gdhsr is a statement at $Q^2 = 0$ the Bjorken
and the Ellis-Jaffe sum rule are predictions at infinite $Q^2$.\\

J.~D.~Bjorken derived his sum rule in 1966 which is the very same year
that the \gdhsr was proposed. Initially, Bjorken himself
disqualified the sum rule in his own publication:
\begin{quotation}
Something may be salvaged from this {\em worthless} equation \ldots
\end{quotation}
However, he reconsidered his sum rule in 1970 ``in light of the
present experimental and theoretical situation''. He was referring to
an article~\cite{Bloom:1969kc} on inelastic electron scattering
results from \textsc{Slac} that showed that ``for high excitations
the cross section shows only a weak momentum-transfer
dependence''. Today, this phenomenon is called Bjorken scaling and is
discussed in about all modern text books on particle physics.

The representation of the Bjorken sum rule in the original
article~\cite{Bjorken:1966jh}~(Eq.~(6.16) and (6.17) therein) shows
the similarity of it to the \gdhsr for real photoabsorption:
\par\nobreak\noindent
\begin{equation}
\lim_{q^2 \to -\infty} \lim_{E\to\infty} \int_0^\infty
\frac{d\nu'}{\nu'} \left[ \frac{d\sigma_{3/2} -
d\sigma_{1/2}}{dq^2d\nu'} \right] = 
- \frac{8\pi\alpha^2}{q^4 E} Z_N \quad \text{with} \quad Z_p - Z_n =
\frac{1}{3} \left( \frac{g_A}{g_V} \right)
\end{equation}
\par\nobreak\noindent
${g_A}/{g_V}$ is the ratio of the phenomenological weak $\beta$-decay
coupling constants. Bjorken in his article only briefly states that
the values of the constants $Z_N$ are unknown but that $SU(6)$
symmetry would lead to $Z_p = 5/9$ and $Z_n = 0$ (see also
Sec.~\ref{sec:ejsr}). 
Like in the derivation of the \gdhsr by means of the equal-times
current algebra in Sec.~\ref{sec:EqTsub} the starting point for the
derivation of the Bjorken sum rule also is the vanishing of the
quark-model equal-time commutator for the space components (see
Eq.~(\ref{eqn:eqt-comm})). Bjorken's proof also relies on a
``No-subtraction'' hypothesis. Again like the \gdhsr, the
Bjorken sum rule also connects static properties of the nucleon with
its dynamic response. In today's nomenclature the Bjorken sum rule is
usually represented like 
\par\nobreak\noindent
\begin{equation}
\label{eqn:BjorkenSR}
\lim_{Q^2 \to \infty} \left[\, \Gamma_1^p(Q^2) - \Gamma_2^n(Q^2)
\,  \right] 
= \frac{1}{6} \left| \frac{g_A}{g_V} \right|
\end{equation}
\par\nobreak\noindent
with the definition for the first\footnote{The common notation ``first
moment'' of the structure function in the literature is at odds with
the usual mathematical naming scheme where it would be called zeroth
moment.
}
moment of the proton or the neutron
structure functions $g_1$
\par\nobreak\noindent
\begin{equation}
\label{eqn:g1first}
\Gamma_1(Q^2) = \int\limits_0^1 g_1(x,Q^2) \, dx \quad .
\end{equation}
\par\nobreak\noindent
Through the neutron decay parameter of $n
\to p e^- \bar{\nu}_e$ this ratio is known very
accurately~\cite{Eidelman:2004wy}: $\lambda \equiv g_A / g_V = -1.2695
\pm 0.0029$.  

Hence, the right hand side of
the Bjorken sum rule Eq.~(\ref{eqn:BjorkenSR}) is known with a relative
precision of about $2.3 \cdot 10^{-3}$. In comparison to the dynamic
observables this is an impressive precision already. On the other
hand, the ratio of the magnetic moment of the proton $\mu_p$ to the nuclear
magneton $\mu_N$ that leads to the anomalous magnetic moment in the
\gdhsr 
has been measured to an even higher precision~\cite{Mohr:1999}:
$\mu_p/\mu_N = 2.792847351 \pm 0.000000028$ which is a relative precision
of $10^{-8}$. 
For the time being there seems to be no chance to come even close to
these accuracies with the verification of these sum rules as the
measurements of the dynamic observables and cross sections have a
systematic error of the order of several percent originating for
example from the determination of the polarizations of target and beam.

However, the dominant limitation of a verification of the Bjorken sum
rule stems from the $Q^2$ evolution which is necessary since $Q^2 \to
\infty$ is not reachable experimentally. At finite values of $Q^2$
radiative QCD corrections are important. Beyond leading order the
corrections also depend on the renormalization scheme and the number
of flavors taken into account. At $Q^2 = 10~\text{GeV}^2$ for example
the correction is about a factor of $7$. Also the experimental data
obtained at fixed beam momentum need to be evolved to a common
$Q^2$ in order to perform the integration in $x$ of
Eq.~(\ref{eqn:g1first}) at fixed $Q^2$. Both these $Q^2$ evolutions of
the 
theoretical Bjorken sum rule prediction and the experimental data
impair the fundamental character the sum rule originally has at $Q^2 =
\infty$. Hence an experimental verification of the Bjorken sum rule
cannot claim to be a test of its fundamental principles like it is the
case with the \gdhsr but is rather a check of our understanding
of QCD, higher order corrections and of scaling violation.

Several experiments have performed verifications of the $Q^2$-evolved
Bjorken sum rule. Most recently the \textsc{Hermes}-Collaboration at
\textsc{Desy} reported~\cite{Ackerstaff:1999ey} an agreement within
the experimental error of about 12~\% at a $Q^2=
2.5~\text{GeV}^2$. The
experimental error does not include an estimate of the error of
extrapolation to unmeasured regions in Bjorken-x.
The Spin Muon Collaboration (SMC) at \textsc{Cern} has combined their
own data~\cite{Adams:1997tq} with the data from their precursor
experiment EMC~\cite{Ashman:1989ig} also at \textsc{Cern} and the
E80/E130~\cite{Alguard:1976bm,Alguard:1978gf,Baum:1983ha} and E142/E143
~\cite{Abe:1994cp,Abe:1995mt} experiments at \textsc{Slac}. They find
agreement with the Bjorken sum rule at $Q^2 = 10~\text{GeV}^2$ within
19~\% experimental uncertainty and at $Q^2 = 5~\text{GeV}^2$ within
11~\% experimental error including estimates of the uncertainties
arising from the Bjorken-x extrapolation.

All in all, there is no hint that the Bjorken sum rule may be wrong
and it seems that the QCD $Q^2$-evolution is well under control to the
level of about 10~\%.

\subsubsection{Ellis-Jaffe sum rules}
\label{sec:ejsr}
J.~R.~Ellis and R.~L.~Jaffe~\cite{Ellis:1973kp} in 1974 have derived
similar sum rules
like the Bjorken sum rule. The Ellis-Jaffe sum rules are statements
for the proton and the neutron individually. They are obtained by
assuming exact SU(3) flavor symmetry and a sea contribution
from strange quarks without a resulting polarization:
\par\nobreak\noindent
\begin{equation}
\int\limits_0^1 g_1^p(x)\, dx = \frac{9F -D}{18} \qquad
\int\limits_0^1 g_1^n(x)\, dx = \frac{6F - 4 D}{18}
\end{equation}
\par\nobreak\noindent
The constants $F$ and $D$ are SU(3) invariant matrix elements of the
axial vector current where for the neutron beta decay $F+D =
g_A/g_V$~\cite{Jaffe:1989jz}. 

Several experiments have reported a violation of the Ellis-Jaffe sum
rules. Most prominently, already in 1989 the European Muon
Collaboration (EMC) claimed a disagreement with the Ellis-Jaffe sum
rule for the Proton~\cite{Ashman:1989ig}. In the na\"ive parton model
the results lead to the conclusion that the total quark spin
constitutes only a small fraction of the spin of the proton. This
finding lead to the so-called ``spin crisis'' which sparked a whole
series of spin physics experiments. Today, the combined data from the
\textsc{Slac} experiments E80/E130 and E142/E143 and from EMC and SMC
at \textsc{Cern} show a discrepancy of about 2 standard deviations for
the proton and about three standard deviations for the deuteron at
an evolved $Q^2 = 10~\text{GeV}^2$. 

The origin of this discrepancy may be a polarization of the
strange quark content. The \textsc{Hermes} experiment at \textsc{Desy}
has performed the first direct experimental extraction of the separate
helicity densities of the light quark
sea~\cite{Airapetian:2004zf}. For strange sea quarks in a leading
order QCD analysis the results do not fully explain the discrepancies
found for the Ellis-Jaffe sum rules as the strange quark polarization
appears rather small. 

On the other hand a sizable gluon
polarization could also change the interpretation of the structure
functions at finite photon virtualities. In this case the polarized
structure function $g_1$ does not only represent the polarization of
the quarks like at $Q^2 = \infty$ where we have
Eq.~(\ref{eqn:g1_naive}). Instead, the quark and the gluon spin
content both are important for $g_1$ at intermediate virtualities. It
is one of the main goals of the \textsc{Compass} experiment at
\textsc{Cern} and the \textsc{Rhic} spin program at BNL to determine
the gluon polarization~\cite{Hedicke:2003pz,Bland:2004um}. 

\subsection{Extension of the GDH Sum to finite photon virtuality}
There are several ways to generalize the integral on left hand side of
the \gdhsr i.e. the integral over all energies. For an overview
see for example 
Ref.~\cite{Drechsel:2000ct}. Amongst these choices experimentalists
sometimes favor a version which is a straight forward generalization
of the \gdhsr in terms of the polarized photoabsorption cross
section:
\beqn
\label{eqn:Igdh}
I_{\text{GDH}}(Q^2)& = & {\boldsymbol -} \int\limits_0^\infty d\nu\, 
\frac{\sigma_{3/2}(\nu,Q^2) -
\sigma_{1/2}(\nu,Q^2)}{\nu} 
 = \frac{8\pi^2\alpha}{m^2} \int\limits_{0}^{\infty} d\nu
   \frac{G_1(\nu,Q^2) - \frac{Q^2}{m\nu} 
   \, G_2(\nu,Q^2)}{\sqrt{\nu^2+Q^2}} \nonumber\\
& = & \frac{16\pi^2\alpha}{Q^2} \int\limits_0^1 dx
   \, \frac{g_1(x,Q^2) - \xi\,
   g_2(x,Q^2)}{\sqrt{1+\xi}} \quad \text{with } \xi = \frac{4m^2x^2}{Q^2}
\eeqn
This definition is close to the measured observables.
On the other hand, theorists often prefer the following integral
related to the first moment $\Gamma_1$ because it is related to  only
a single structure function namely $g_1$ (see Eq.~(\ref{eqn:g1first})): 
\be
I_1(Q^2) = \frac{2m^2}{Q^2} \Gamma_1(Q^2) = \frac{2m^2}{Q^2}
\int\limits_0^1 dx \ g_1(x,Q^2) = \int\limits_0^\infty
\frac{d\nu}{\nu} \ G_1(\nu,Q^2) 
\ee
In the real photon limit we obtain the limits for both generalized
integrals from the \gdhsr 
\be
I_\text{GDH}(0) = - \frac{2\pi^2\alpha}{m^2}\kappa^2 \quad \text{and}
\quad I_1(0) = \frac{m^2}{8\pi^2\alpha} I_\text{GDH}(0) = - {1 \over
4} \kappa^2
\ee
while in the scaling limit both generalized integrals coincide
\be
Q^2\to\infty:~ I_\text{GDH}(Q^2) = I_1(Q^2) \quad \text{with the
Bjorken sum rule} \  
 \left[ I_1^p(Q^2) - I_1^n(Q^2) \right] = {m^2
\over 3Q^2} \, {g_A \over g_V} \  .
\ee

Recently, Ji and Osborne~\cite{Ji:1993mv,Ji:1999mr} have developed a unified
formalism to describe the generalized GDH integral $I_1$ with respect
to the doubly-virtual Compton forward scattering (VVCS) process. One
considers the forward scattering of a virtual photon with space-like
four-momentum $q^2 = q_0^2 - \vec{q}\,^2 = -Q^2 < 0$. The VVCS
amplitude for forward scattering of virtual 
photons generalizes Eq.~(\ref{eqn:CompForw}) by introducing an
additional longitudinal polarization vector $\vec{\hat{q}}$,
\be
T(\theta=0,\nu,\,Q^2) = f_1 \ 
\vec{\epsilon_2}^{\ast}\cdot\vec{\epsilon_1} +
f_2 \  
i\vec{\sigma}\cdot(\vec{\epsilon_2}^{\ast}\times\vec{\epsilon_1}) +
f_3 \   +
f_4 \   i\,(\vec{\epsilon_2}^{\ast}-\vec{\epsilon_1})\cdot(\vec{\sigma} \times \vec{\hat{q}} \,)
\label{eqn:CompForwVirtual}
\ee
$f_{1,2}(\nu,Q^2)$ are now functions of $\nu$ and $Q^2$ and coincide
with those of Eq.~(\ref{eqn:CompForw}) at $Q^2=0$ while the functions
$f_{3,4}(\nu,Q^2)$ are due to the longitudinal polarization components
of the virtual photon.
To connect the VVCS amplitudes with the nucleon structure functions
one writes Eq.~(\ref{eqn:CompForwVirtual}) in a covariant form and
separates the spin independent $T^{\{\mu\nu\}}$ and spin dependent
amplitudes $T^{[\mu\nu]}$ with $T^{\mu\nu} = T^{\{\mu\nu\}} +
T^{[\mu\nu]}$. We are interested in the spin dependent part which reads
\be
T^{[\mu\nu]}(\nu,Q^2,\theta=0) \  = \  
\frac{i}{2} \, \epsilon_{2_{\mu}}^{\ast}\epsilon_{1_{\nu}}
\, \varepsilon^{\mu\nu\alpha\beta} \, 
\left\{ q_{\alpha} 
s_{\beta}\, S_1(\nu,Q^2)
+ \frac{1}{m^2}\,q_{\alpha} (P \cdot q\
s_{\beta}-s\cdot q\ P_{\beta})\, S_2(\nu,Q^2) \right \} \ .
\ee
$\varepsilon_{\mu\nu\alpha\beta}$, $P$ and $s$ are defined as before with
Eq.~(\ref{eqn:AntiSymmHadron}). We are mainly interested in the
forward scattering amplitude $S_1(\nu,Q^2)$  which is connected to
$f_2(\nu,Q^2)$ in the real photon case (see
Eqs.~(\ref{eqn:CompForw}) and (\ref{eqn:CompForwVirtual})).
From general principles (causality and unitarity) as well as an
assumption about the large-$\nu$ behavior of $S_1(\nu,Q^2)$ -- similar to
the ``No-subtraction'' hypothesis discussed in the context of $f_2$ --
one can now write down a dispersion relation
\be
S_1(\nu,Q^2) = \frac{16 \pi \alpha}{m^2} \,  \int\limits_0^\infty 
\frac{\nu' d\nu' \; G_1(\nu',Q^2)}{\nu'^2 - \nu^2}
\ee
where we have used the optical theorem
$\text{Im } S_1(\nu,Q^2) = 2\pi G_1(\nu,Q^2)$.

While $G_1(\nu',Q^2)$ is difficult to calculate it can be measured
experimentally. On the other hand, $S_1(\nu',Q^2)$ is hard to measure
experimentally but it can be calculated theoretically in terms of the
VVCS process. Again like with the derivation of the \gdhsr one
takes the limit $\nu\to 0$:
\be
\label{eqn:jiosb}
\bar{S}_1(0,Q^2) = \frac{16 \pi \alpha}{m^2} \int\limits_{\nu_0}^\infty
\frac{d\nu'}{\nu'} \  \bar{G}_1(\nu',Q^2) =
\frac{16 \pi \alpha}{m^2}\, \bar{I}_1(Q^2)
\ee
were $\bar{S}_{1,2} = S_{1,2} - S^\text{el}_{1,2}$ are the amplitudes
without the elastic intermediate state and $\bar{G}_1,\  \bar{I}_1$
the like, $\nu_0$ is the inelastic threshold. 

\begin{figure}
\includegraphics[width=.5\textwidth]{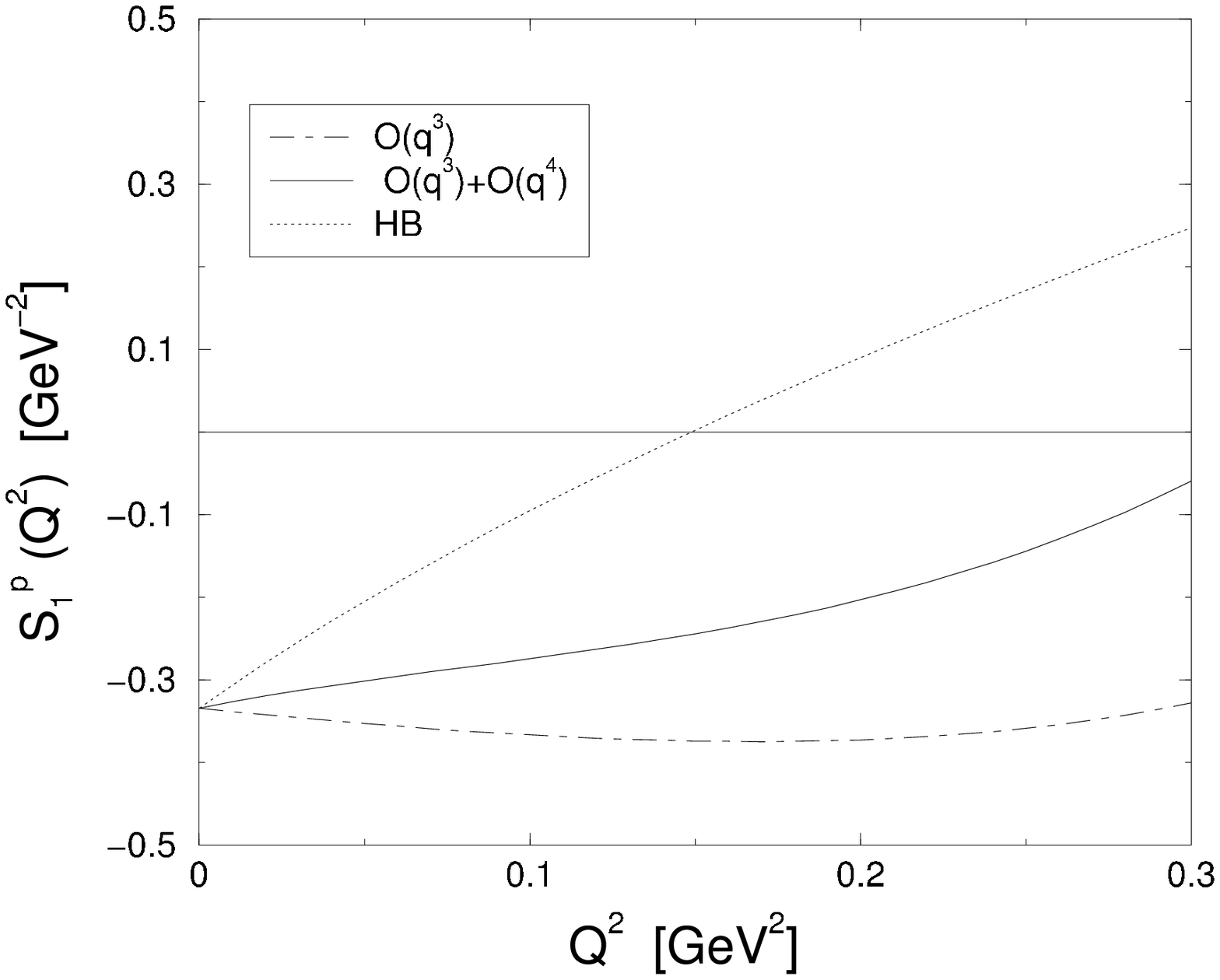}
\includegraphics[width=.5\textwidth]{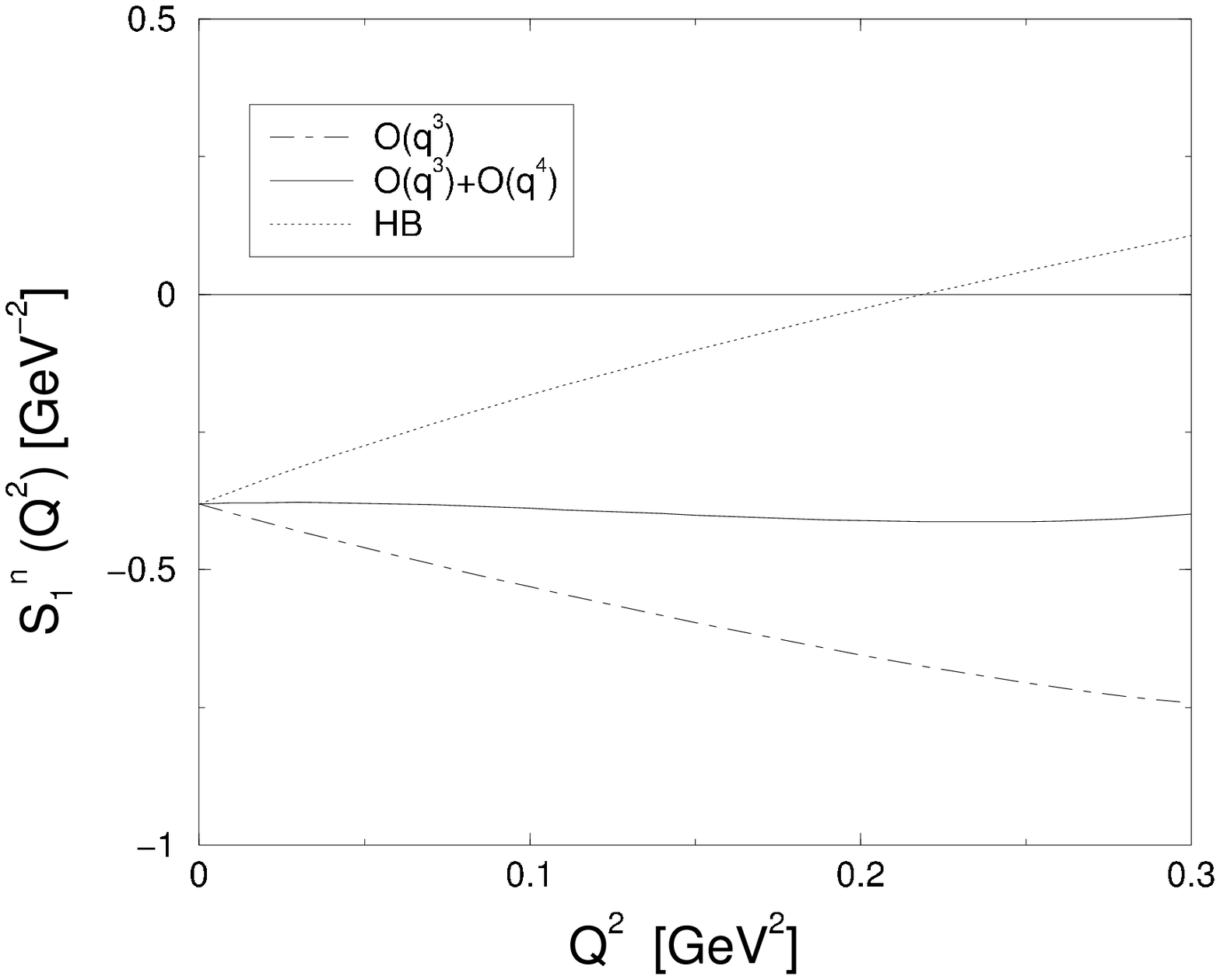}
\caption{
          Chiral loop contribution to the structure function $\bar{S}_1
          (0,Q^2)$ with the elastic contribution subtracted. 
          The solid (dot-dashed) line gives the result of
          the calculation~\protect\cite{Bernard:2002bs} to order $q^4$ ($q^3)$ in comparison
          to the heavy baryon result of \protect\cite{Ji:1999pd} (dotted line).
          Left (right) panel: Proton (neutron).
	  Source: Ref.~\protect\cite{Bernard:2002bs}
}
\label{fig:s1pn}
\end{figure}

At $Q^2 > 1$~GeV$^2$ QCD operator product
expansions should yield the value of 
$\bar{S}_1$ while at $Q^2 < 0.1$~GeV$^2$ chiral perturbation theory
calculations are used. However, it turns out that the chiral calculations
have not yet converged. A comparison of calculations in the
heavy baryon approach by Ji, Kao and Osborne~\cite{Ji:1999pd} with
calculations done by Bernard, Hemmert and
Meissner~\cite{Bernard:2002bs} shows that both approaches do not agree
and moreover, that the chiral expansion has not yet
converged (see Fig.~\ref{fig:s1pn}). The
level of uncertainty already at $Q^2 = 0.1$~GeV$^2$ for
$\bar{S}_1(Q^2)$ is of the order of 50~\% while the value of $\bar{S}_1$
at $Q^2=0$ is already taken from the \gdhsr prediction. 

\begin{figure}
\begin{center}
\includegraphics[width=0.85\textwidth]{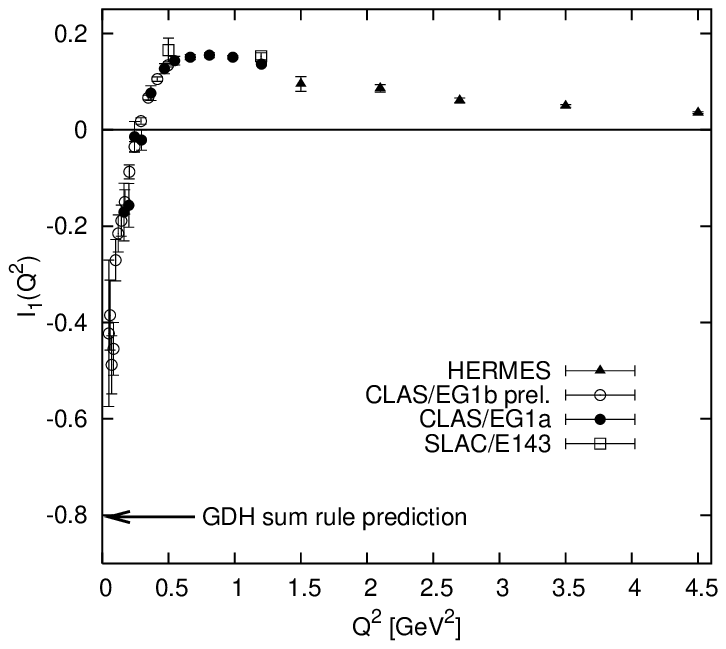}
\end{center}
\caption{The generalized GDH integral $I_1(Q^2)$ as defined in the
text. Shown are data from the \textsc{Hermes} experiment at
\textsc{Desy}~\protect\cite{Airapetian:2002wd} (filled triangles),
from the E143 experiment at \textsc{Slac}~\protect\cite{Abe:1996ag}
(open squares), from the EG1a experiment using the \textsc{Clas}
detector at \textsc{JLab}~\protect\cite{Fatemi:2003yh} (filled
circles) and preliminary data from the EG1b
experiment~\protect\cite{Prok:GDH2004} (open circles).}
\label{fig:I1p}
\end{figure}

Despite, there is no stringent rule to these integrals established at
finite $Q^2$, one can study the transition from hadronic degrees of
freedom to partonic structure. Fig.~\ref{fig:I1p} shows this
transition in terms of the generalized GDH integral $I_1(Q^2)$. At
large $Q^2$ with $Q^2 > 2$~GeV$^2$ one observes a $1/Q^2$ behavior
which is due to Bjorken scaling. Around $Q^2=1$~GeV$^2$ a dramatic
change sets on and at about $Q^2 = 0.25$~GeV$^2$ the sign of the
generalized integral changes toward the negative value of the GDH sum
rule prediction. At even lower momentum transfer the generalized
integral shows a very steep slope. 

However, even at the lowest
momentum transfer accessible nowadays of $Q^2 \simeq 0.05$~GeV$^2$
one is still about a factor of 2 away from the \gdhsr
prediction. Taken together with the observation of the steep slope in
Fig.~\ref{fig:I1p} it appears hopeless to estimate the GDH integral at
the real photon point with a reasonable precision and it appears
imperative to measure the GDH sum rule at the real photon point with
exactly $Q^2 = 0$. This measurement at the real photon point is what
we will focus on in the following. 

\newpage

\section{The \gdhex at ELSA and MAMI}
The magnetic moment of the proton in nuclear magnetons $µ_p/µ_N$, is
the ratio of the spin axis precession frequency of a proton in a
magnetic field to the frequency of the proton's orbital motion in the
same field, called the cyclotron frequency. In this ratio the
typically dominant experimental error, the magnetic field strength,
cancels mostly. Frequencies are amongst the most precisely measurable
quantities in physics. Consequently, the proton anomalous magnetic moment is
known today with a relative precision of $10^{-8}$. Also the mass of
the proton is known with that precision. That is why the \gdhsr
is a very stringent prediction.

To verify the \gdhsr the integral in photon energy over the
polarized total cross sections -- i.e. the left hand side of
Eq.~(\ref{eqn:gdh}) -- has to be determined experimentally.

\subsection{Experimental concept}
The dynamic observables on the left hand side of the
Gerasimov-Drell-Hearn sum rule (Eq.~(\ref{eqn:gdh})) need to be
measured in a large energy range to ensure that contributions from
unmeasured energy regions only represent minor uncertainties. 
The \gdhcol\footnote{For a member list of the
\gdhcol see for example Ref.~\cite{Dutz:2003mm}} has chosen
to perform the measurement of the integrand of the sum rule at two
accelerators: \elsa\footnote{\elsa: \underline{El}ectron
\underline{s}tretcher \underline{a}ccelerator} 
in Bonn and \mami\footnote{\mami: \underline{Ma}inz \underline{mi}crotron} in 
Mainz, Germany. This covers the energy
range from pion threshold\footnote{The data, however,  in the energy range
from 140~MeV through 200~MeV are currently still under analysis with
respect to the total photoabsorption cross section.} at 140~MeV up to 3~GeV. 
The measurements at \mami are dedicated to the lower energy part up to 800~MeV
while, with an overlap, the measurements at \elsa address
photon energies of 600~MeV through 3~GeV. In total this allows to
cover the whole resonance region and to reach the onset of the Regge
regime. The resonances allow to study the hadronic spin structure in
detail while the Regge regime ultimately provides a description of the
part of the integration up to infinite energies that is not
accessible experimentally.

The photons needed to study the photoabsorption cross section are
produced by brems\-strahlung of the primary electrons from the
accelerators (Sec.~\ref{sec:tagging}). At both accelerator sites a
tagging spectrometer is used to identify the photon energy and to
determine the photon flux.

The relative helicity states of photon and proton of $3/2$
(parallel) and $1/2$ (antiparallel) are obtained by a fixed polarized solid
state target (Sec.~\ref{sec:poltarg}) and by means of a
polarized electron beam (Sec.~\ref{sec:epol}). The polarization of the
electrons is (partially) transfered to the photons in the
bremsstrahlung process (see Sec.~\ref{sec:tagging}). The degree of
polarization of the electron beam is obtained by M{\o}ller polarimetry
(Sec.~\ref{sec:moller}).

Finally, the cross sections for the different spin configurations of
the \gdhsr have to be determined. As discussed in
Secs.~\ref{sec:OptTheo} and~\ref{sec:LowTheorem} it is sufficient to
focus on photoabsorption which is experimentally more
convenient 
than measuring the total cross section including elastic
contributions. The lowest energy with a non-vanishing photoabsorption
cross section is the pion threshold at 140~MeV photon energy in the
nucleon rest frame.
The photoabsorption cross sections are determined by hadronic final
states with two detectors: The \gdhdet at 
\elsa and \textsc{Daphne} at \mami together with additional components
in forward direction (see Sec.~\ref{sec:detectors}). Since the difference
$\sigma_{3/2}-\sigma_{1/2}$ is only about a 0.1~\% effect compared to
unpolarized total event rates -- given the experimental conditions like
effective polarization and background from unpolarized material --
these detectors need to be capable of determining the absolute cross
sections very reliably.

\begin{figure}
\begin{center}
\includegraphics[width=\textwidth]{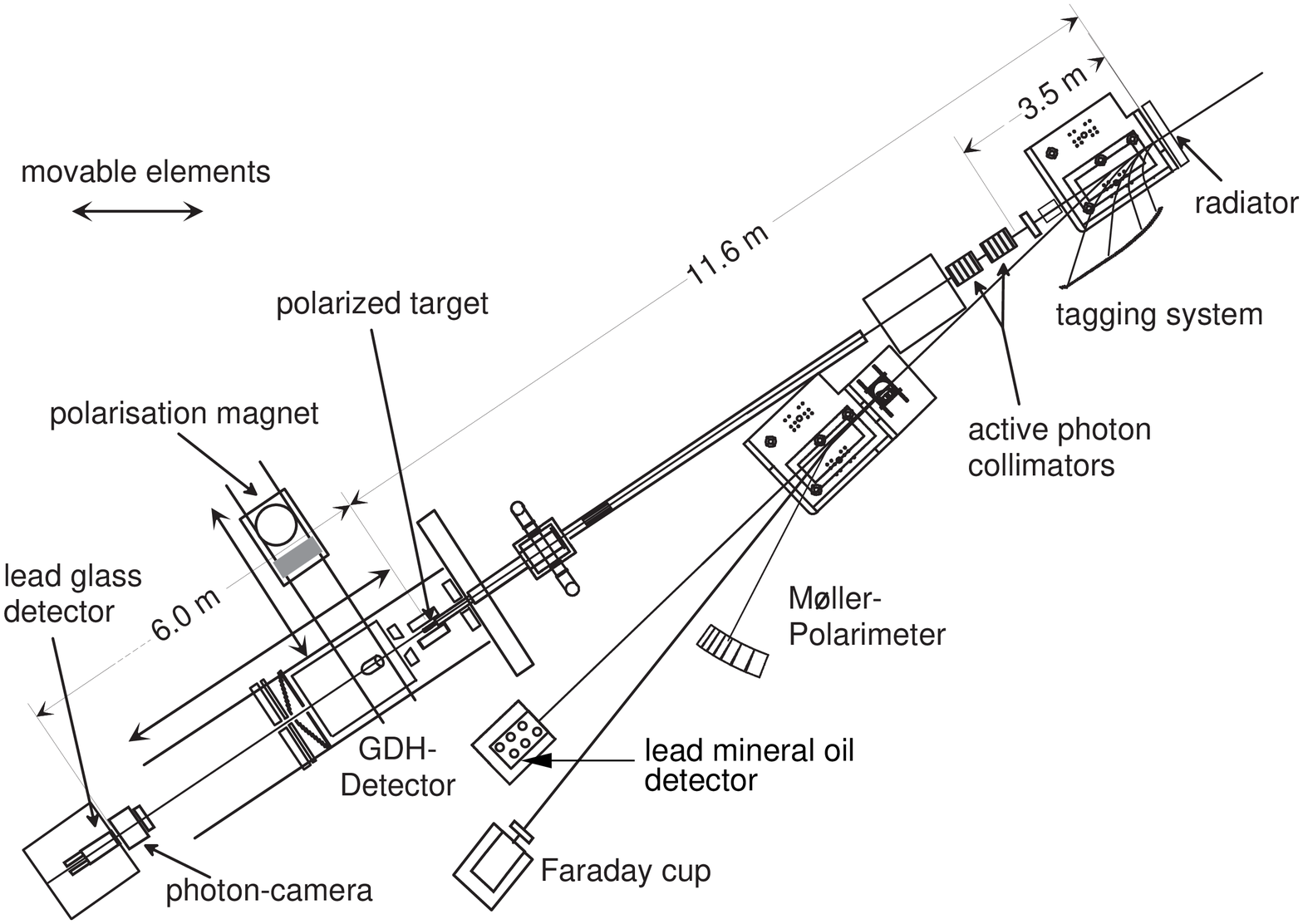}
\end{center}
\caption{Experimental setup of the \gdhex at \elsa}
\label{fig:GDHelan}
\end{figure}
\begin{figure}
\begin{center}
\includegraphics[width=\textwidth]{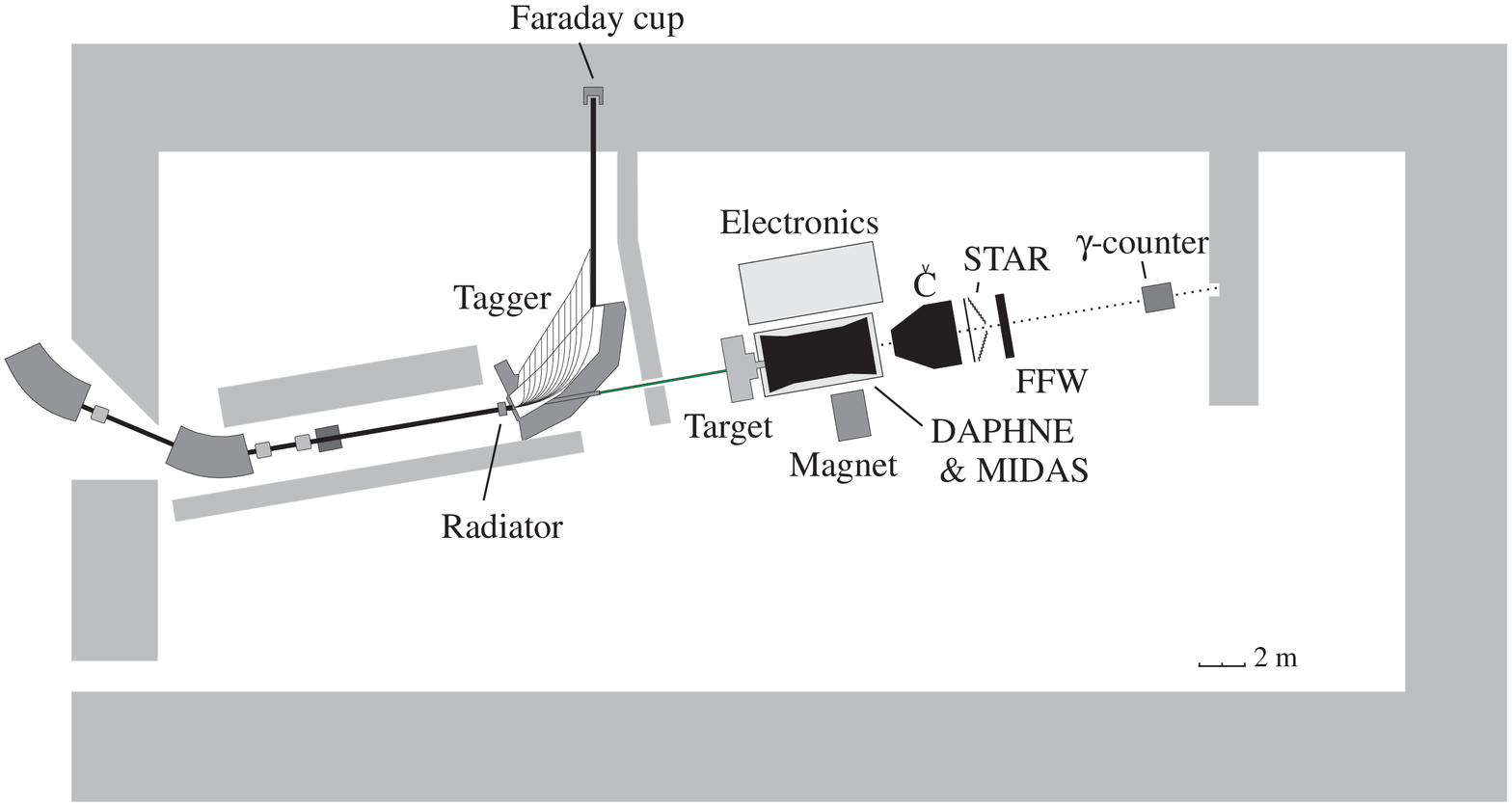}
\end{center}
\caption{Experimental setup of the \gdhex at \mami}
\label{fig:GDHa2}
\end{figure}
Fig.~\ref{fig:GDHelan} shows the experimental setup of the
\gdhex at \elsa. The setup at \mami is shown in
Fig~\ref{fig:GDHa2}. The setup at \mami is very
similar to the one at \elsa with the exception of the \molpol and the
subsequent devices that are not present at \mami. At \mami
Møller polarimetry was incorporated into the tagging system (see
Sec.~\ref{sec:moellermami}). In both cases the electron beam first
impinges on the bremsstrahlung radiator. At \elsa the primary
electrons then reach the \molpol. The photon beam is collimated and
guided through a vacuum system to the polarized target. The polarized 
target is hermetically surrounded by a detector (the
\gdhdet or \daphne) which determines the total cross section. At \elsa
both the lead glass detector and the lead mineral oil detector serve
as vetos for background processes. The beam dumps for photon and
electron beams contain beam diagnostic devices.

\subsection{Electron beam polarization}
\label{sec:epol}

\subsubsection{Polarized electrons at MAMI}
\begin{figure}
\begin{center}
\includegraphics*[width=0.6\textwidth]{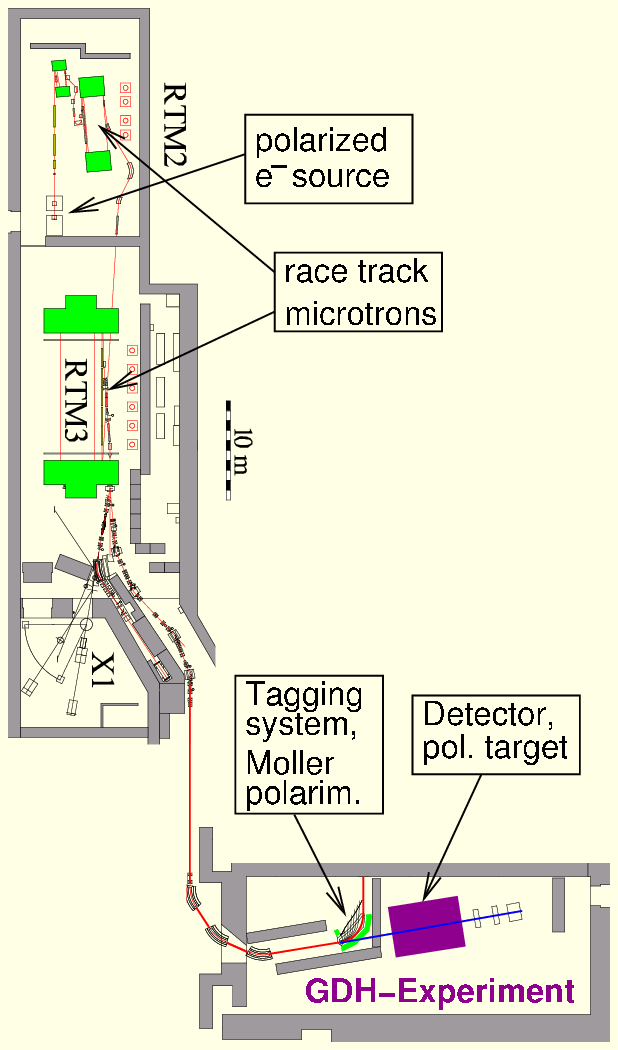}
\end{center}
\caption{The \gdhex at \mami. 
}
\label{fig:mami}
\end{figure}

The electron accelerator \mami B is operated by the institute
for nuclear physics of Mainz university. It serves experiments with
electrons (virtual photons) and real photons. Polarized electrons
can be accelerated up to a maximum energy of 855~MeV. A sketch of
\mami B including the experimental area of the \gdhex
is represented in Fig.~\ref{fig:mami}. The pulse frequency of the
accelerator is 2450 MHz, which corresponds to a bunch distance of
about 400~ps. 
 
Polarized electrons are produced by photoelectric effect at a gallium
arsenide crystal~\cite{Steffens:1993ih}. A ``strained layer'' of a 
GaAs$_{0.95}$P$_{0.5}$ photocathode is exposed to circularly polarized
laser light.  The obtained electron current is over 10 $\upmu$A with a
polarization degree of approximately 75~\%. In the magnetic dipole
fields of the accelerator the spin of the electrons rotates faster than
the angular frequency because of the g-factor anomaly. 
The beam polarization orientation at the radiator for bremsstrahlung
depends on the beam energy. 
The injection system of the polarized electron source is too compact
to incorporate a spin rotating system to compensate for this. Instead,
the longitudinal orientation of the polarization
at the bremsstrahlung radiator is achieved by fine tuning of the exact
energy gain of the microtron and the input energy. 
Two electron beam energies 855~MeV and 525~MeV were used for the \gdhex.

For the \gdhex the spin orientations parallel and
antiparallel to the target spin are used. To obtain these two
orientations and to minimize systematic effects the helicity of the
laser light at the polarized source is changed every two seconds.

\subsubsection{Polarized electrons at ELSA}

\begin{figure}
\begin{center}
\includegraphics*[angle=90,height=0.95\textheight]{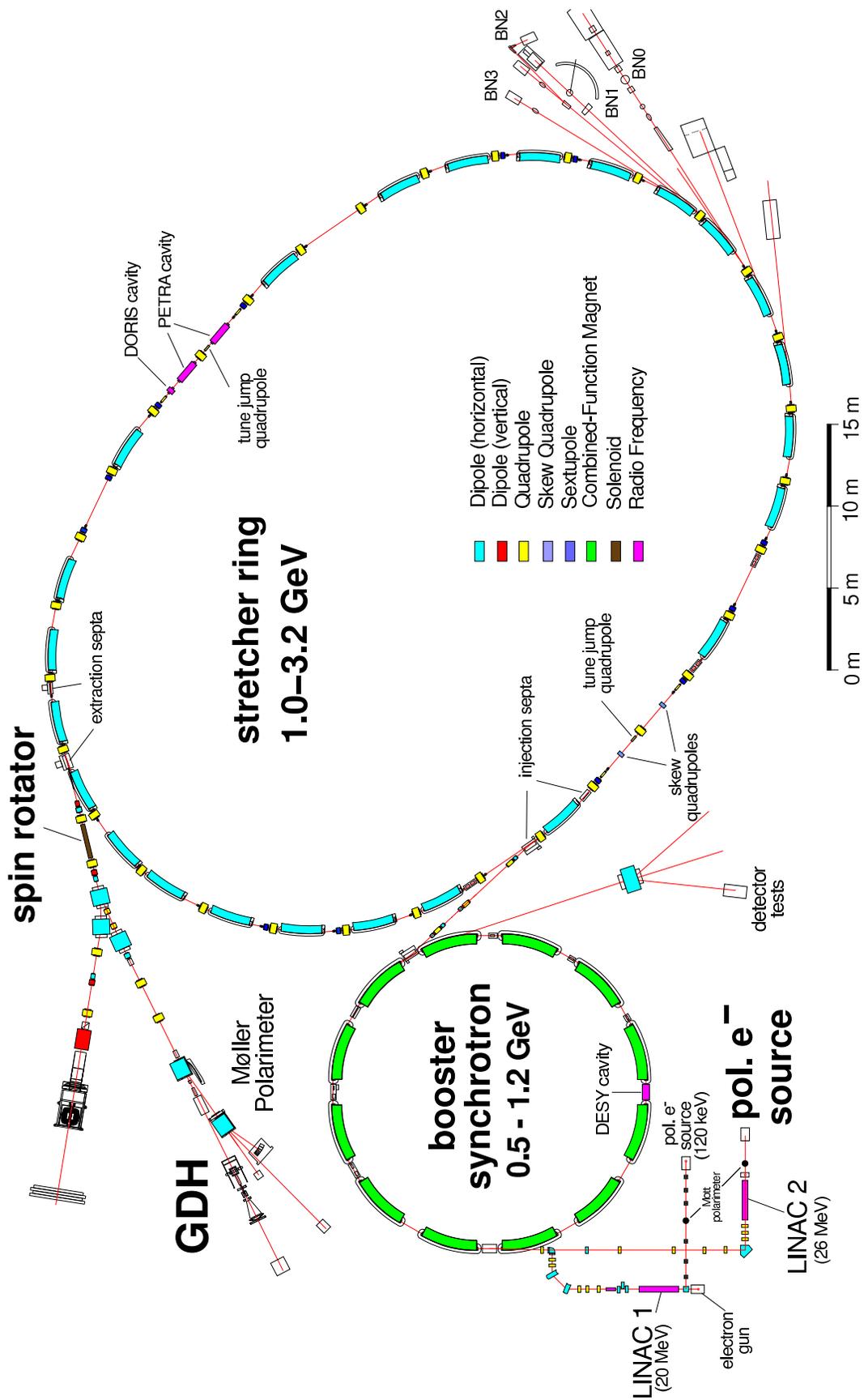}
\caption{The \gdhex at \elsa. Shown in bold face are the
components crucial for the electron beam polarization and the GDH
experiment. } 
\label{fig:elsa}
\end{center}
\end{figure}

The electron accelerator \elsa is operated by the physics
institute of the university of Bonn.
Fig.~\ref{fig:elsa} shows the general layout of the electron
accelerator \elsa which consists of 2 alternative \textsc{Linac}s with
corresponding electron sources, a synchrotron and the stretcher
ring. Electrons coming from one of the three available electron
sources (2 polarized and 1 unpolarized) are pre-accelerated in Linac 1
(120 keV polarized source or thermionic gun) or Linac 2 (50 keV
polarized source)~\cite{Hillert:2000xu}, respectively. This quasi continuous electron beam is
converted by a prebuncher into a 50 Hz pulsed beam before it is
injected into the synchrotron. In the synchrotron the electrons are
accelerated up to a maximum 
energy of 1.6 GeV. For the \gdhex 1.2 GeV were used. The electrons are then
injected into the stretcher ring. Up to 28 shots of the synchroton,
which corresponds to an injection time of 480 ms guarantee a
homogeneous filling of the stretcher ring.
After injection, the electrons are further accelerated (up to
3.5~GeV). The accelerating cavities are operated at 500~MHz which
corresponds to a bunch time structure with a period of 2~ns. The
electron bunches have a width of 50 ps. 
The electrons in the stretcher ring can either be stored inside the
ring for experiments with synchrotron light or they can be extracted
to external experiments. 
For the extraction of the electrons from the storage ring the spatial
distribution of the electrons is increased by magnetic
quadrupoles. Electrons at the edge of the beam are deflected into the
external beam line 
by 2 septum magnets. 

Polarized electrons at \elsa are available up to
3.2~GeV~\cite{Nakamura:1998fs,Hoffmann:2000xd} with an intensity 
of up to 2~nA at the experiment and a duty-cycle of up to 95~\%. 
A beam polarization of up to $73~\%$ parallel to the magnetic
field in the dipoles of the stretcher ring has been achieved. 
During acceleration in a ring accelerator with non-deterministic
particle tracks only the vertical polarization component is
conserved. Since the experiment requires longitudinally polarized
electrons, the electron spin has to be rotated in the external beam
line. By means of a super conducting solenoid magnet the vertical spin
is rotated around the longitudinal axis into the horizontal plane. 
In the adjacent dipole magnets the spin is rotated around the vertical
axis due to Thomas-precession into the longitudinal direction.

This process, however, can be incomplete. At 2.46~GeV, the maximum
field strength of the solenoid magnet is reached and a vertical spin
component remains. This effect together with the occurrence of
depolarizing resonances due to imperfections of the magnetic field of
the accelerator was the motivation to build a 
\molpol that allows to study all three spatial spin
components as fast as possible at \elsa (Sec.~\ref{sec:moellerelsa}). 

The electron beam helicity is randomly reversed at the source every
few seconds  to give access to the different relative spin
orientations.

\subsection{M{\o}ller polarimetry}
\label{sec:moller}
Mott polarimeters are employed to determine the degree of polarization
at low energies. At \elsa and at \mami, Mott
polarimeters are suited to monitor the performance of the electron
source before the acceleration process can have an impact on the
polarization.

Compton  and \mol polarimeters are used to measure the
polarization of high energy electron beams. While Compton polarimeters
are widely employed to measure the electron polarization of high
intensity beams for example in storage rings or linear accelerators,
{\molpol}s are the natural choice for low intensity
electron beams due to the large cross section. 

\subsubsection{M{\o}ller polarimetry at ELSA}
\label{sec:moellerelsa}
\begin{figure}
\begin{center}
\includegraphics[width=0.9\textwidth]{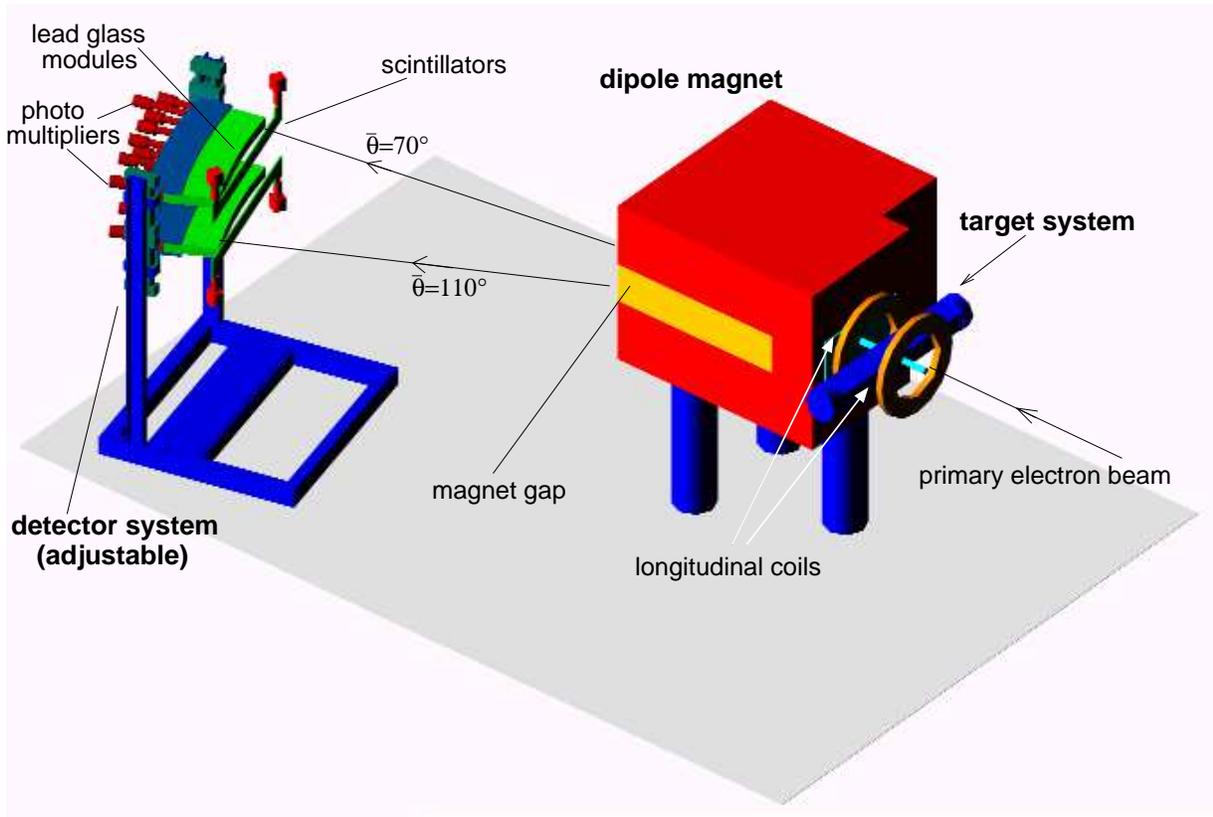}
\end{center}
\caption{The GDH-M{\o}ller-Polarimeter}
\label{fig:moeller}
\end{figure}
\begin{figure}
\begin{center}
\includegraphics[width=0.7\textwidth]{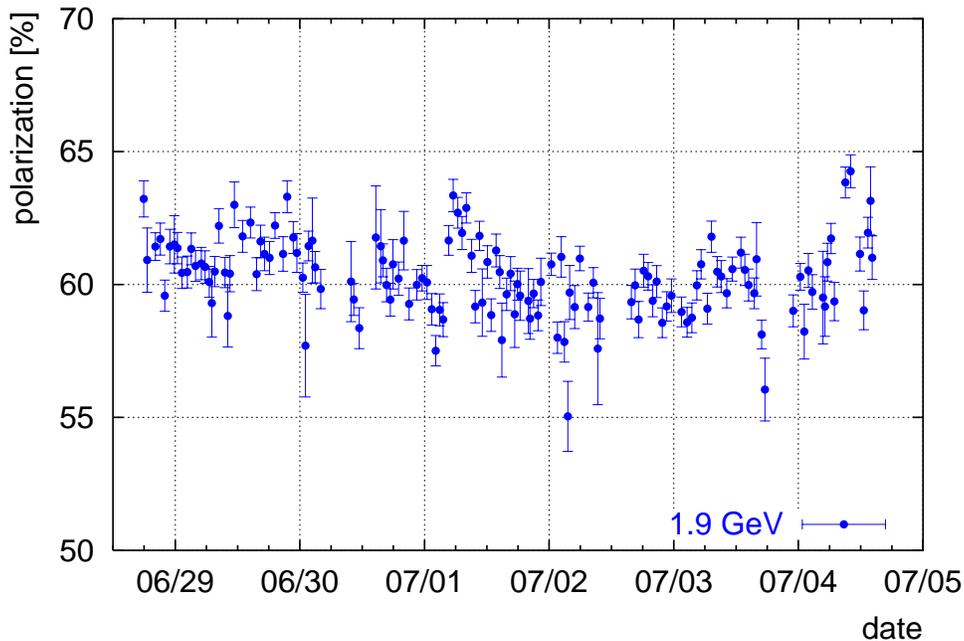}
\end{center}
\caption{Longitudinal electron polarization monitored during a beam
time at 1.9 GeV at \elsa in June/July 2001.}
\label{fig:PolTime}
\end{figure}
For the \gdhex at \elsa a dedicated \molpol has
been designed~\cite{Speckner:2004kt}. It is situated in the primary electron beam
approximately 5~m behind the tagging system (see Fig.~\ref{fig:GDHelan})
and permanently measures the electron polarization during data taking
with the \gdhdet.  
The polarimeter employs a dipole magnetic spectrometer and lead glass
counters to detect both \mol scattered electrons in
coincidence (see~Fig.~\ref{fig:moeller}). The scintillators are used
in coincidence with the lead glass detectors to further improve
background rejection. A target system consisting
of three different pairs of coils provides the field to
magnetize Vacoflux\footnote{Vakuumschmelze GmbH, Hanau, Germany} foils
in all three different space-orientations. This 
allows to measure all beam polarization components. The variable
geometry of the detector system enables adjustments to the kinematic
conditions between 0.8 and 3.5 
GeV electron beam energy. A large center of mass acceptance of
$\overline{\theta}=[65^o; 115^o]$ provides the measurement of the
longitudinal electron beam polarization for example at 1.9~GeV
with  a statistical precision of $1~\%$ within 10 min ($I_e=500$
pA). Fig~\ref{fig:PolTime} shows the absolute degree of polarization
versus time for a typical run period.
The systematic error of the polarization determination with the
GDH-\mol-Polarimeter is about 2~\% with the dominant source being the
uncertainty in the polarization of the magnetic foils.
Unprecedented systematic studies of M{\o}ller polarimetry as well as
direct measurements of the transversal beam polarization components
have been performed with this device. 
The polarimeter has been used extensively to determine the electron beam
polarization during accelerator tuning and the data-taking for the
verification of the Gerasimov-Drell-Hearn data taking period at \elsa.

\subsubsection{M{\o}ller polarimetry at MAMI}
\label{sec:moellermami}
With a deterministic race track beam polarization transport is
substantially less problematic at \mami than at \elsa.
Also, the polarization at the bremsstrahlung radiator was always
alinged longitudinally. Hence, no dedicated detector components for
the identification of M{\o}ller 
electrons were built. Instead, the existing plastic
scintillators of the photon 
tagging system were used for M{\o}ller polarimetry, which were read
out in coincidence by means of additionally installed electronics. 
Thus, the tagging spectrometer 
simultaneously served the purpose of photon energy
measurement and as a long-term polarimeter. The acceptance of the \molpol
was limited by the geometry of the vacuum system and by the
distance of the poles of the deflecting dipole magnet only.

Three to four hours are needed to achieve a statistical error of
1.5~\%. In order to adjust the spin orientation at the beginning of a
beam time instead of the \molpol a fast qualitative examination of the
polarization degree was done with a Compton
polarimeter~\cite{Drescher:1995,Fuerst:1998}. This polarimeter was
operated at much higher electron currents than usable during the
regular data acquisition.  

\subsection{Photon beam preparation}
\subsubsection{Photon polarization}
\label{sec:tagging}
\begin{figure}
\begin{center}
\includegraphics[width=0.65\textwidth]{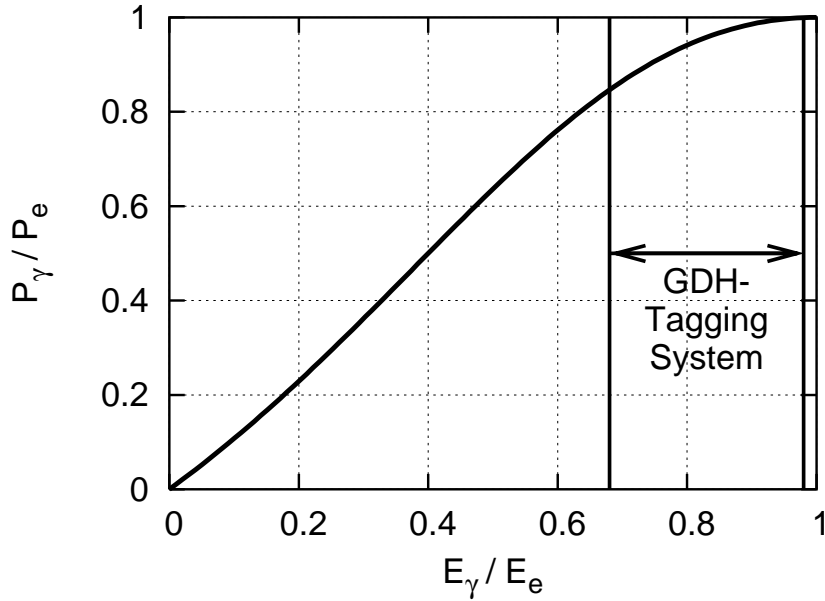}
\caption{Fraction of electron polarization transfered from the electron beam to the photon
beam according to Eq.~(\ref{eqn:helicity_transfer}.)
}
\label{fig:heltrans}
\end{center}
\end{figure}
The helicity transfer $h(k)$ connects the degree of circular
polarization $P_{\gamma, {\rm circ}}$ transfered to the
bremsstrahlung photon beam to
the longitudinal electron polarization $P_{e, {\rm long}}$ of the
relativistic electron beam \cite{Olsen:1959}: 
\begin{equation}\label{eqn:helicity_transfer}
h(k)=\frac{P_{\gamma,{\rm circ}}}{P_{e, {\rm long}}} = \frac{ k (3 + ( 1 - k))}
                       { 3 - 2(1-k) + 3( 1-k)^2}.
\end{equation}
$k=E_{\gamma}/E_0$ denotes the fraction of the energy $E_{\gamma}$ of
the photon produced by the primary electron with energy $E_0$. 
With the knowledge of the electron beam polarization $P_{e, {\rm
long}}$ and the photon energy $E_{\gamma}$ -- as determined by the
tagging system --  the circular polarization $P_{\gamma,{\rm circ}}$
of the energy tagged photon beam can be
calculated. Fig.~\ref{fig:heltrans} shows that the helicity is
transfered most efficiently at high energies. Therefore, the
GDH-Tagging-System described in Sec.~\ref{sec:GDHTagger} only uses the
upper third of the bremsstrahlung spectrum for tagging purposes. To cover
the energy range from about 680~MeV through 2.9~GeV at \elsa 
in total 7~primary electron energy settings were used: 1.0,
1.2~\footnote{used for measurements of the neutron cross sections
only}, 1.4, 1.9, 2.4, 2.9 and 3.0~GeV. At \mami two energy
setting were used to address reactions starting at the pion threshold
through 800~MeV: 525~MeV and 855~MeV. 

\subsubsection{Photon tagging}
\label{sec:GDHTagger}

The \gdhex requires the determination of the absorption cross section
of real photons.
A common and well established method of generating such a high-energy
photon beam utilizes the bremsstrahlung process:
A primary electron beam of energy $E_0$ impinges on a thin metal foil.
At  intermediate energies of the scattered electrons the
bremsstrahlung process dominates over electron-electron (\mol)
scattering, so most electrons that suffer any significant energy loss
in the radiator foil radiate a photon which is then used in a real
photon experiment. 

The residual electron energy $E_{\rm residual}$ is detected by
a magnetic spectrometer. Neglecting the small energy transfer
to the nucleus this allows to determine the photon
energy $E_\gamma$:

\begin{equation}\label{brems_energy}
 E_\gamma = E_0 - E_{\mathrm{residual}}
\end{equation}

The combination of a bremsstrahlung radiator foil and
a magnetic spectrometer is called a {\em tagging system}.
Such systems have become a standard component of
real photon experiments in the GeV range
\cite{Detemple:1992fs,Sober:2000we}. The tagging system
provides three essential parameters to the experiment:
\begin{itemize}
\item the energy of each photon impinging on the target,
\item the flux of photons of a given energy reaching the target and
\item the time information of each photon reaching the target.
\end{itemize}

\begin{figure}
\includegraphics[angle=-90,width=\textwidth]{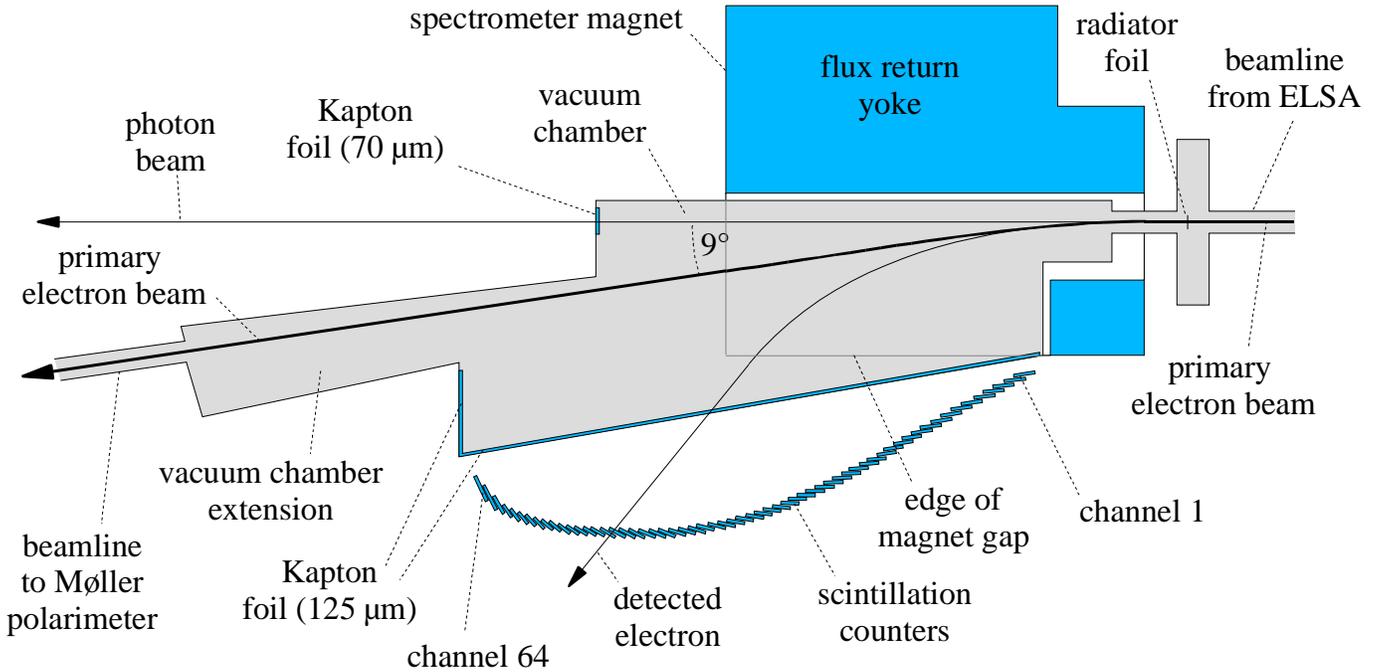}
\caption{The GDH tagging system at \elsa (top view of the focal plane)}
\label{fig:GDHTagger}
\end{figure}
\begin{figure}
\begin{center}
\includegraphics*[bb=135 515 415 750,width=0.75\textwidth]{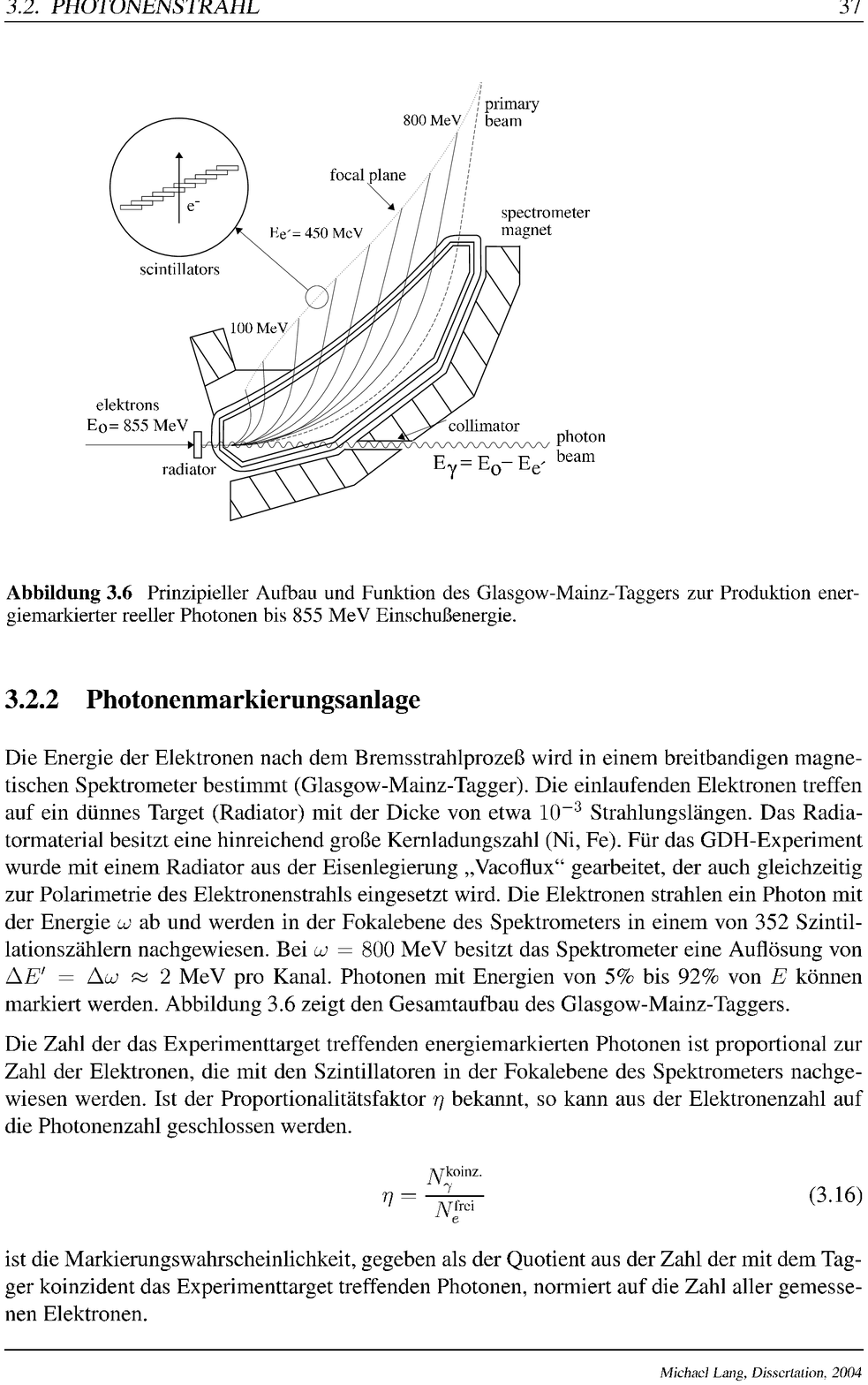}
\end{center}
\caption{The Glasgow-Mainz tagging system at \mami (top view of the focal plane)}
\label{fig:A2Tagger}
\end{figure}
Fig.~\ref{fig:GDHTagger} shows the GDH tagging system operated in
Bonn at \elsa~\cite{Naumann:2003vf}.
The system consists of a C-type dipole
magnet and a hodoscope of 65 scintillation counters
which are operated pairwise in coincidence to form 64 tagging
channels. Photons can be tagged in a range of
68-97~\% of the primary electron energy $E_0$. The energy
resolution ranges from 0.2 to 0.6~\% of $E_0$. The
time resolution is better than $\sigma = 165$~ps. The system was
operated at rates up to $5\cdot10^6$~photons/s in the full tagged
energy range. 
The {\em Glasgow-Mainz-Tagger}~\cite{Anthony:1991eq}
at \mami tags photons in the range of
5-92~\% of the primary electron energy and is shown in
Fig.~\ref{fig:A2Tagger}. Its 352 scintillation counters 
provide an energy resolution of about 0.2~\% 

\subsubsection{Collimation of the photon beam}
Due to the natural divergence of the bremsstrahlung process and the
emittance of the primary electron beam the photon beam has to be
limited in its divergence and in its transverse size by
collimation. The collimators are usually made of a block of lead
with a hole in the  center, which will be called passive collimators
in the following. This type of collimation was used in the GDH setup
at \mami. 

The collimators have to absorb a sizable proportion of the photons, i.e. 
10~\% to 90~\%, depending on the emittance and the required dimensions
of the beam at the hadron target. The following problem arises from
passive collimation: A high energy  
photon which is tagged by the tagging spectrometer may interact with
the collimator material and, by pair-production and bremsstrahlung, 
produces a shower of secondary electrons, positrons and photons. 
Part of this shower may not be absorbed in the collimator but pass the
holes of the collimation system.
While the 
charged particles are deflected away from  the beam with a so called
sweeping magnet, secondary photons may proceed along with the
beam to the target. These photons reach the  
hadron target with less energy than indicated by the tagging system.
With a hadron detector without kinematical over-determination,
which cannot determine the photon energy independently of the tagging
system, one cannot reject the corresponding events. 
Additionally, it is
imperative to detect and to suppress these secondary photons for a
precise determination of the tagging photon definition probability
(see Sec.~\ref{sec:pgamma}) and of the photon flux.  

Since the \gdhdet at \elsa has only moderate particle identification
capabilities this effect has to be
considered. Consequently, the active collimation technique described
in the following is employed at \elsa. At \mami the central detector
has particle identification and tracking. So, this problem is not
relevant for the measurements performed at \mami.

\begin{figure}
\includegraphics[width=\textwidth]{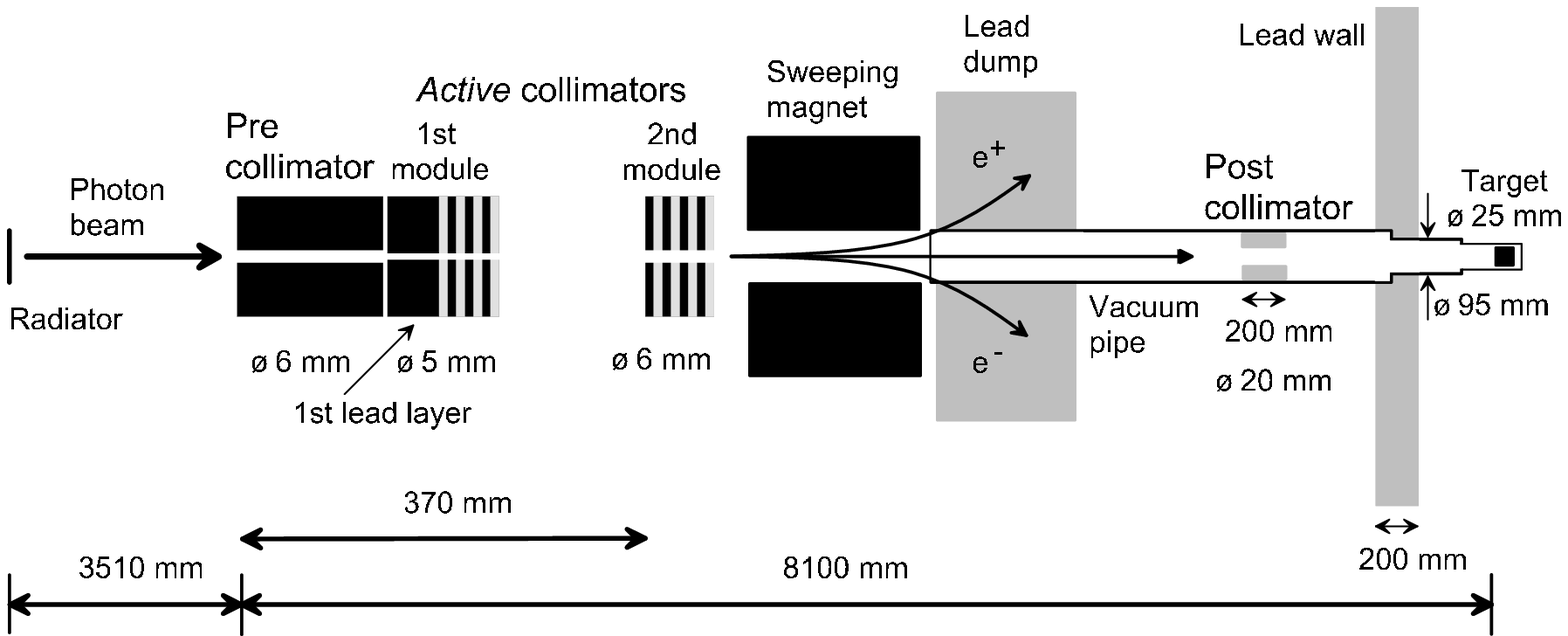}
\caption{Collimator setup and photon beamline at the \gdhex at 
\elsa (not to scale).}
\label{fig:ActColl}
\end{figure}

The idea of an active collimator is to produce a signal each time a photon
of the beam hits the collimator and interacts with its material~\cite{Helbing:1997}.
The photons are identified by their interaction with 
the collimator medium itself (active collimation). This 
can be realized with a sandwich of scintillator and lead layers. When
the photon hits the collimator medium it produces a shower in which a
secondary charged particle gives rise to a signal in the subsequent
scintillators. Fig.~\ref{fig:ActColl} shows the setup of
active collimation and passive shielding for the \gdhex the \elsa as
optimized by extensive simulations and test
measurements~\cite{Zeitler:2001wq}. 
The geometry with pre-collimation allows to operate this active
collimator system at rates of the tagging system exceeding 1~MHz.
The collimator system is able to
reject secondary low-energy photons with an efficiency of more that
99.9~\% for primary photon energies above 500~MeV. 

\subsubsection{Monitoring the photon definition probability}
\label{sec:pgamma}
An important property of a tagged photon beam is the photon definition
probability $\eta$ (sometimes also called {\em tagging efficiency}). It relates the
flux measured in each channel $i$ of the tagging system to the photon
flux in the tagged energy range at the experimental target. The precise
knowledge of $\eta$ is therefore vital for the correct determination
of the photon flux. $\eta_{i}$ for each tagging channel $i$ is defined by:
\begin{equation}
\eta_{i} = \frac{\dot{N}_{{\rm tag},i\wedge {\rm targ}}}
                     {\dot{N}_{{\rm tag},i}}
\end{equation}
$\dot{N}_{{\mathrm{tag}},i}$ is the count rate of tagging channel
$i$ and $\dot{N}_{{\mathrm{tag}},i\wedge {\mathrm{targ}}}$ is the
photon count rate at the experimental target in coincidence with
the tagging channel $i$.

In case of the \gdhex at \elsa $\dot{N}_{{\rm tag},i\wedge {\rm
targ}}$ was determined by a detector at the photon beam dump. A
totally absorbing lead glass \cer counter was used. This detector and
its photomultiplier with the 
corresponding circuitry has been specifically optimized to withstand
the high count rates right in the photon beam, even with regular
data-taking intensities. This allows a permanent measurement of the
photon definition probability during data-taking. The lead glass was
exchanged after about every two weeks of time in beam to avoid
inefficiencies due to the aging of the glass induced by the beam
radiation. 

In contrast, at \mami the lead glass \cer counter was only used for a
short time about every 24 hours during data-taking to obtain an
absolute calibration of $\eta$. For permanent monitoring a detector
consisting of 3 scintillators and a copper converter was used. The
copper converter was placed between the first two scintillators.
The second and third scintillators were operated in anticoincidence to
the first one in front of the converter -- thus counting the pair
creation rate of the photon beam. Under the assumption of constant
relative photon definition probabilities of the individual tagging
channels this allows to monitor overall drifts of $\eta$. 

The systematic uncertainty of the photon definition probability at
\mami and at \elsa is less than 2~\% for all primary energies. 

\subsubsection{Beam position monitoring}
The photon definition probability and hence the efficiency of the
tagging process depend critically on the alignment of the electron
and photon beam. At both experimental sites at \elsa and \mami beam
profile monitors at the location of the bremsstrahlung radiator were able
to determine the electron beam position to within fractions of a
millimeter. At \elsa also devices were installed at the photon
and electron beam dump as well as at the \mol target. 
These position monitors are combinations of wires and small
detectors. The wires are moved through the beam by stepping motors and
the produced secondary particles are detected subsequently.

In addition a camera observing a scintillator following a converter in
the photon beam was used for fast beam position
monitoring~\cite{Krimmer:2002yu} at \elsa and \mami. Due to the more involved electron
beam extraction procedure from \elsa its signal was fed back into the
accelerator control system to stabilize the beam position and
intensity. 

With the exception of the monitors at the bremsstrahlung radiators, all
these monitors were used online during the data taking.

\subsection{Polarized target}
\label{sec:poltarg}
One of the reasons, why doubly polarized experiments
were not feasible until recently and the \gdhsr has not been measured
since the 1960ies is the complexity of the polarized target and the
complication of its integration with a detector with almost complete
solid angle coverage.

\begin{figure}
\includegraphics[width=\textwidth]{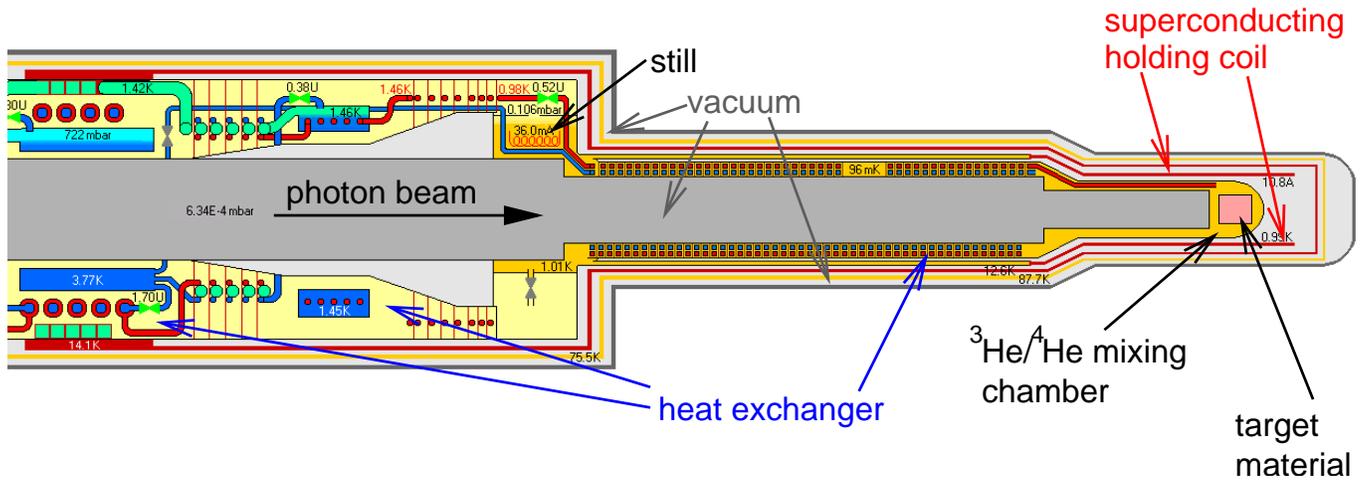}
\caption{Cryogenic system of the polarized frozen-spin target}
\label{fig:cryo}
\end{figure}
For the \gdhex a new ``frozen-spin'' target has been developed with the
$^3$He/$^4$He dilution refrigerator installed horizontally
along the beam axis \cite{Bradtke:1999zg}. Frozen-spin mode
implies that the polarization is maintained at low temperatures of
typically less than 100~mK by only a low magnetic field strength of
about 0.5~Tesla. 
The refrigerator includes an internal superconducting holding coil to
provide this longitudinal field for nucleon polarization in the
frozen-spin mode. The superconducting wire of this holding coil is
wound on the inner cooling shield of the vertical dilution
refrigerator. Fig.~\ref{fig:cryo} shows the cryogenic system of the
polarized target. The horizontal alignment of the cryostat together
with the small internal holding coil minimizes the solid angle
affected by these components and distortion of particle tracks due to
the magnetic field can be neglected. 

The initial polarization of the nucleons is obtained in a
slightly different mode. 
The temperature is raised to a few hundred milli-Kelvin to increase
the speed of electron spin relaxation. At the same time a magnet
external to the $^3$He/$^4$He cryostat increases the field to about
5~T. This magnet takes the place of the detector during the polarization
phase. Detector and polarizing magnet are movable on rails.
The proton or deuteron polarization in thermal equilibrium is very
small. However, the electrons with a much higher magnetic moment ($\mu_e
\simeq 660\mu_p$) are almost completely polarized.
The high polarization of the electrons is transfered to the protons or
deuterons by the {\em dynamic nuclear polarization} (DNP)
technique. Basically, a simultaneous spin flip of an electron and a
proton or deuteron transfers the spin polarization from the electrons
to the protons. This is achieved applying microwaves of
appropriate frequency. A nuclear magnetic resonance (NMR) system measures the
degree of polarization of the protons or deuterons.

However, most electrons of the intact molecules of the target material
are paired. Only electrons of chemical radicals can be used for this
polarization transfer. 
Butanol (C$_4$H$_9$OH) has been used as a polarized proton target at
\elsa and at \mami. For the polarized deuteron target at \mami
deuterated butanol (C$_4$D$_9$OD) was used while lithium-deuteride
($^6$LiD) was used for the \gdhex at \elsa. The butanol and deuterated
butanol have been chemically prepared for the DNP process with the
highly stable free radical porphyrexid
(\,(CH$_3$)$_2$C$_3$(NO)(NH)(NH)$_2$\,). The paramagnetic centers in
the $^6$LiD were created by irradiating the target material with the
20~MeV electron beam of the Bonn injection linac.

\begin{figure}
\begin{center}
\includegraphics[width=0.65\textwidth]{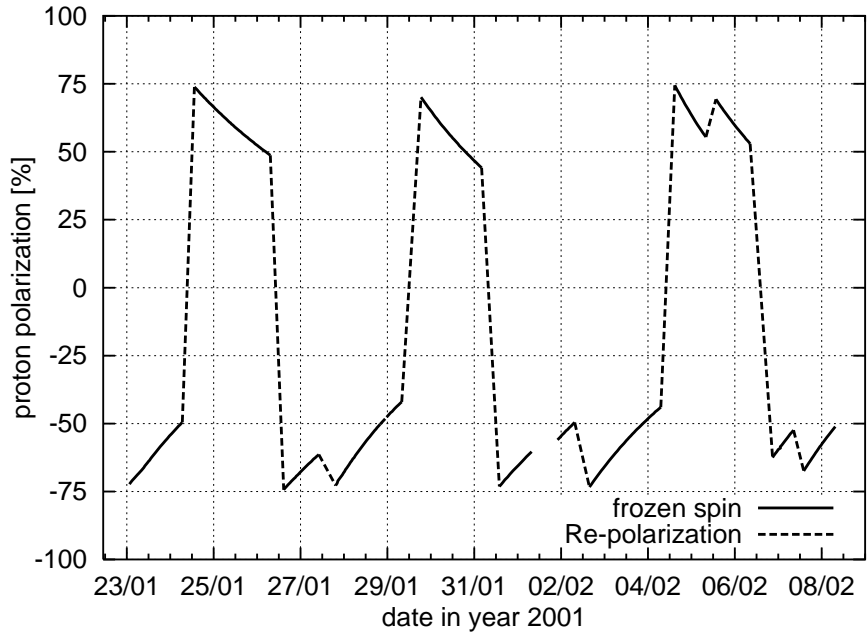}
\caption{Proton polarization buildup, repolarization and relaxation
during a GDH run at \elsa with a regular butanol target providing
polarized protons.}
\label{fig:PTrelax}
\end{center}
\end{figure}

Typical proton polarization values of 70-80~\% and
deuteron polarization values of 60~\% have been obtained.
Polarization relaxation times of 200~h (protons) and 100~h (deuterons)
have been measured under ideal conditions with butanol. The average deuteron
polarization for the $^6$LiD was 27~\% with a relaxation time in
excess of 1000~h. Fig.~\ref{fig:PTrelax} shows the typical cycle of
polarization buildup, repolarization and relaxation during data-taking
periods from the proton.

While the interpretation of the polarization as determined with the
NMR system for the proton is straight forward the situations is more
involved for the deuteron since the nuclear binding modifies the naive
picture of two separate nucleons without interaction. For the
\gdhsr investigations, however, the individual polarizations of the
protons and neutrons are relevant\footnote{A \gdhsr for the deuteron
also exists but photo-disintegration of the deuteron is the dominant
process contributing to the integral. Since this has not yet been
measured in a doubly polarized experiment from the disintegration
threshold to the pion threshold we will focus on the sum rules for the
proton and the neutron.}. In the naive picture the proton and the
neutron inside the deuteron have parallel spin in the same direction
as the deuteron. The dominant correction for the deuteron is
the D-state orbital angular
configuration~\cite{Melnitchouk:1994tx,Rondon:1999da}. This correction
reduces the net nucleon polarization in the deuteron to 93~\% of that
of the deuteron.

To a first approximation the $^6$Li can be regarded as being composed
of a spinless $\alpha$ particle and a proton and a neutron carrying the
total spin-1 of the nucleus. In this picture the target material
$^6$LiD has almost 50~\% of the nucleons' polarized i.e. a polarization
dilution factor of $1/2$. This is to be
contrasted to butanol with a dilution factor of $10/74$.
Unlike for the deuteron the valence nucleons of the $^6$Li can be
found also in a P-state. The net nucleon polarization of the unpaired
neutron and proton is 87~\% of the $^6$Li
polarization~\cite{Rondon:1999da}.
In principle both the polarization of the deuterons and the $^6$Li
need to be measured separately. These are however connected by the
{\em equal spin temperature} (EST)
concept~\cite{Abragam:1976}. Basically all nuclear species in the
target material share the same spin temperature i.e. the Zeeman
occupation numbers of each system can be calculated from the
respective Brillouin function. Owing this correlation the NMR system
was only used to monitor the polarization of the deuterons of the
$^6$LiD target.

Small contaminations of the $^6$Li by $^7$Li and of the deuterium by
regular hydrogen ($^1$H) both for the $^6$LiD target as well as for the
deuterated butanol have been taken into
account~\cite{Ball:2003vb}. For the two butanol targets also the 
admixture of the radical prophyrexid dissolved in water representing
a 5~\% contribution by weight are relevant for the computation of the
densities of polarizable nucleons. 

The target container length was $l = 18.8$~mm for the setup at \mami
and $l = 28.8$~mm for the setup at \elsa. The diameter of this PTFE
container immersed in the $^3$He/$^4$He mixing chamber was 2.0~cm at
\mami and 2.6~cm at \elsa. The butanol was prepared in frozen beads
(ball shaped)
with 2~mm diameter while the $^6$LiD material has the shape of
granules (flat blanks). The stacking inside the target container has
been determined  in detailed investigations~\cite{Rohlof:2004ec} to
arrive at the target column density.

\subsection{Photoabsorption detectors}
\label{sec:detectors}
Two detector arrangements are used to meet the specific requirements for
the different energy ranges: The \daphne detector with additional
components in forward direction at \mami and the \gdhdet at \elsa. 

\textsc{Daphne}~\cite{Audit:1991gq} is designed for charged
particle detection and for the identification of low multiplicity final
states. It is essentially a charged particle tracking detector having a
cylindrical symmetry. In addition it has a useful detection efficiency for
neutral pions. In forward direction a silicon microstrip device
called \textsc{Midas}~\cite{Altieri:1999yr} extends the acceptance for
charged particles.

The \gdhdet~\cite{Helbing:2002kk} has been specifically
designed for measurements of total cross sections and is designed
for situations where the contributing reaction channels are not well
known and extrapolations due to unobserved final states are not
advisable. On the other hand, the identification of all individual
processes is not required for the determination of the GDH integrand.
The concept of the \gdhdet is to detect at least one
reaction product from all possible hadronic processes with almost
complete acceptance concerning solid angle and efficiency. The
acceptance for hadronic processes is better than 99~\%. 

\begin{figure}
\includegraphics[width=\textwidth]{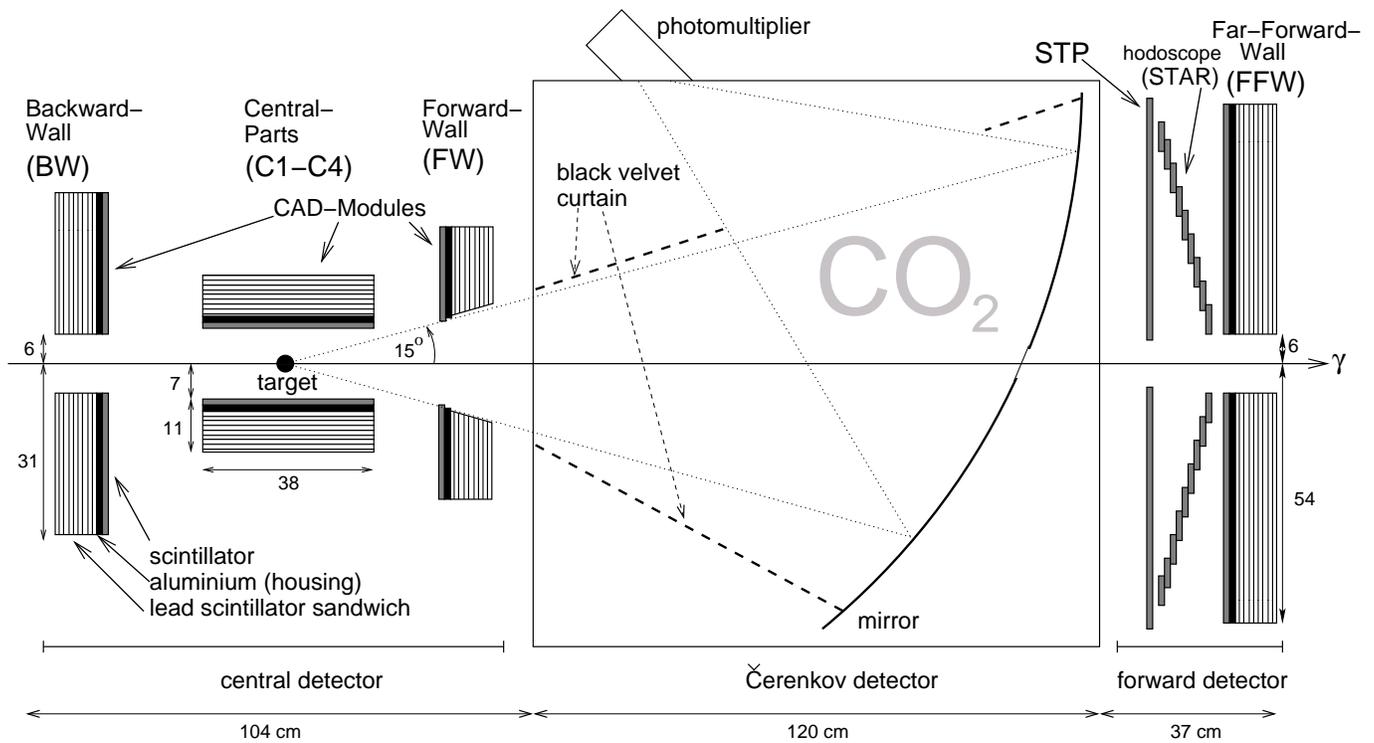}
\caption{The \gdhdet used at \elsa for energies above 680~MeV}
\label{fig:gdhdet}
\end{figure}

\begin{figure}
\begin{center}
\includegraphics[width=\textwidth]{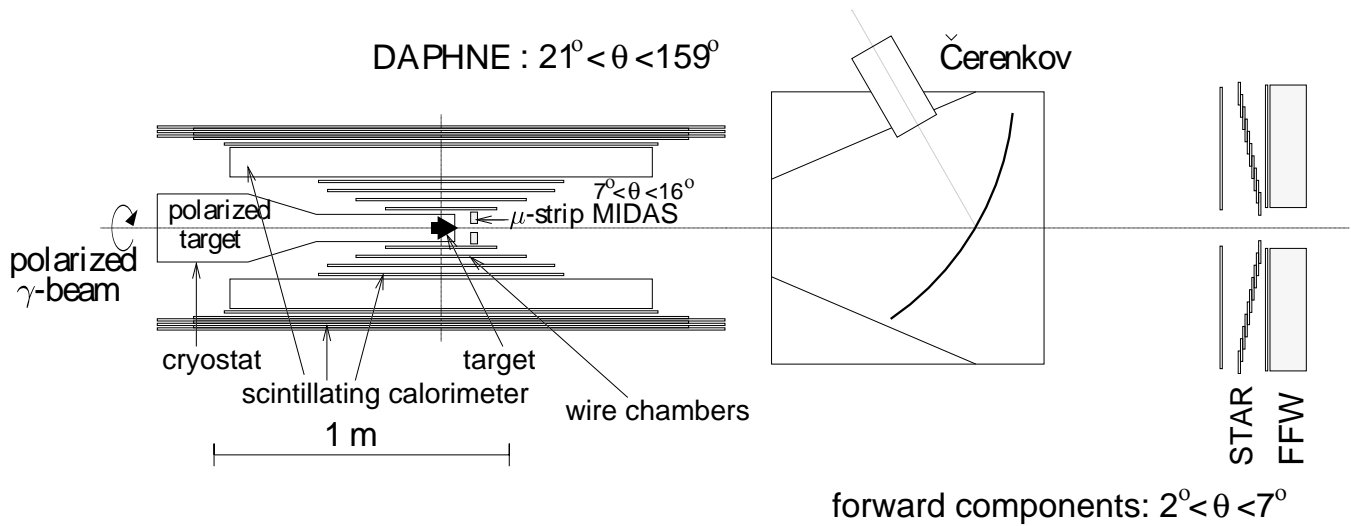}
\caption{The \daphne detector setup with forward components at \mami
for energies below 800~MeV}
\label{fig:daphne-setup}
\end{center}
\end{figure}

Both detection systems have similar components in forward direction
(see Figs.~\ref{fig:gdhdet} and~\ref{fig:daphne-setup}).
The electromagnetic background is suppressed by about 5 orders of
magnitude by means of a threshold \cer
detector~\cite{Helbing:2002kk}. The \cer detector is followed by
the \textsc{Star} detector component~\cite{Sauer:1996gg} (a
scintillator hodoscope to resolve forward polar angles) and the
Far-Forward-Wall~\cite{Helbing:2002kk} (a component similar to the
central parts of the \gdhdet) to complete the solid angle coverage.


\subsubsection{The \gdhdet}
A cut through of the \gdhdet~\cite{Helbing:2002kk} is shown in
Fig.~\ref{fig:gdhdet}. Apart from the reliable detection of charged
particles a high efficiency for neutral decay modes of mesons is
necessary in order to increase the total detection probability of the
respective photoabsorption channel. 

The detection of the charged particles is achieved by plastic scintillators; the
decay photons are observed in lead scintillator sandwich detectors behind
it. The modules combining these two tasks have been named
CAD-Modules\footnote{Charged particle And Decay photon detector}. The
CAD-Modules are arranged to cover a solid angle of 99.6~\%$\ \times \
4\pi$. The central detector including the backward and
forward wall covering the polar angles $15^\circ \leq \theta \leq
174^\circ$ and also the far-forward wall covering $1.6^\circ \leq
\theta \leq 15^\circ$ are CAD-Modules. In all the hadron detection of
the \gdhdet provides a solid angle acceptance of $4\pi$ -- with only
0.02~\% solid angle leakage in forward direction and 0.4~\% in
backward direction. 

Charged particles are detected with almost 100~\% probability in the
front scintillator of the CAD-Modules.
The sandwich structure behind is comparable to a calorimeter; however,
it was designed with special regard to the efficient detection of
low-energetic decay photons and good time resolution. Fig.~9 of
Ref.~\cite{Helbing:2002kk} shows that a probability of
88~\% was reached for the detection of photons 
with energies above 50~MeV. Altogether for the
detection of the decay of e.g. a $\pi^0$-mesons an efficiency
of at least 98~\% results since the two decay photons independently
produce a shower. 

Wavelength shifters embedded in plastic light guides are used to minimize the
number of photomultipliers needed to read out the CAD-Modules. We have
obtained a time resolution of 500~ps even for the 1.2~m$^2$ large
far-forward CAD-Module and an energy resolution for photon energies above
60~MeV of 
\begin{equation} 
\frac{\Delta E}{E} = \frac{9.7~\%}{\sqrt{E/\mbox{GeV}}} \label{sigE} \quad ,
\end{equation}
has been achieved. The spatial resolution was found to be 15~cm for
both charged particles and decay photons.

\begin{figure}
\begin{center}
	\includegraphics[width=0.7\textwidth,height=9.7cm]{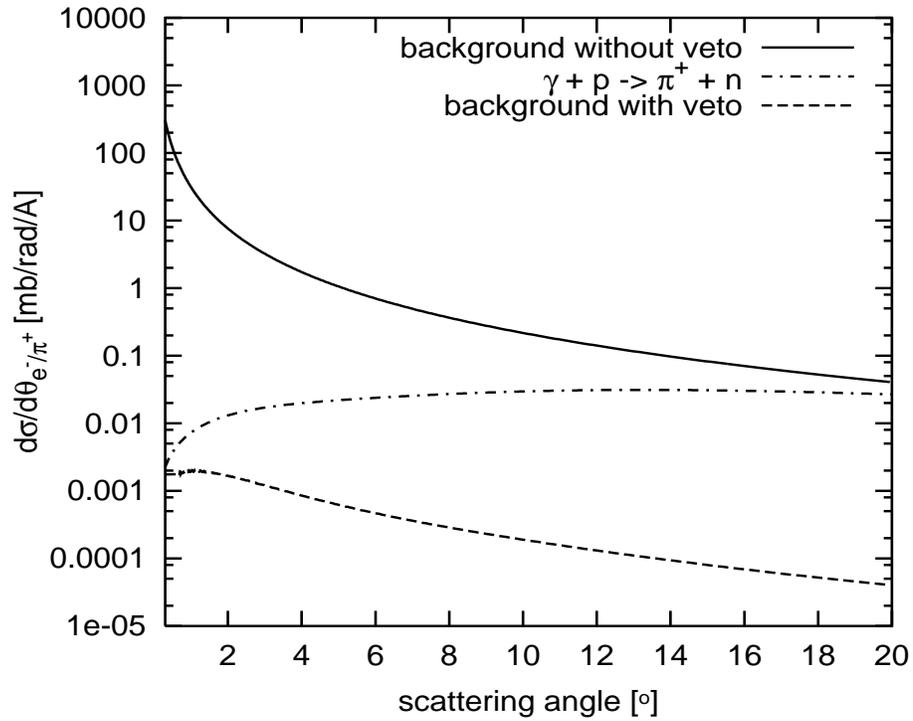}
	\caption{Simulated angular distribution of electrons originating from
	electromagnetic background (normalized to the number of
	nucleons) with and without \cer veto
	compared to a typical hadronic process ($\gamma p \to \pi^+ n$) at 1500~MeV photon energy.} 
\label{fig:elmagangle}
\end{center}
\end{figure}

However, the interactions of photons with the butanol
target are predominantly of electromagnetic nature: pair production of
electrons and 
positrons in the Coulomb field of the atomic nuclei and Compton
scattering off orbital electrons of the atoms. 
The angular distribution of this background as obtained in detailed
\textsc{Geant} simulations is shown in
Fig.~\ref{fig:elmagangle} in comparison to an important partial
channel of the photoabsorption process.
One observes that the background has to be suppressed by several orders of
magnitude to gain access to the hadronic cross section. However, a separation of the
background is possible on the basis 
of the angular and the momentum distributions which significantly deviate from that
of the hadronic final states. 
Also, it suffices to detect and veto one of the leptons of a created
electron-positron pair to veto an electromagnetic process. 

A threshold gas \cer detector installed in forward direction is used to detect
particles with a Lorentz factor $\gamma$ larger than a certain
threshold value. The \cer threshold needs to be higher than the
Lorentz factor the most energetic pions (the lightest hadrons) can reach.
The appropriate threshold for the \gdhex at \elsa is $\gamma > 22$ . We
have used CO$_2$ with a threshold of $\gamma = 36.2$ at standard
conditions as a \cer medium.
Detailed simulations have shown that a \cer detector with an efficiency of
99.9~\% for highly  relativistic electrons within a polar half opening
angle of 15$^\circ$ is well suited to suppress the
background. Fig.~\ref{fig:elmagangle} illustrates the effectiveness of
this veto detector. 

The quantity of produced \cer light is proportional to the length of the
radiator. A radiator length of 90~cm is sufficient to achieve the
required efficiency of 99.9~\%. An ellipsoidal
mirror with short and long axis of 95.3~cm and 110~cm respectively 
is inclined by $20^{\circ}$ with respect to the vertical plane and focuses the light
onto a photomultiplier (see Fig.~\ref{fig:gdhdet}).
The mirror also reflects the ultraviolet part of the \cer light. The
reflecting layer consists of aluminum (200~nm thickness), coated with a
protective layer of magnesiumfluorid (250~nm thickness, MgF$_2$) on a Plexiglas
carrier (3~mm thickness). 
To avoid that the photon beam produces background in the mirror of
the \cer detector -- which obviously would not be suppressed by the
\cer veto itself -- the mirror has a central hole with 5~cm diameter. The hole
is covered by a reflecting aluminum coated mylar foil.
In test measurements the \cer detector was found to have indeed an
efficiency of 99.990~\% for electrons of 1.2~GeV energy. This actually
exceeds the required efficiency by one order of
magnitude~\cite{Helbing:2002kk}. 

\begin{figure}
\begin{center}
\includegraphics[width=0.8\textwidth]{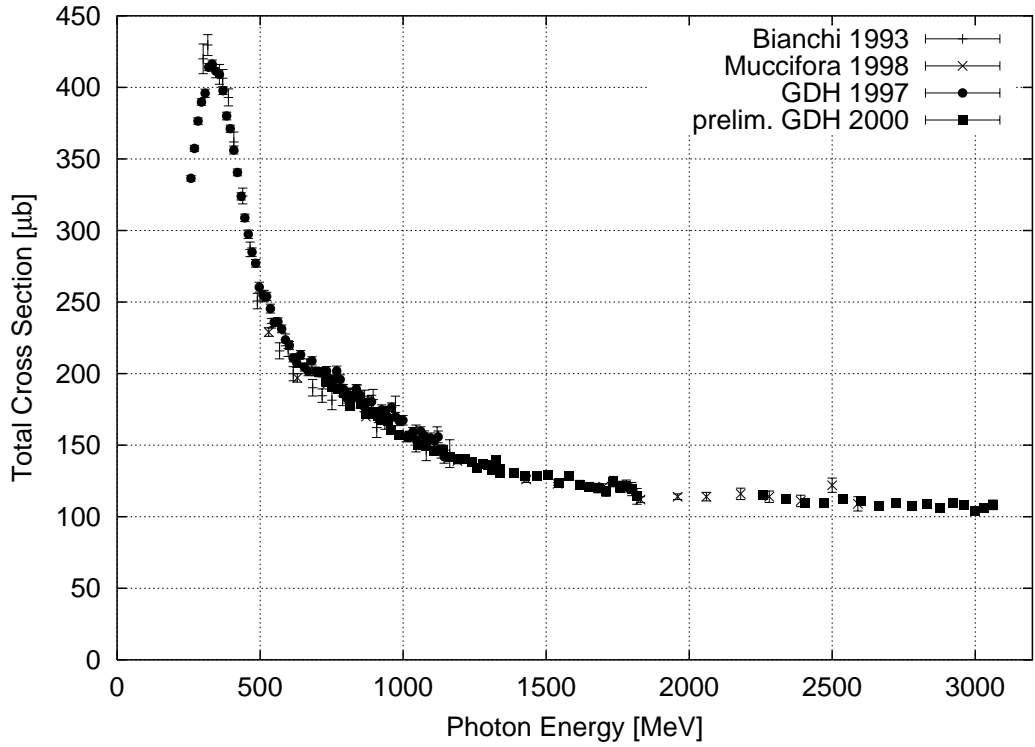}\\
\includegraphics[width=0.8\textwidth]{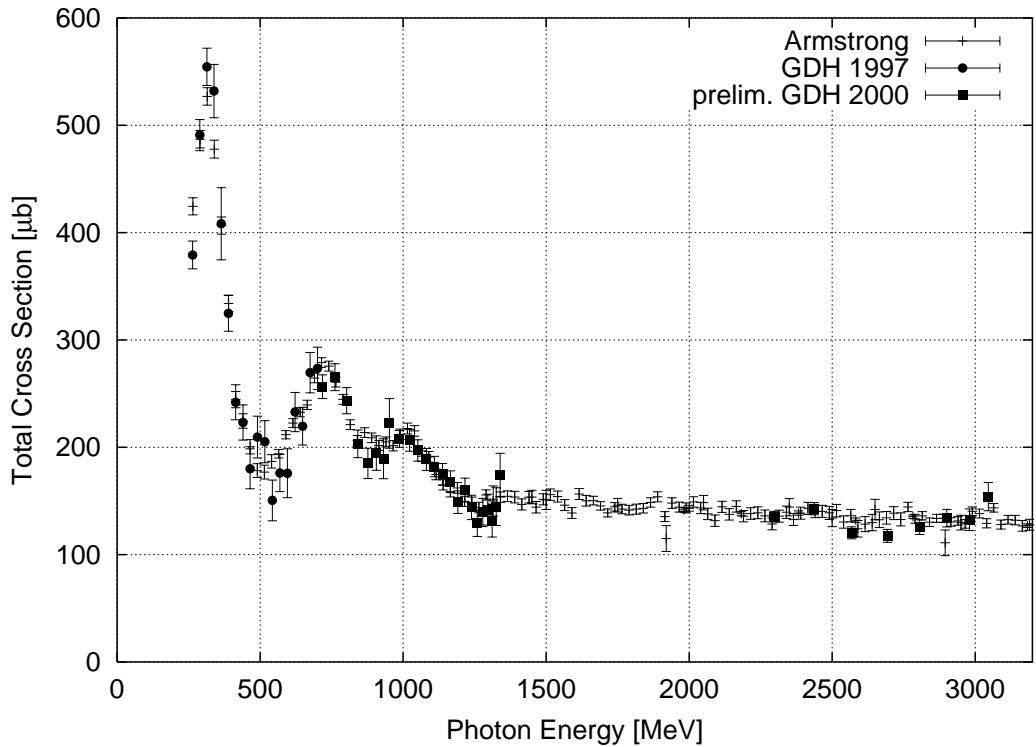}
\end{center}
\caption{Top: unpolarized total cross section of carbon;
Bottom: unpolarized total cross section of hydrogen obtained as
difference of cross sections $(\sigma_{\text{CH}_2} -
\sigma_\text{C})/2$. Both cross sections are compared to published
data~\protect\cite{Armstrong:1972sa,Bianchi:1993nh,Muccifora:1998ct}.\protect\footnotemark}
\label{carbon}
\label{hydrogen}
\end{figure}

For the GDH integral a difference of polarized total cross sections is
to be measured. This is more difficult than the measurement of an
asymmetry as is typically done in
polarization experiments because the absolute normalization of all count
rates and acceptances is essential. The GDH setup was tested with
measurements of unpolarized cross sections~\cite{Helbing:1997,Michel_promo}. For
different primary energies (1.0, 1.4, 1.9, 3.2~GeV) the
photoabsorption cross section of carbon (see Fig.~\ref{carbon}) and
of CH$_2$ was measured\footnote{amongst a variety of other nuclei}. To
verify the capability to measure meaningful differences of cross sections
the hydrogen photoabsorption cross section was extracted from the
difference of the total cross section of the C-target and of the 
CH$_2$-target (see Fig.~\ref{hydrogen} on the left). The agreement
with the literature
data~\cite{Armstrong:1972sa,Bianchi:1993nh,Muccifora:1998ct}, the data
obtained in the pilot experiment~\cite{Helbing:1997,Sauer} at
\textsc{Phoenics} and the matching of the values at the boundaries
between primary energy settings shows the reliability of the
entire system and of the analysis procedure. The data for carbon are
unparalleled in systematic and statistical accuracy.  
\footnotetext{The shown error bars in all following figures are
statistical only unless otherwise stated.} 

\subsubsection{The \daphne setup}
In the energy range from the pion threshold through 800~MeV, determined
by the \mami maximum energy, only a handful of partial channels
contribute to the photoabsorption process of the proton and the
neutron with the final states $p\,\pi^0$, $n\,\pi^+$, $n\,\pi^0$,
$p\,\pi^0\pi^0$, $n\,\pi^0\pi^0$, $p\,\pi^+\pi^-$,  $n\,\pi^+\pi^-$, $p\,\pi^-\pi^0$,
$n\,\pi^+\pi^0$, $n\,\pi^+\pi^-$,  $n\,\eta$ and $p\,\eta$.

The central component of the detector setup at \mami is the detector
\daphne~\cite{Audit:1991gq}. It is a large-acceptance 
hadron detector capable of handling few-particle final states. 
\daphne is roughly cylindrically symmetric \footnote{each coaxial layer
is subdivided into 16~sectors with respect to the azimuthal angle} and
covers the polar angles from $21^\circ$ through $159^\circ$ which permits
particle identification within 94~\% of the total solid angle. The
detector was essentially built for charged particle identification. 

The target in the center of the detector is surrounded by three layers
of multiwire proportional chambers with cathode readout. Up to 5
charged particle tracks can be distinguished with these wire
chambers. The spatial resolution is of the order of 2~mm which
translates into angular resolutions of $\Delta\theta \le 1^\circ$
and  $\Delta\phi \le 2^\circ$ (FWHM).

Further outside, the detector
consists of 6 scintillator layers. In front of the last three
scintillators metal plates (Al, Fe, Pb) are attached which serve
as converters for photons. This 
leads to a useful detection efficiency also for neutral pions of approximately
20~\%. The efficiency for charged particles is in the range of
85-95~\%. The spatial resolution of the calorimetric component is
10~cm. 

The incomplete polar angle acceptance is accounted for by extrapolations
based on simulations of the detector and the respective partial
channel. For channels with particles in the final state that are not
predominantly emitted in forward direction like $\gamma p\to n \pi^+$
these corrections are small. So, although we disregard neutron
detection in \daphne for obtaining the inclusive results, the
systematic error for the total cross section of this 
partial channel due to the extrapolation is only 2~\%. As another
example, the total cross section of $\gamma p \to p \pi^0$ can be
determined from the photons of the $\pi^0$ decay. A determination of
the differential cross section is complicated due to the forward boost
of the proton. 

In order to improve the forward angle coverage \daphne is complemented
by the detector components \textsc{Midas}~\cite{Altieri:1999yr},
\textsc{Star}~\cite{Sauer:1996gg} and the
Far-Forward-Wall~\cite{Helbing:2002kk}. A \cer detector similar to the
one developed for the \gdhdet is us as well.
Here this threshold \cer veto detector is filled with dry nitrogen
and a 5~cm thick layer of aerogel at the entrance window instead of
the CO$_2$ used at \elsa to account for the lower energies at \mami. 
We will not discuss the forward components in further detail since 
the results obtained so far are based on the analysis of \daphne only.

\begin{figure}
\begin{center}
\includegraphics[width=0.65\textwidth]{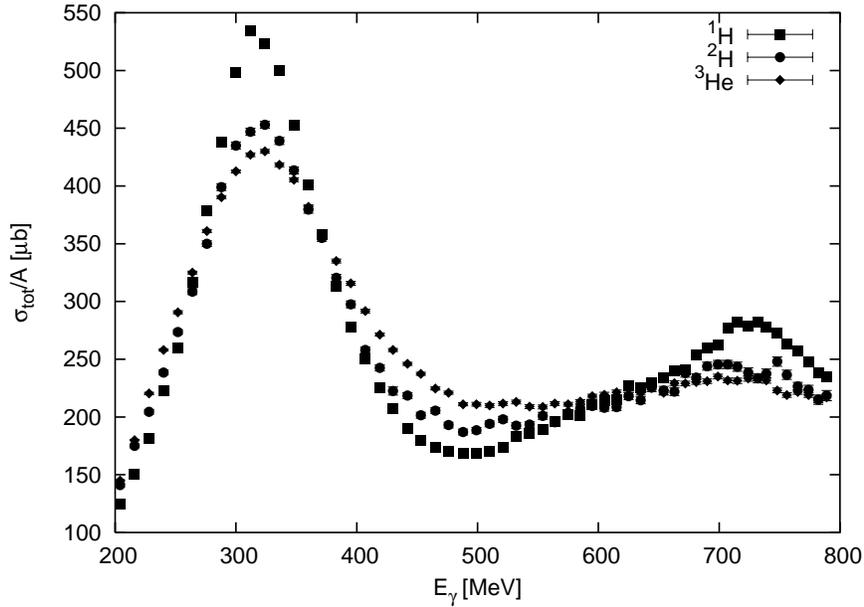}
\caption{Total cross sections per nucleon of hydrogen, deuterium and
helium with \daphne at \mami.} 
\label{fig:DAPHNE_unpol}
\end{center}
\end{figure}

Like for the \gdhdet the setup at \mami has demonstrated its
performance by determining unpolarized cross sections with
unprecedented quality. Fig.~\ref{fig:DAPHNE_unpol} shows several such
cross sections~\cite{MacCormick:1996jz}. The $^3$He data are the first
of its kind for this nucleus. 

\newpage

\section{Results from MAMI and ELSA}
\label{sec:results}
\subsection{Data analysis}
The cross section difference $\Delta\sigma(\nu) =
\sigma_{3/2}(\nu)-\sigma_{1/2}(\nu) $ is obtained by:
\be
  \label{eqn:sigmadiff}
\sigma_{3/2}(\nu)-\sigma_{1/2}(\nu)=
\frac{Y_{3/2}(\nu)-Y_{1/2}(\nu)}{\eta(\nu) \cdot f_\text{t} \cdot
P_\text{t} \cdot P_\text{circ}(\nu,E_{0})} 
\ee
with the photon definition probability $\eta$ (see
Sec.~\ref{sec:pgamma}), the column density of polarized nucleons
$f_{t}$, the target polarization $P_\text{t}$ (see
Sec.~\ref{sec:poltarg}) and the photon beam polarization
$P_\text{circ}(\nu,E_{o})$. $Y$ is the hadronic yield for the
respective spin configurations. This hadronic yield needs further
explanation as it is obtained with different methods from the detector
setups at \elsa and \mami.

With the \gdhdet at \elsa obtaining $Y$ is straight forward. For
any hadronic reaction at least one particle of the final state is
observed. Hence extrapolations to unobserved angles and corrections for
incomplete efficiency or acceptance are obsolete. This is also true
for the photoabsorption of the neutron. While the detection efficiency
for neutrons is moderate only, charged or uncharged mesons
are observed with at least 98~\% efficiency.
Hence, the hadronic yield $Y_{3/2,1/2}(\nu)$ is simply determined by
the hadronic count rate of the \gdhdet in each spin configuration
normalized to the photon flux. A hadronic event is identified by a
signal in at least one of the 15 detection units of the \gdhdet
provided it is time-correlated with a signal of the tagging system. To
avoid double counting, events where two or more units have detected a
reaction product are counted with a corresponding
weight. Electromagnetic background events are suppressed by the veto
detectors. Only very low energy electrons and positrons produced by
untagged photons are not 
identified by the veto detectors. These background events can be
suppressed in the analysis by their random character and their lower
energy deposition compared to hadronic events. The application of an
energy threshold minimizes their effect on the statistical
error. A
prerequisite for such a handling is the observed negligible fraction
of less than $10^{-3}$ of events which occurred as coincident with the
tagging system in one module and as randomly correlated in
another. Also veto dead-time effects are accounted for.

Obtaining the hadronic yield with \daphne is quite different due to
the need to account for acceptance.
The largest part of the total hadronic yield $Y$ for the total
photoabsorption cross section from Eq.~(\ref{eqn:sigmadiff}) is
identified by charged particles detected inside \daphne. 
Charged pion channels without a neutral pion in the final state need
to be corrected for the fraction of events emitted into the angular
regions outside the \daphne acceptance. 
The second largest part is determined from neutral pion events without
accompanying charged particle. This number is scaled by the $\pi^0$
detection efficiency as determined by detector simulations. The
efficiency is finite for all energies and angles of the neutral
pion. Hence, no extrapolation is needed for partial channels with at
least one $\pi^0$ in the final state. 

\subsection{Systematic studies}
\begin{figure}
\includegraphics[width=0.5\textwidth]{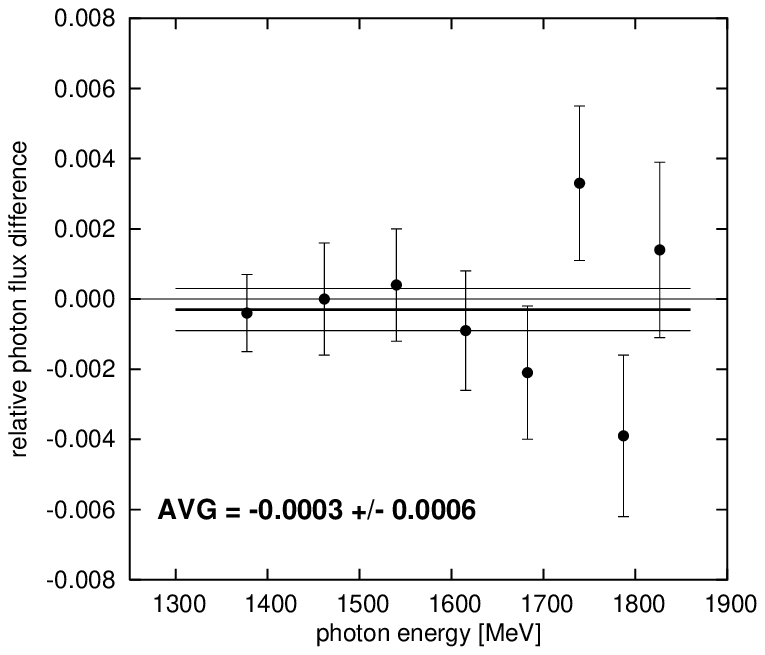}
\includegraphics[width=0.5\textwidth]{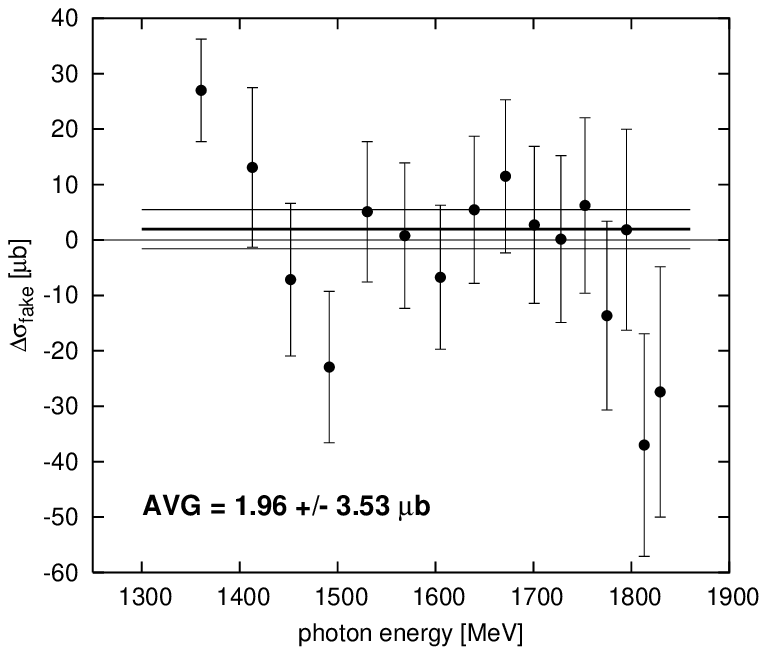}
\caption{Left: relative photon flux difference for the two helicity
states of the polarized source at \elsa; Right: Fake asymmetry with
unpolarized butanol target with target holding field reversal at \elsa.}
\label{fig:SystElsa}
\end{figure}
In order to assure that the measured polarized cross section differences
are generated exclusively by polarized photons scattering off the polarized
protons, deuterons or lithium nuclei several systematic studies have
been performed. So-called {\em false or fake asymmetries} could arise from
correlations of the photon flux with the helicity or from the magnetic
holding field of the target but also from misconceptions of the
trigger logic electronics with respect to processing the
helicity signal of the accelerator.

The left hand side of Fig.~\ref{fig:SystElsa} shows the photon flux
difference as determined with the tagging system for the two helicity
states of the polarized source at \elsa relative to the total photon
flux. One observes that the flux is independent of the helicity.
On the right hand side of Fig.~\ref{fig:SystElsa} the fake asymmetry
of the cross section difference for scattering off an unpolarized
butanol target is shown i.e. no microwaves for the dynamic nuclear
polarization (DNP, see \ref{sec:poltarg}) have been applied. As
expected, even with the reversal of the holding field no significant
fake asymmetry arises.
The energy dependence of fake asymmetries using unpolarized targets
has been studied as well. The resulting upper limits for such
asymmetries have been found to be well below 1~$\upmu$b for all energies.

\begin{figure}
\begin{center}
\includegraphics[bb=14 21 322
203,width=0.75\textwidth]{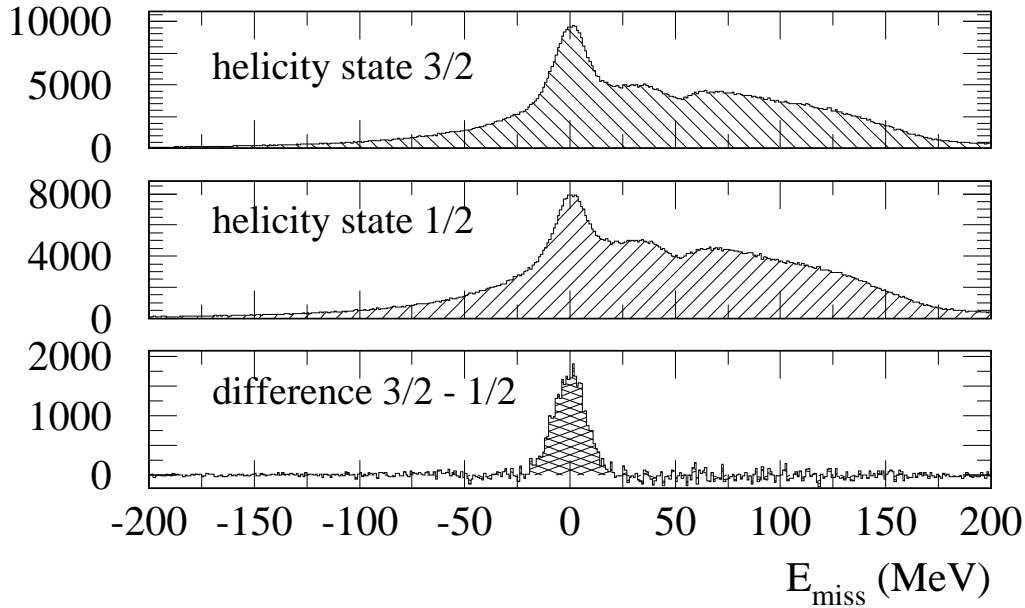}
\caption{Missing energy spectra for the reaction 
$\vec{\gamma}\vec{p} \to p\pi^0$ at \mami
under the assumption that the proton originated from a
reaction on a free proton. The spectra are shown for both helicity
states and their difference. }
\label{fig:SystMami}
\end{center}
\end{figure}
Fig.~\ref{fig:SystMami} shows the analysis of data taken at \mami with
polarized butanol target. 
The missing energy for the reaction $\vec{\gamma}\vec{p} \to p\pi^0$
is plotted with the assumption that the proton originated from a
reaction on a free proton i.e. the photon scattered off 
a polarized hydrogen nucleus. 
Background contributions from unpolarized
carbon and oxygen nuclei of the butanol target are still present in
the two separate helicity state diagrams. However, this background
cancels nicely in the difference of the two helicity states. 

\subsection{Spin dependent total photoabsorption cross sections}
\subsubsection{Results on the proton} 
\begin{figure}
\includegraphics[width=\textwidth]{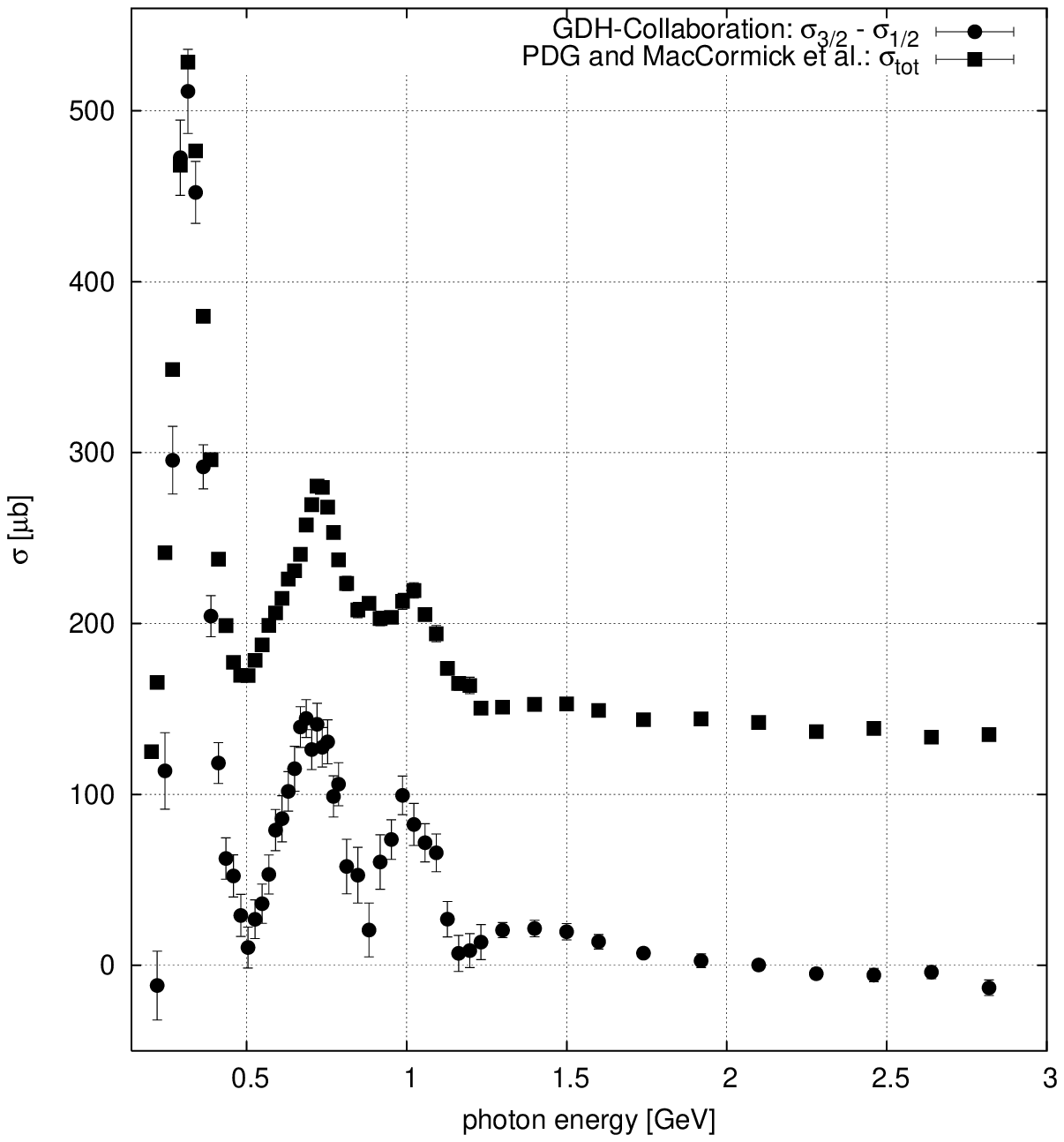}
\caption{Difference of the polarized total photoabsorption cross
sections of the proton~\protect\cite{Ahrens:2001qt,Dutz:2003mm,Dutz:2004mm} in
comparison to the unpolarized cross
section~\protect\cite{Eidelman:2004wy,MacCormick:1996jz}.}  
\label{fig:gdh_all}
\end{figure}
Fig.~\ref{fig:gdh_all} shows the final doubly polarized results for
$\sigma_{3/2} - \sigma_{1/2}$ on the proton. For comparison also the
unpolarized cross section is
plotted~\cite{Eidelman:2004wy,MacCormick:1996jz}. These proton data
are published in
Refs.~\cite{Ahrens:2001qt,Dutz:2003mm,Dutz:2004mm}. 
The three major resonances
known from the unpolarized total cross section are present in the
difference as well - they are even more pronounced.  A large
background of the order of 150~$\upmu$b of helicity insensitive
background disappears in the cross section difference. Also an
indication of a 4th resonance can be seen in the polarized data (see
Sec.~\ref{sec:Resonances}).  

\begin{figure}
\includegraphics[width=\textwidth]{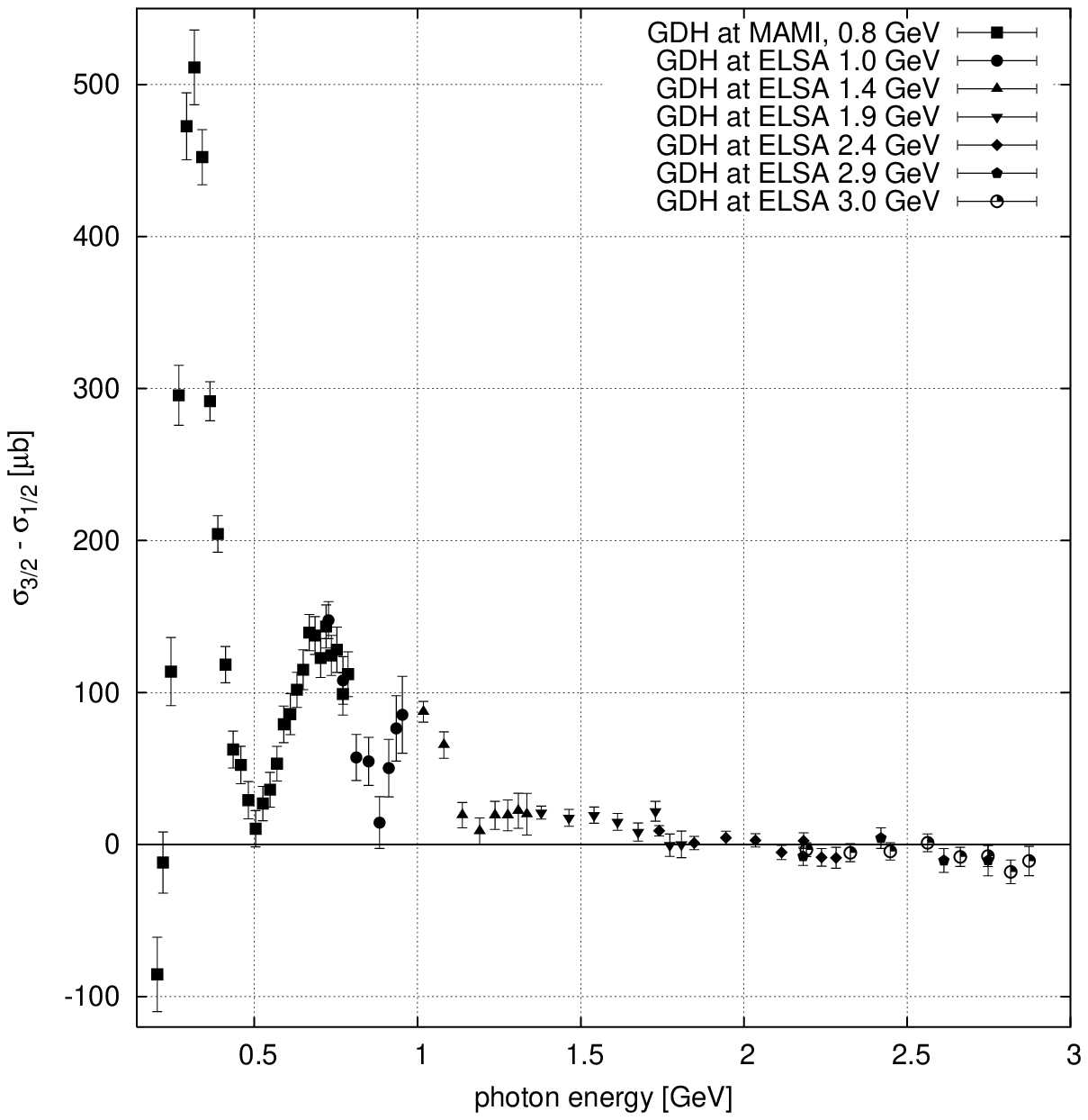}
\caption{Comparison of individual data sets for all energy
settings for the polarized total photoabsorption cross
section difference of the proton without rebinning and merging.}
\label{fig:gdhall}
\end{figure}

To demonstrate that the results for the different energy settings at
the two accelerators match each other very well Fig.~\ref{fig:gdhall}
shows a comparison of all the individual data sets.

\subsubsection{Results on the deuteron and the neutron} 
\begin{figure}
\includegraphics[width=\textwidth]{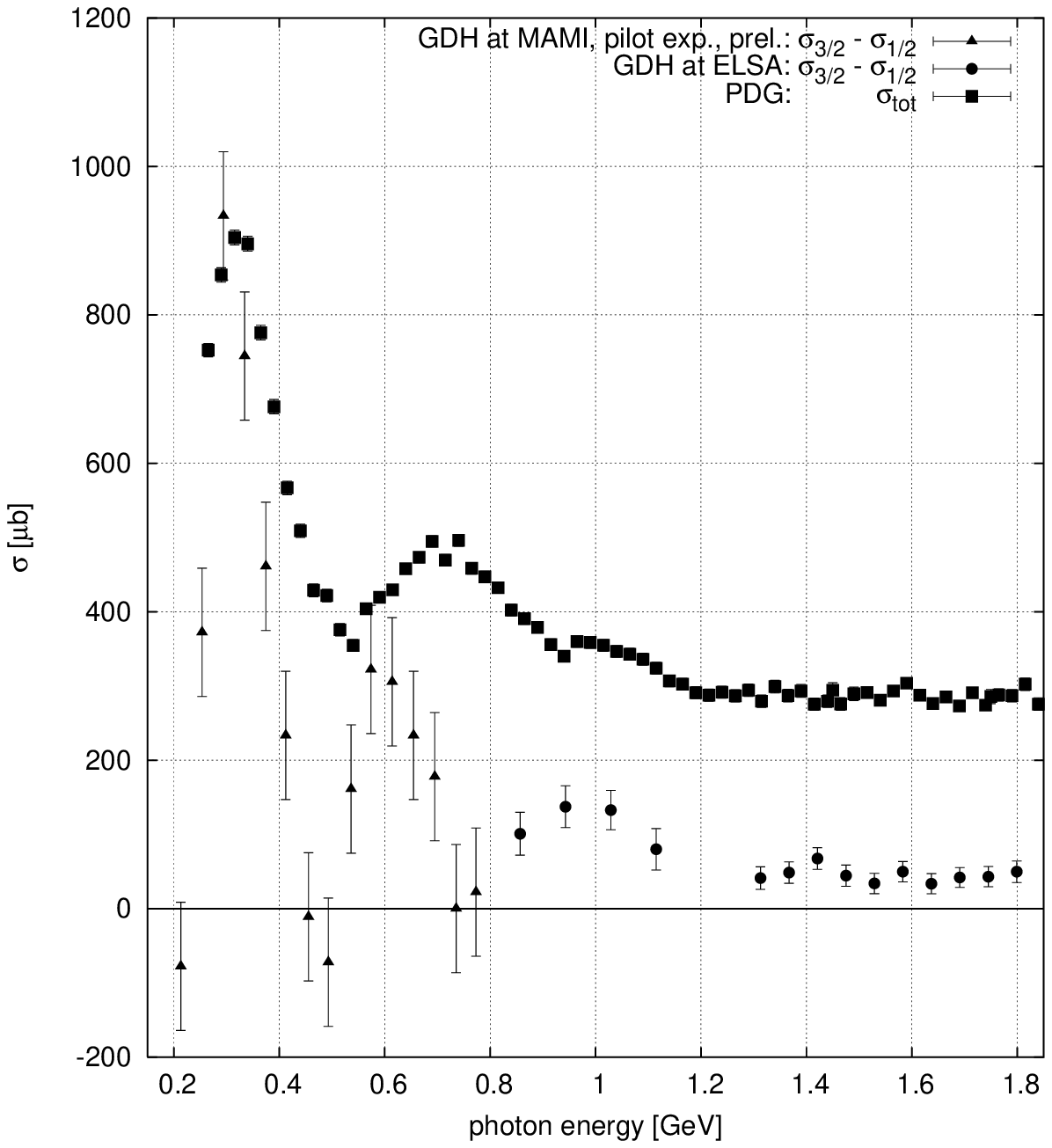}
\caption{Difference of the polarized total photoabsorption cross
section of the deuteron in comparison to the unpolarized cross section.}
\label{fig:DeuteronAll}
\end{figure}
Fig.~\ref{fig:DeuteronAll} shows the results obtained so far for the total
photoabsorption cross section of the deuteron again in comparison with
the unpolarized cross section~\cite{Eidelman:2004wy}. The data from
\mami~\cite{Rostomyan:2005,Jahn:2004} is still under analysis and of
preliminary nature. The data were obtained in 
a pilot run carried out in 1998 at \mami. The high statistics
data of the 2003 deuteron run at \mami is currently under analysis. On
the other hand, the
polarized results from \elsa are of final nature~\cite{Dutz:2005ns}.

Qualitatively, the comparison of the polarized to the unpolarized data
leads to similar observations like those for the proton. The
resonances again appear even more pronounced while a
helicity-independent background of the order of 300~$\upmu$b
disappears in the cross section difference. 

\begin{figure}
\centering
\includegraphics[width=0.75\textwidth]{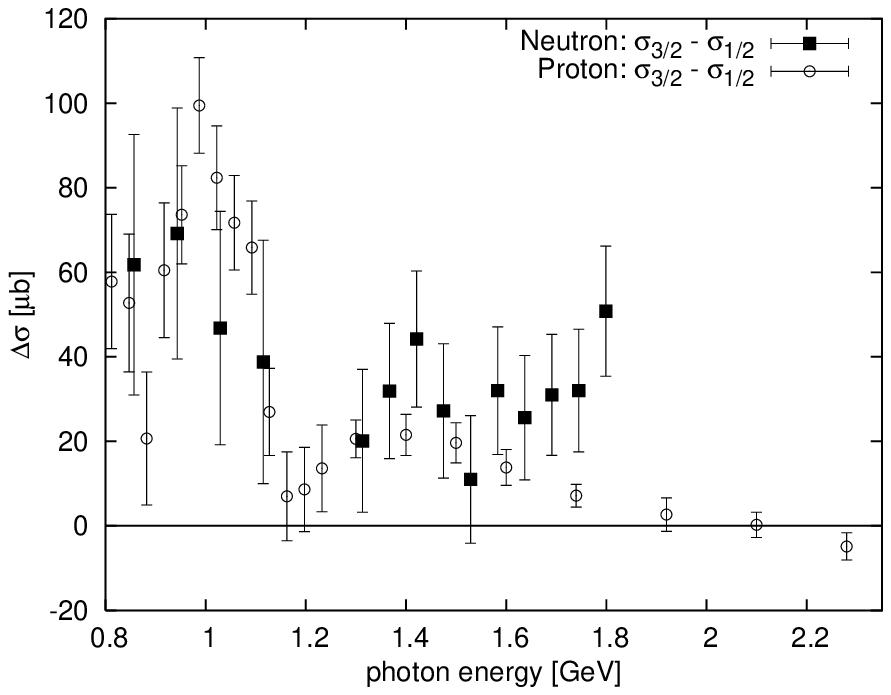}
\caption{Comparison of the proton and neutron polarized cross section differences.}
\label{fig:PolNeutron}
\end{figure}

To compute the neutron cross sections from the $^6$LiD data taken at
\elsa we have accounted for nuclear effects and chemical admixtures
that modify the neutron polarization relative to the measured
polarization of the free deuteron (see Sec.~\ref{sec:poltarg}). 
The simple further decomposition of $\Delta \sigma_\text{D} = \Delta
\sigma_\text{p}+ \Delta \sigma_\text{n}$ is based on recent
calculations~\cite{Arenhovel:2004ha} which show that coherent contributions
present at lower energies can be neglected in the \elsa energy range.
The resulting cross section of the neutron based on the \elsa data is
shown in Fig.~\ref{fig:PolNeutron}. While the response of the neutron
at photon energies above 800~MeV in general is very similar to that of
the proton one observes a discrepancy of the two cross sections above
1.5~GeV where the proton cross section is headed for a sign change
while the neutron data does not show this trend. 
We will discuss the implications of this
further in Secs.~\ref{sec:HighEnergies} and~\ref{sec:GDHintNeutron}.

\subsection{Resonance structure}
\label{sec:Resonances}
It is beyond the scope of this review to summarize the wealth of
information obtained for the individual resonances with respect to all
partial channels studied. More details of the results can be found
in~\cite{Ahrens:2000bc,Ahrens:2002gu,Ahrens:2003na,Ahrens:2003bp,Ahrens:2004pf}.
Here, we want to give an overview and the general picture of the current
level of understanding of the resonance structure with respect to this data.

\begin{figure}
\centering
\includegraphics[width=\textwidth]{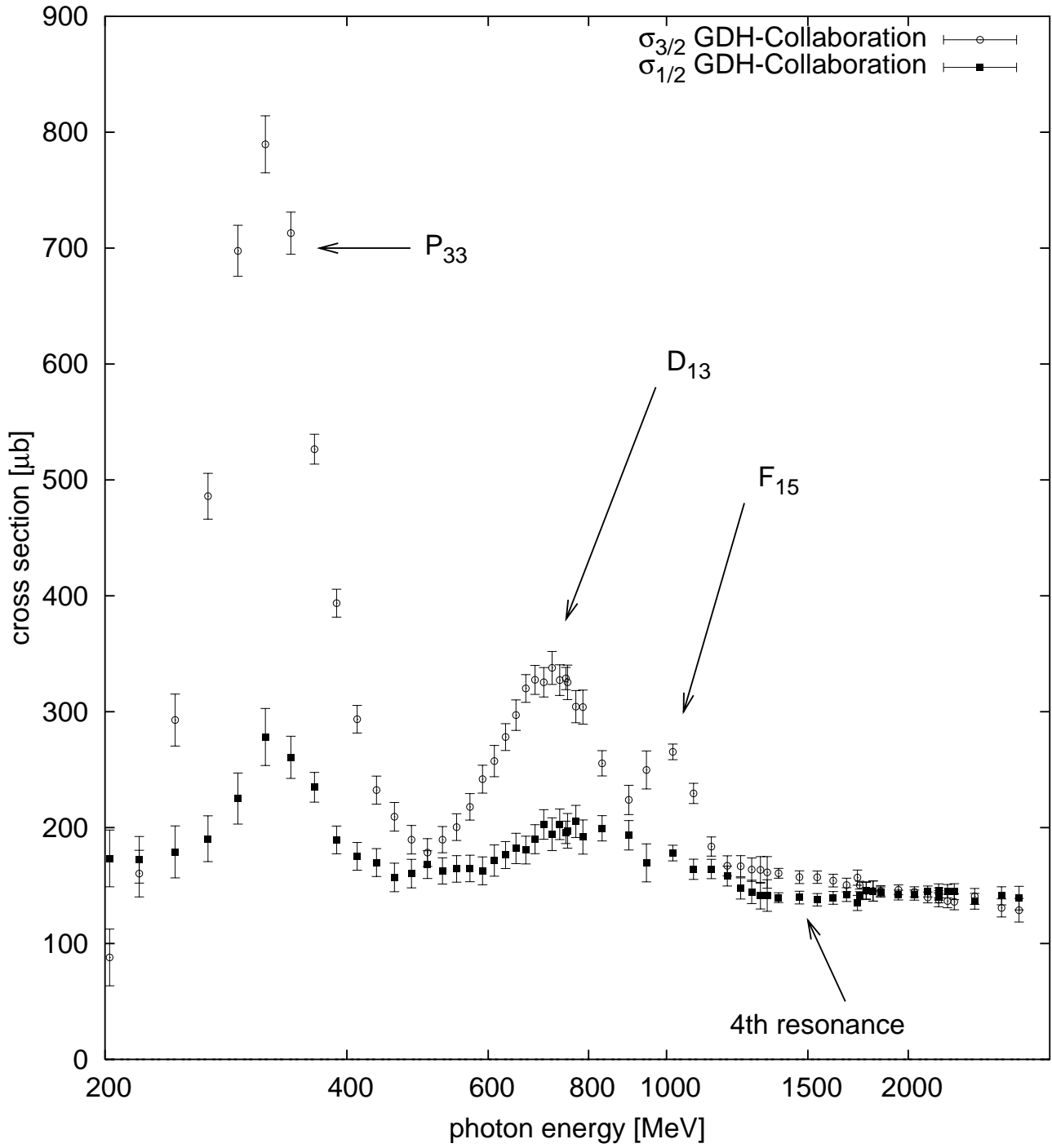}
\caption{Separate helicity state total cross sections $\sigma_{3/2}$
and $\sigma_{1/2}$ of the proton.}
\label{fig:resonances}
\end{figure}

Fig.~\ref{fig:resonances} shows the separate helicity contributions to
the total cross section. The separated helicity states are obtained by
adding resp. subtracting the polarized cross section difference from
the unpolarized data. Clearly, most of the resonance strength of
the first three resonances originates from the $3/2$ helicity
channel. This can be understood intuitively as all major resonances contributing to
the cross section have $J \ge 3/2$.
The situation appears to be different for the 4th resonance. This structure has not
been observed before in unpolarized total cross section data. Here the
structure stems at least partially from the drop in the strength of the
contribution from helicity $1/2$. This structure might be due to the
$F_{35}$ and the $F_{37}$ resonances.

\subsubsection{First resonance: $\mathbf{\Delta}(1232)$}
\label{sec:delta}
In the low energy region around the $\Delta(1232)$-resonance only
the two single pion production channels contribute to the total cross
section of the proton. This resonance has been studied extensively
over decades in pion scattering and pion photoproduction. Mostly
unpolarized data have been taken but also single polarization
observables (e.g.~\cite{Dutz:1996uc}). Hence, the main multipoles are
well determined. 

\begin{figure}
\includegraphics[width=0.49\textwidth]{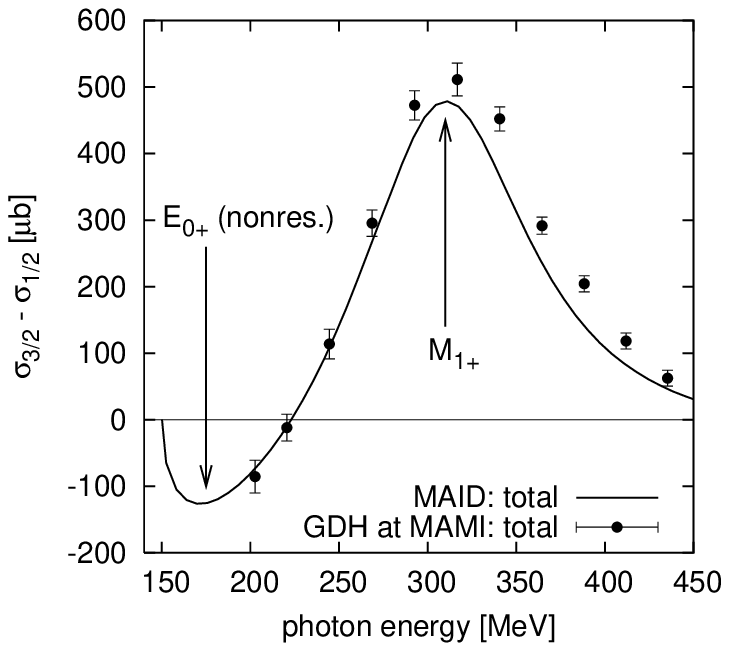}
\includegraphics[width=0.49\textwidth]{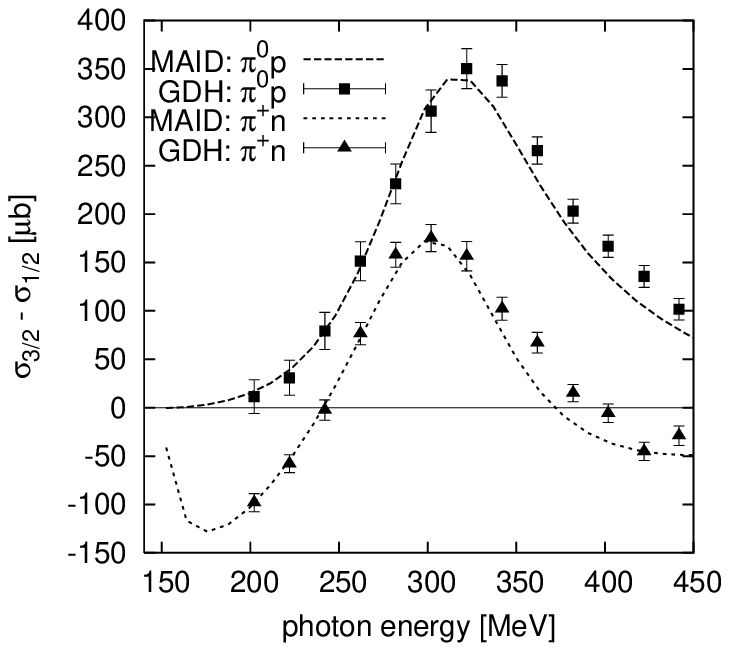}
\caption{Helicity dependence of $\gamma p \to N \pi$ below
450~MeV in comparison to the multipole
parameterization \maid~\protect\cite{Drechsel:1998hk}. Left: The total cross
section difference; Right: The two 
contributing single pion partial channels.}
\label{fig:delta}
\end{figure}

Indeed, Fig.~\ref{fig:delta} shows that also the doubly
polarized data are reasonably well described by the existing
parameterizations. Shown for comparison is the parameterization of
\maid~\cite{Drechsel:1998hk}. This approach incorporates resonance
multipole contributions but also Born terms in a unitary isobar model.
An alternative parameterization called \textsc{Said} can be found in 
Ref.~\cite{Arndt:1995ak} which works equally well.
The $\Delta$-resonance at the real photon point is primarily of the
spin-flip M1 (M$_\text{1+}$) magnetic transition type which dominates
the electric quadrupole amplitude E2. One observes that at threshold
the E1 (E$_\text{0+}$) contributes only to the $\pi^+ n$ channel. This
can be understood in a simple picture: in the $\pi^0 p$ system the
charge is at the center of gravity and hence the dipole moment is
small compared to the $\pi^+ n$ system. On the other hand the E1
amplitude is the only one that can be excited right at the threshold
as it is the one with orbital angular 
momentum zero of the produced pion nucleon system. For the evaluation
of the GDH integral this E1 amplitude is of particular importance (see
Sec.~\ref{sec:GDHintegral}). 

\subsubsection{Second resonance: $\mathbf{D_{13}(1520)}$}
\begin{figure}
\includegraphics[width=0.49\textwidth]{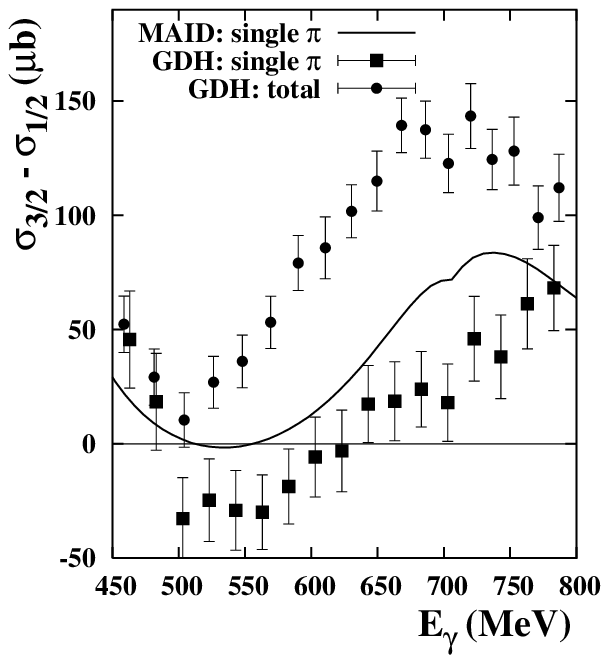}
\includegraphics[width=0.49\textwidth]{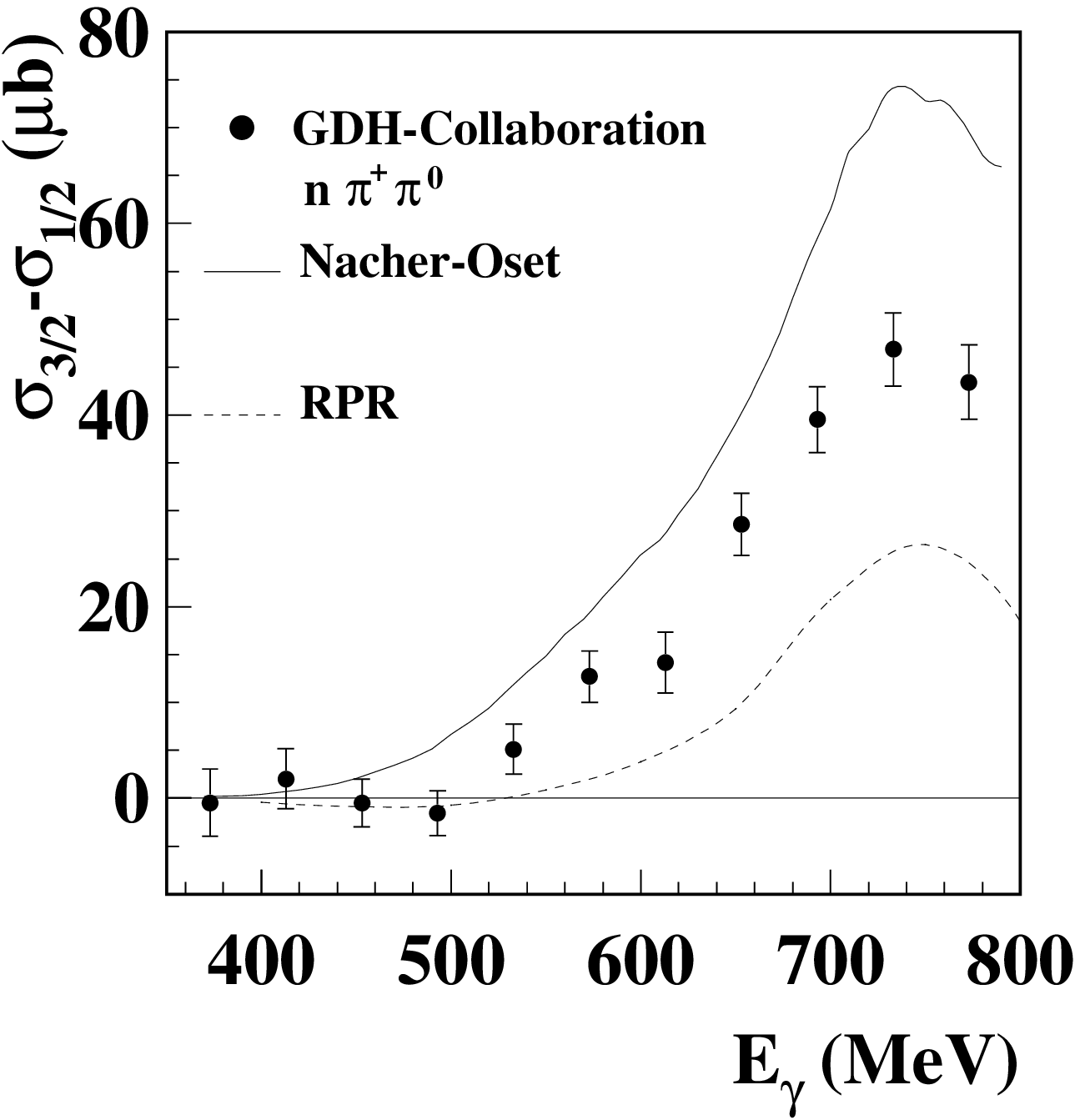}
\caption{Left: Difference of polarized total cross sections and
difference of polarized single pion production cross sections in
comparison with \maid; Right: Difference of polarized double pion
for the channel $\gamma p \to n \pi^+ \pi^0$ in comparison to with two
models~\protect\cite{Nacher:2002,Holvoet:2001}.}
\label{fig:2ndResonance}
\end{figure}

With the appearance of additional channels, namely the double pion
channels, the parameterizations appear overstrained
already. The left hand side of Fig.~\ref{fig:2ndResonance} shows the
discrepancy between the preliminary data on single pion production with the
parameterization of \maid. The results on $\pi^0$
production~\cite{Ahrens:2002gu} indicate that the ratio of multipoles
M$_{2-}/$E$_{2-}$ is rather of the order of 0.56 than the value
of 0.45~\cite{Eidelman:2004wy} obtained before. 

The double pion production channels $n\pi^+\pi^0$,
$p\pi^+\pi^-$ and $p\pi^0\pi^0$ were analyzed separately. The right
hand side of Fig.~\ref{fig:2ndResonance} shows the helicity dependent
difference of the $\gamma p \to n \pi^+ \pi^0$ channel.
Also, models~\cite{Nacher:2002,Holvoet:2001} to describe the double pion
channels have not yet lead to a satisfactory agreement and the
underlying production mechanisms are not fully understood.

\subsubsection{Third resonance}
\begin{figure}
\centering
\includegraphics[width=0.7\columnwidth]{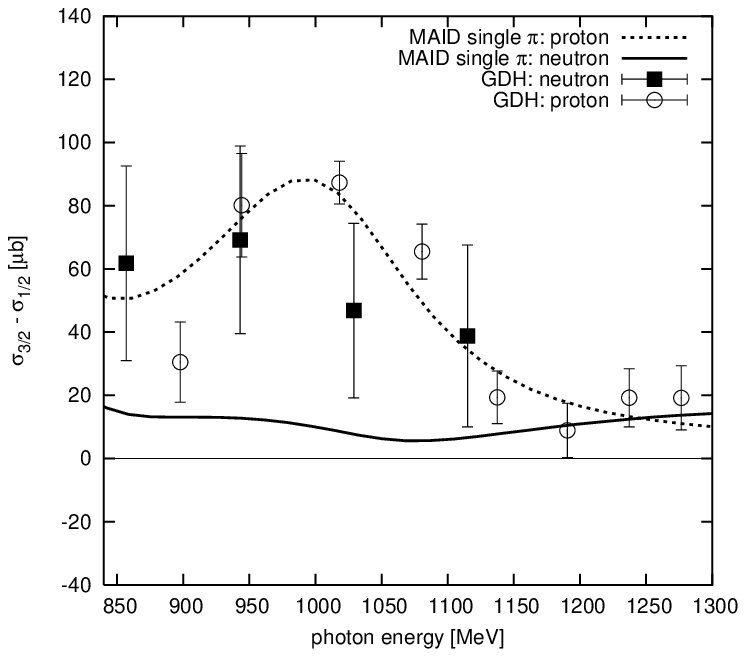}
\caption{Comparison of the polarized total cross sections in the region of the 3rd
resonance for the proton and neutron with the single pion production
prediction of \maid.}
\label{fig:3rdresonance}
\end{figure}
Fig.~\ref{fig:3rdresonance} shows the response of the proton and the neutron to
polarized real photons in the region of the third resonance. The
proton data appear well described by the 
single pion photoproduction alone as predicted by
\maid. This would indicate that this structure
is dominantly generated by single pion production and not by double
pion production or other contributions. 

For the neutron however, \maid in contrast to the data shows no resonance
structure and lies significantly below the experimental results. The
neutron data exhibit a structure in the 3rd resonance region similar
to the proton data. This
could indicate that the scattering mechanism for the neutron is quite
different from that of the proton or that the agreement of \maid with
the proton data is by chance only and the conclusion of single pion
production dominance around 1~GeV is incorrect.
This puzzle has to be resolved by
future experiments with partial channel resolution.

\begin{figure}
\centering
\includegraphics[width=.7\textwidth]{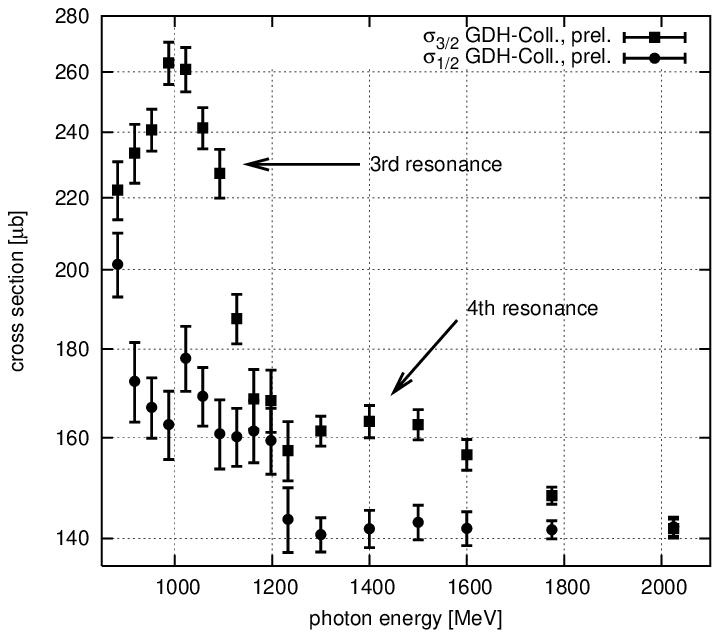}
\caption{Separate helicity state total cross sections $\sigma_{3/2}$
and $\sigma_{1/2}$ for the 3rd and the 4th resonance of the proton.}
\label{fig:4thresonance}
\end{figure}
\subsubsection{Fourth resonance}
Beyond the three resonances well known from the unpolarized
total photoabsorption the polarized data exhibit another structure at a photon
energy of about 1400~MeV that we may call the fourth resonance.
The separate helicity cross sections $\sigma_{3/2}$ and $\sigma_{1/2}$
show a peak difference of about 20~$\upmu$b (see
Fig.~\ref{fig:4thresonance}). One might guess that this structure might
be due to the excitation of the F$_{35}$(1905) or F$_{37}$(1950)
resonances. Another explanation could be a cusp affecting 
$\sigma_{1/2}$ mainly.
Further clarification is needed from future
experiments to identify the exact origin of the observed new structure.

\subsubsection{Comparison with virtual photoabsorption}
\begin{figure}
\centering
\includegraphics[width=.375\textwidth]{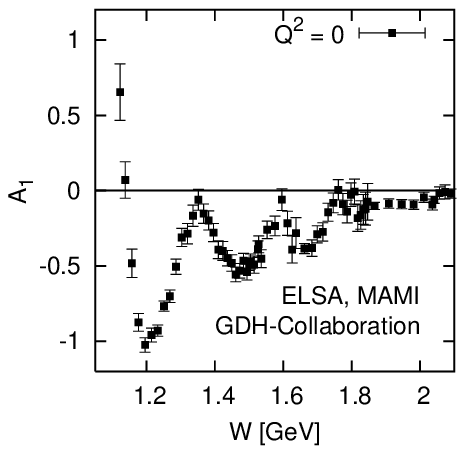}
\includegraphics[width=.615\textwidth]{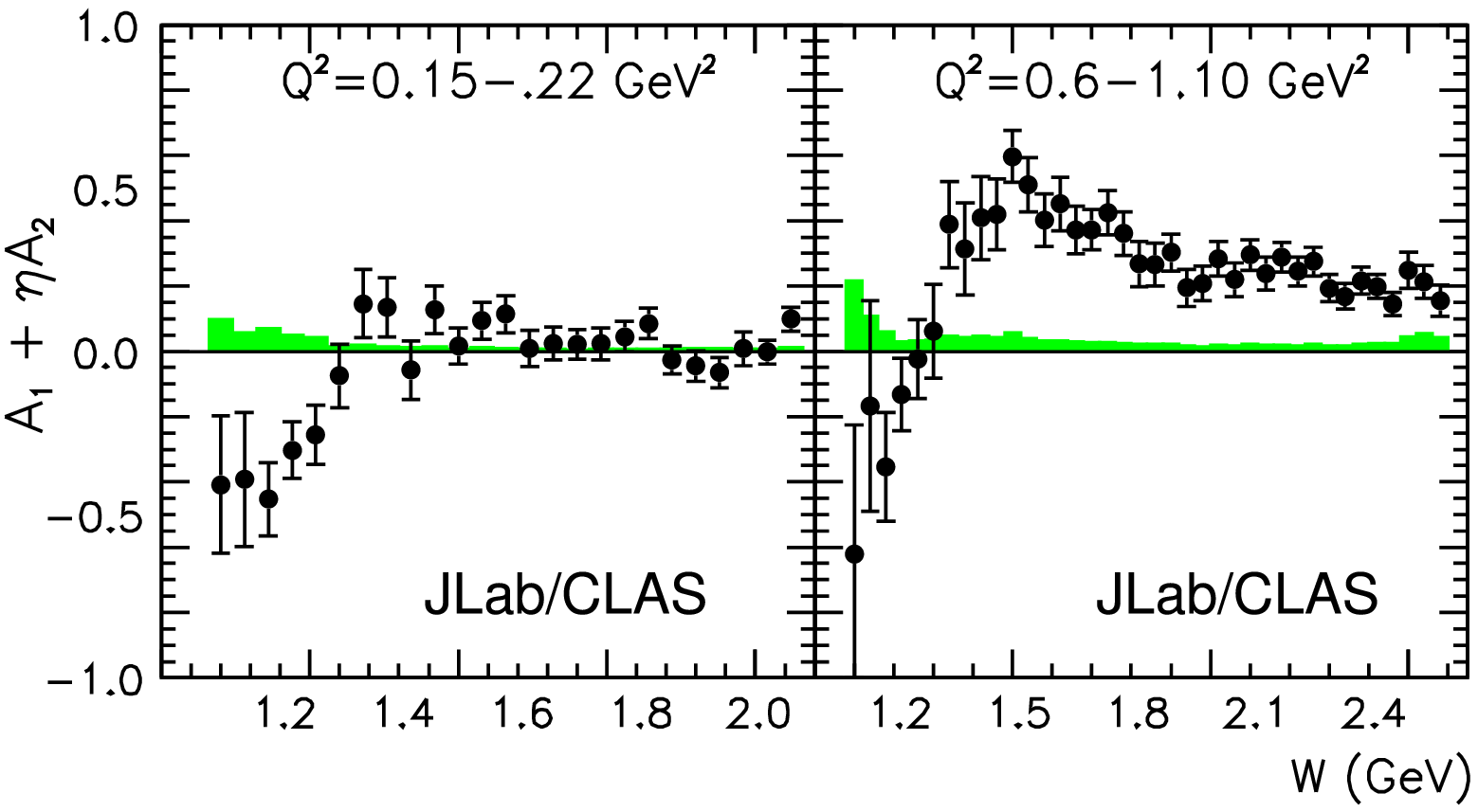}
\caption{Evolution of the asymmetry $A_1$ with photon virtuality $Q^2$.}
\label{fig:a1}
\end{figure}

Fig.~\ref{fig:a1} shows how the asymmetry $A_1=
\left. \left(\sigma_{1/2}-\sigma_{3/2}\right)\right/
\left(\sigma_{1/2}+\sigma_{3/2}\right) 
\equiv \sigma_{TT'}/\sigma_T$ evolves from the
real photon point\footnote{Displayed is actually the experimental
asymmetry $A_\parallel \propto A_1 + \eta A_2$ (see
Sec.~\ref{sec:PolAbs}). At small $Q^2$ however $\eta A_2$ represents
only a small correction to $A_1$ and we will ignore it here.}. The
data in the left panel are from the \gdhcol  and the data on the right
of Fig.~\ref{fig:a1} have been taken by the
\textsc{Clas}-Collaboration in Hall~B at
\textsc{JLab}~\cite{Burkert:2002fs}. One observes that apart from the
Delta-resonance no higher resonances are visible in the Hall~B
data. The Delta-resonance looses strength when going to finite $Q^2$
and the asymmetry above $W>1.5$~GeV even changes sign. 

This dramatic transition already at such a low $Q^2 \simeq
0.18$~GeV$^2$ demonstrates why a verification of the \gdhsr can only
be performed at the real photon point as the main contributions to the
GDH integral obviously come from the resonance region.

\subsection{High-energy behavior}
\label{sec:HighEnergies}
Regge parameterizations incorporate many of the most complicated
confinement aspects of strong interactions. This is today's
justification for using Regge approaches even though it was thought to
be superseded by QCD. 

\begin{figure}\centering
        \includegraphics[bb=35 70 550 740,
        width=0.75\textwidth]{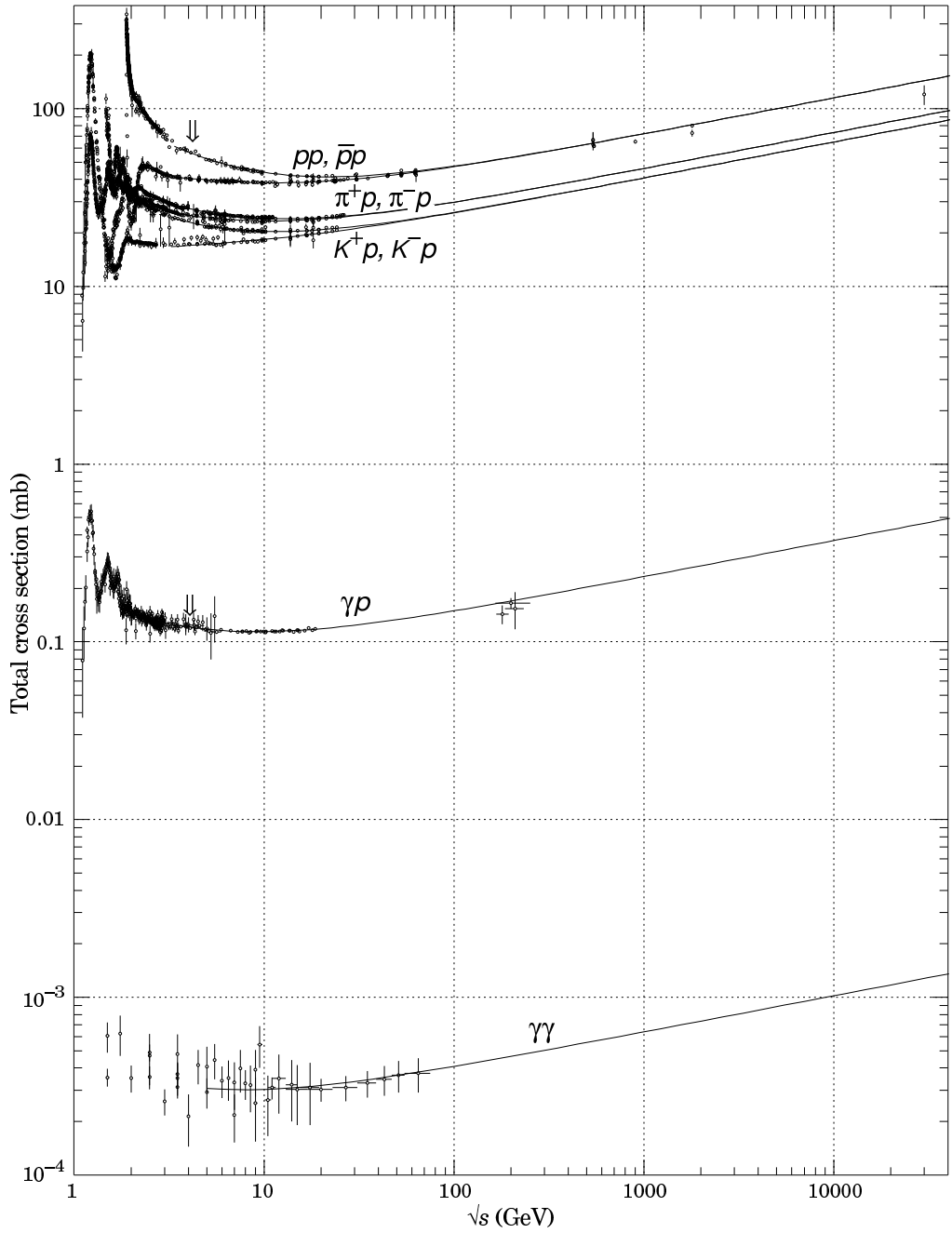} 
        \caption{Summary of hadronic, $\gamma p$ and $\gamma\gamma$
        total cross sections and a universal Regge fit to the
        data~\protect\cite{Comp98,Eidelman:2004wy}.} 
        \label{fig:totcross}
\end{figure}

It has been shown, that the spin averaged total cross section at highest
energies is dominated by the Pomeron while at energies right above the
resonance region it can be described by the $\rho,\omega$
trajectory~\cite{Donnachie:1992ny}:  
\be
\sigma_T(s) \simeq c_1 \cdot
s^{\alpha_{R}(0)-1} + c_2 \cdot s^{\alpha_P(0)-1}
\label{eqn:donn92}
\ee
This equation is a simple power law where $\alpha_{R}(0)$ and
$\alpha_P(0)$ denote the intercepts of the
$\rho,\omega$-trajectory\footnote{These trajectories happen to be
approximately exchange-degenerate.} and the intercept of the
Pomeron-trajectory at Mandelstam $t=0$.  
All unpolarized total cross sections can be parameterized with the
very same values for $\alpha_{R}(0)$ and $\alpha_P(0)$ as can be seen in
Fig.~\ref{fig:totcross}. The values for these intercepts are
$\alpha_{R}(0)=0.53$ and $\alpha_P(0)=1.08$. Observations of this kind
have led to a revival of Regge theory.

\begin{figure}\centering
\includegraphics[width=0.75\textwidth]{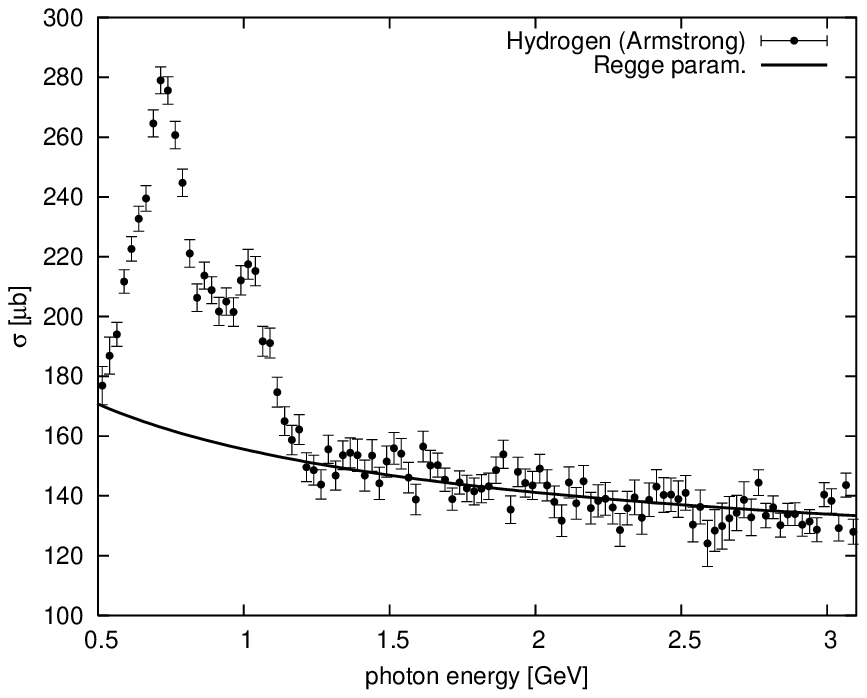}
\caption{Extrapolation of the high-energy Regge
parameterization~\protect\cite{Donnachie:1992ny} of unpolarized total
cross sections to low energies for real photoabsorption in comparison
the experimental data~\protect\cite{Armstrong:1971ns}.} 
\label{fig:ReggeUnpol}
\end{figure}

We now address the question of the lowest energies where the Regge
parameterization may be valid.  
For the real photoabsorption we
have extended the power law fit described above down to low energies
and compared it with unpolarized hydrogen
data~\cite{Armstrong:1971ns}. Fig.~\ref{fig:ReggeUnpol} shows that
indeed the parameterization stemming from data of the several hundreds
of GeV range still matches the unpolarized hydrogen cross section
down to about 1.2~GeV in photon energy just above the third resonance.

\begin{figure}\centering
\includegraphics[width=0.495\textwidth]{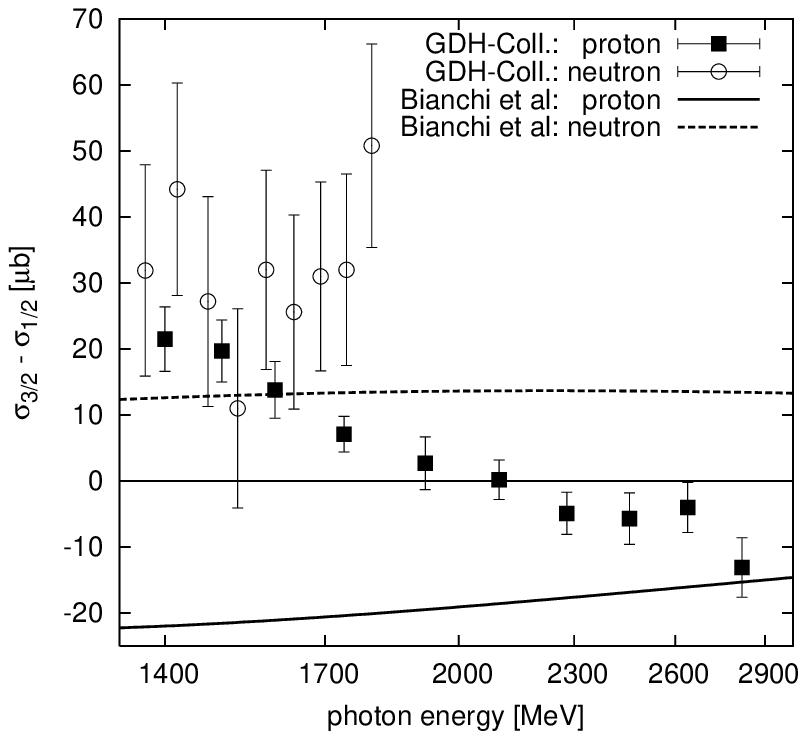}
\includegraphics[width=0.495\textwidth]{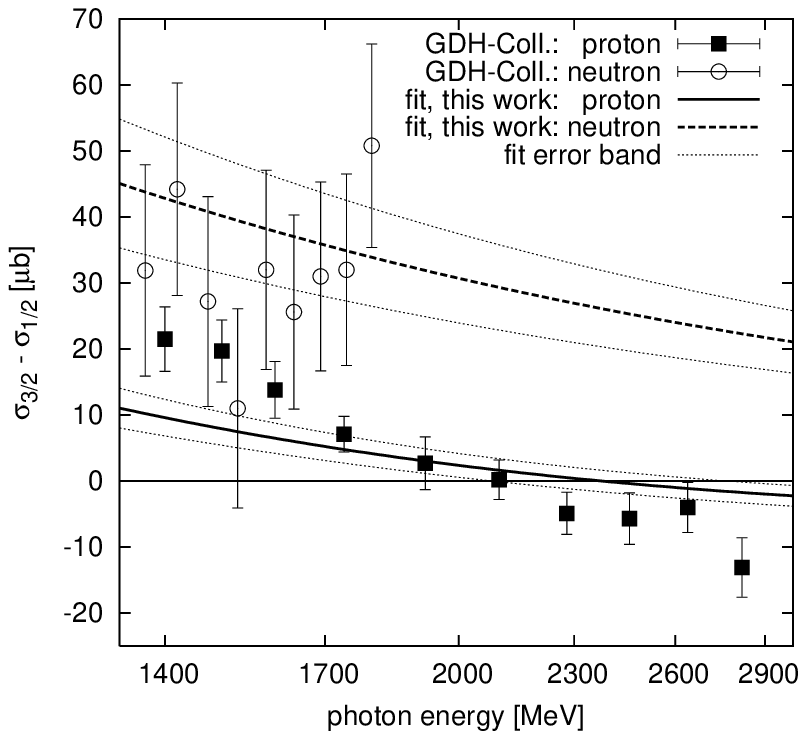}
\caption{Comparison of Regge parameterizations with the data from the
\gdhcol; Left: Parameterization of DIS data extrapolated to $Q^2 =0$
by Bianchi and Thomas~\protect\cite{Bianchi:1999qs}; Right: Our own
parameterization of the GDH data according to Eq.~(\ref{eqn:BassBrisu}).}
\label{fig:BianchiReggeFit}
\end{figure}
For polarized real and virtual photoabsorption Bass and
Brisudov\'a~\cite{Bass:1997fh} have shown that the cross section
difference can be described by the following Regge behavior: 
\be
\sigma_{3/2} - \sigma_{1/2} =  
\left[ c_1 \  s^{\alpha_{a_1}-1} \cdot I +
c_2 \  s^{\alpha_{f_1}-1} +
c_3 \frac{\ln s}{s} + \frac{c_4}{\ln^2 s}
\right]F(s,Q^2)
\label{eqn:BassBrisu}
\ee
where $I$ denotes the isospin of the nucleon and $s$ denotes the
square of the center of mass energy. This represents a
generalization of the earlier ideas by Close and
Roberts~\cite{Close:1994yr} as well as by Bass and
Landshoff~\cite{Bass:1994xb}. Bianchi and Thomas have applied 
Eq.~(\ref{eqn:BassBrisu}) to fit polarized deep inelastic scattering (DIS)
data from scattering off protons and
deuterons~\cite{Bianchi:1999qs}\footnote{Here, we disregard a
parameterization by Simula {\it et al.}~\cite{Simula:2001iy} as it
represents an incoherent addition of Regge behavior and
resonances which is a violation of the idea of duality.}.
The extrapolation to the real photon point of their resulting fits for
the proton and neutron on the left hand side of
Fig.~\ref{fig:BianchiReggeFit} are compared to the high-energy data of
the \gdhcol for the proton and neutron. Obviously, this
parameterization does not describe the data. This could have two
reasons: the extrapolation of this fit from the DIS domain to $Q^2=0$
fails or the data still contains structure like resonances that are not
well averaged by a Regge parameterization. The latter reason would be
a phenomenon unique for the polarized data as the unpolarized data of
this energy region are still well described. The extrapolation to the
real photon point may fail because of the dramatic transitions
observed below $Q^2 = 1$~GeV as seen for example in Figs.~\ref{fig:a1} and
\ref{fig:I1p}.

The logarithmic terms in Eq.~(\ref{eqn:BassBrisu}) are due to Regge cuts and can be neglected at
$Q^2=0$~\cite{Kuti}. Also for the parameterization of Bianchi and
Thomas these terms are insignificant at the real photon
point. $F(s,Q^2)$ simplifies to a constant at the real photon point
and can be absorbed into the constants $c_1$ and $c_2$. $\alpha_{a_1}$ and 
$\alpha_{f_1}$ are the Regge intercepts of the respective meson
trajectories. Hence in the case of real photons the expression for the
Regge behavior simplifies considerably to
\begin{equation}
\label{eqn:regge}
\sigma_{3/2} - \sigma_{1/2} = \tilde{c}_1 \
s^{\alpha_{a_1}-1} \cdot I +  
\tilde{c}_2 \ s^{\alpha_{f_1}-1}
\end{equation}
The intercept $\alpha_{f_1}$ of the $f_1$ trajectory is relatively well defined by
deep inelastic scattering data and usually found to be about -0.5. The situation is less
clear with the intercept of the $a_1$ trajectory where the values from different fits range
from about -0.2 to +0.9. 

The right hand side of Fig.~\ref{fig:BianchiReggeFit} shows the
fit of Eq.~(\ref{eqn:regge}) to both the polarized proton and the polarized 
neutron data at photon energies above 1.45~GeV via $\tilde{c}_{1}$ and
$\tilde{c}_{2}$ as well as via $\alpha_{a_1}$ and $\alpha_{f_1}$. For the
intercepts we obtain $\alpha_{a_1} = 0.42 \pm 0.23$ and $\alpha_{f_1}
= -0.66 \pm 0.22$ which is in reasonable agreement with the findings of DIS
fits. The coefficients turn out as $\tilde{c}_{1}=-34.1\pm5.7~\upmu$b
and $\tilde{c}_{2}=209.4\pm29.0~\upmu$b where we have used $s$ in
units of GeV$^2$.

The resulting parameterization of the polarized real photoabsorption
is in better agreement with the data than the parameterization of
Bianchi and Thomas. This may be by construction of course.
The result for the proton indicates a sign change at photon energies
above 2~GeV as does the data. The fit below 1.7~GeV deviates
from the proton data. This may be a consequence of the fourth resonance
structure previously discussed. A fit to the proton data alone does
not exhibit this feature~\cite{Helbing:2003zs} which illustrates the
significant impact of the inclusion of the neutron data in the fit.
The polarized neutron cross section below 1.7~GeV appear slightly
lower than the result of the 
fit. Since there is no polarized data for the neutron at energies
above 1.9~GeV it is not clear whether this is a significant
discrepancy. 

The Regge description of the neutron might also be
impaired by a 4th resonance.
The statistical error of the fit for the neutron is of the order of
10~$\upmu$b while the departure from the fit of the proton data due to
the 4th resonance is also of about this size. Hence, the
systematic error due to the ignorance of a possible 4th resonance in
the neutron case is not dominant. An extrapolation of the high-energy
behavior based on this parameterization appears reasonable. However, a
verification of this parameterization with data at higher energies on
the proton and also on the neutron would represent a very valuable
cross-check. 

\subsection{The GDH integral and the validity of the \gdhsr}
\label{sec:GDHintegral}
\subsubsection{The \gdhsr for the proton}
\begin{figure}\centering
\includegraphics[width=0.8\textwidth]{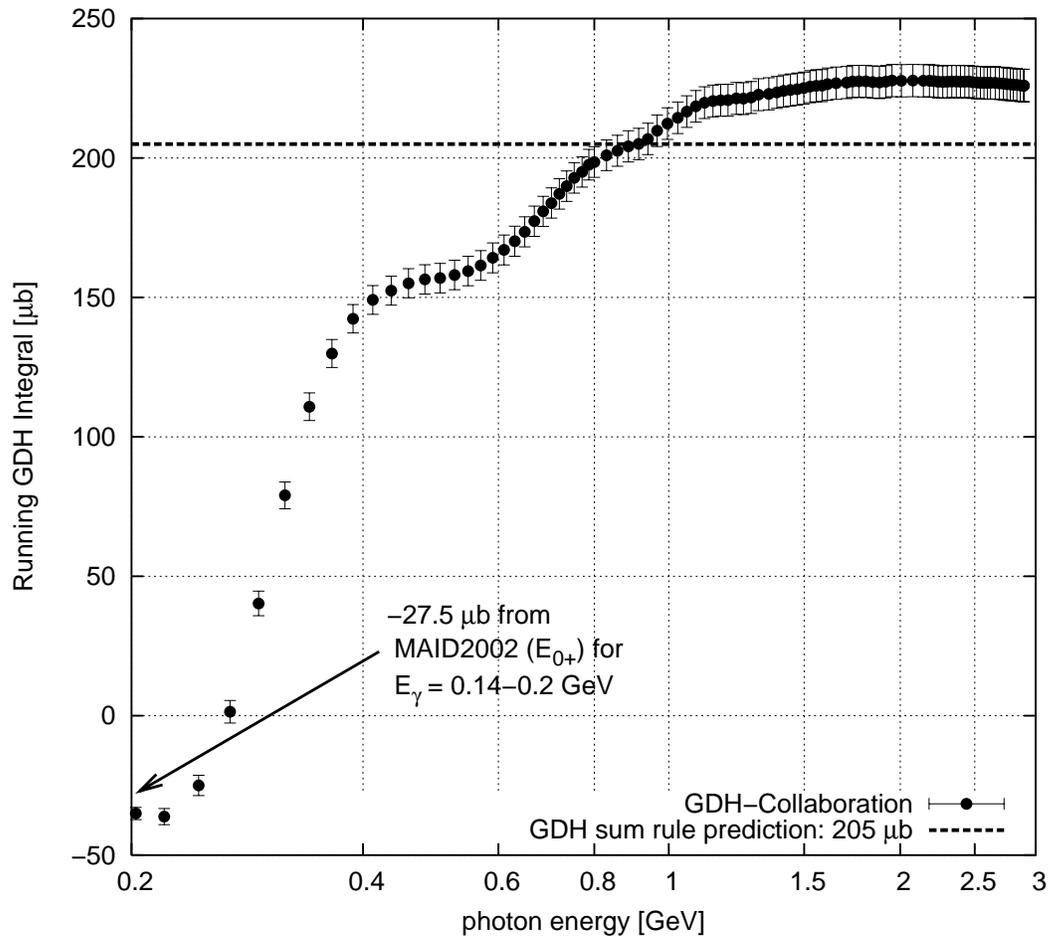}
\caption{Measured ``running'' GDH integral up to 2.9~GeV including the
threshold contribution.
\label{fig:Running}}
\end{figure}

The \gdhsr prediction for the proton amounts to 205~$\upmu$b. 
In Fig.~\ref{fig:Running} this prediction is compared to the
experimental results. The diagram shows the so-called ``running'' GDH integral
where the infinite integral on the left hand side of Eq.~(\ref{eqn:gdh}) is
replaced by an integration up to a certain energy $E_\text{run}$:
\be
I_\text{run} (E_\text{run}) = \int\limits_0^{E_\text{run}} \frac{d\nu}{\nu} 
\left[ \sigma_{3/2}(\nu) - \sigma_{1/2}(\nu) \right]
\ee
The experimental data on the proton range from 0.2 through 2.9~GeV in
photon energy while the first contribution for photoabsorption at low
energies comes from the pion threshold at 0.14~GeV. Between 140 and
200~MeV the two single pion channels $p\pi^0$ and $n\pi^+$ are the
only relevant channels. Since \daphne is not used to 
detect neutrons alone, the minimum photon energy providing access to
the channel $\gamma p \to n \pi^+$ is about 200~MeV when the pions
have enough energy to enter the detector. As a consequence, also the
total polarized photoabsorption is known only above this energy.

In Sec.~\ref{sec:delta} we have discussed that at threshold the main
contribution has to come from the single E$_{0+}$ multipole. Also, we
have seen that the characteristics of the $\Delta$-resonance are well
described by the multipole parameterizations \maid and \said,
especially with regard to low energies (see
Fig.~\ref{fig:delta}). Hence this contribution is very well defined:
$-27.5 \pm 3~\upmu$b is the estimate from \maid~\cite{Tiator:2002zj}
and $-28~\upmu$b that of \said~\cite{Arndt:2002xv}. Due to the simple
nature of the excitation at threshold this can be regarded to be a
reliable estimate.  

Fig.~\ref{fig:Running} includes this low-energy contribution from the unmeasured
region. One observes how all three major resonances contribute to the
running GDH integral. With the contribution of the third resonance at
about 900~MeV the integral overshoots the GDH prediction and stays
significantly above the \gdhsr value. The value of
the GDH sum up to 2.9~GeV is (see Ref.~\cite{Dutz:2004mm})
\be I_\text{run} (2.9~\text{GeV}) = 226 \pm 5_\text{stat} \pm
12_\text{syst}\ \upmu\text{b}~. 
\ee

The integrand $\sigma_{3/2} - \sigma_{1/2}$ remains positive from
about 230~MeV on up to about 2~GeV as seen in Fig.~\ref{fig:gdh_all}. 
The sign change of the integrand at higher energies within the
measured energy range  
only marginally improves the agreement of the running GDH
integral and the \gdhsr prediction for the proton.
However, our Regge parameterization discussed in
Sec.~\ref{sec:HighEnergies} of the polarized proton and neutron data
indicates a negative contribution from the unmeasured high energy
region above 2.9~GeV for the proton.  
We obtain a contribution of -15.3~$\upmu$b to the GDH integral above 2.9~GeV. This fit
result from the polarized data almost coincides with those based on DIS data. The
parameterization by Bianchi and Thomas~\cite{Bianchi:1999qs} gives
-14~$\upmu$b and another one by Simula {\it et al.}~\cite{Simula:2001iy}
gives -13~$\upmu$b.  

With only 3~$\upmu$b the statistical error of the proton extrapolation
based on our parameterization is rather
small. This is because the errors of the fit parameters are largely
anticorrelated. However, the systematic error appears larger as long
as it has not been verified with real photon data at higher
energies. We estimate it to be 10~$\upmu$b as a conservative estimate
considering that the different parameterizations are
close together with respect to this high energy contribution.

Including this extrapolation to high energies one obtains for the
experimental value of the GDH integral 
\be \boxed{\mathbf{
I_\text{run} (\infty) = 212 \pm 6_\text{stat} \pm
16_\text{syst}\ \upmu\text{b}~}.}
\ee
This is in good agreement with the \gdhsr prediction. The level of
precision obtained for the verification of the \gdhsr for the
proton is about 8~\% including the systematic uncertainties. The dominant
sources of the systematic error are uncertainties of beam and target
polarization as well as the high energy extrapolation.
\begin{center}
\bf This result
represents the first verification of the \gdhsr ever!
\end{center}

\subsubsection{The \gdhsr for the neutron and the isovector case}
\label{sec:GDHintNeutron}
The \gdhsr prediction for the neutron is 233~$\upmu$b which is
almost 30~$\upmu$b higher than the value for the proton. Moreover, the
contribution below 200~MeV due to the $E_{0+}$ amplitude is
-50~$\upmu$b~\cite{Drechsel:1998hk} i.e. even 22~$\upmu$b lower than for
the proton. The cross section difference in the $\Delta$-resonance as
predicted by \maid is very similar to that of the proton. The single
pion contribution as described by this multipole parameterization in
the region of the third resonance is negligible. Taken together this
seems to indicate a failure of the \gdhsr for the neutron. 

Prior to the experimental results of the \gdhcol a rich literature was addressing
the possible failure of the \gdhsr for the isovector
case~\cite{Karliner:1973em,Workman:1991xs,Arndt:2002xv}. In the
isovector case even the sign of these estimates turns out opposite to
the \gdhsr prediction as the following table shows:

\begin{center}
\begin{tabular}{lccc}\vspace{2pt}
                                & $I^p_\text{GDH}$   & $I^n_\text{GDH}$   & $I^{p-n}_\text{GDH}$       \\ \hline
\bf\gdhsr                          & 205          & 233          & -28
\\ \hline
Karliner \cite{Karliner:1973em}           & 261          & 183          & 78           \\
Workman, Arndt \cite{Workman:1991xs}      & 260          & 192          & 68           \\
Sandorfi et al. \cite{Sandorfi:1994ku}    & 289          & 160          & 129          \\
Drechsel-Krein \cite{Drechsel:1998fk}     & 261          & 180          & 81           \\
\end{tabular}
\end{center}

The preliminary results~\cite{Rostomyan:2005,Jahn:2004} from \mami for
the polarized cross section up to 800~MeV for the deuteron within the
statistical uncertainties are similar to twice the proton cross
section. The GDH integral from 200~MeV through 800~MeV for the
deuteron amounts to about 420~$\upmu$b. Given the large statistical
uncertainties of the data, we will neglect a discussion of nuclear
effects for the energy domain from 200 through 800~MeV and simply
assume that the deuteron data is the incoherent sum of the proton and
neutron. To arrive at an estimate for the neutron we account
200~$\upmu$b for the integral in this energy interval.

The experimental result at \elsa for the GDH integration of  neutron
data in the energy region 815~-~1825~MeV is $33.9 \pm 5.5_\text{stat} \pm
4.5_\text{syst}\ \upmu\text{b}$~\cite{Dutz:2005ns}. This contribution
previously was assumed to be zero while it turns out to be even larger
than for the proton in the respective energy domain. It is one of the
two major missing pieces that explain why the validity of the \gdhsr
also for the neutron and hence the isovector case is likely. 

The other missing piece is the
high energy part. Most often this was not accounted for either, like
in the analyses summarized in the table. 
Here we obtain +41~$\upmu$b with our own Regge parameterization as compared to
only -15~$\upmu$b for the proton. The statistical error of this
contribution to the neutron integral is of the order 10~$\upmu$b. 
The parameterization of Bianchi and Thomas~\cite{Bianchi:1993nh}
results in +30~$\upmu$b which is compatible.
In total, with the threshold contribution, we obtain an estimate of
225~$\upmu$b for the neutron GDH integral. This is in good agreement
with the \gdhsr prediction of 233~$\upmu$b. The systematic and
statistical errors are large however.

Considering the isovector case the situation is even more
accentuated. The largest contributions to the GDH integral come from the
behavior at threshold and at energies above about 1~GeV. The
+22~$\upmu$b up to 200~MeV are more than compensated by about
-60~$\upmu$b in the range above 800~MeV. 
The energy range at \mami only gives
rise to $\sim$26~$\upmu$b which is compatible with zero given the large
statistical and systematic uncertainties of the preliminary data
analysis and the ignorance of nuclear effect of the deuteron in the
discussion here. The
estimate for the total integral for the isovector case amounts to
about -10~$\upmu$b. This is to be compared to the \gdhsr
prediction of -23~$\upmu$b. Within the large systematic uncertainties
this again represents a good agreement. Also, this estimate shows
that most of the strength in the isovector case comes from high
energies and not from the resonance regime. Since this part has been
neglected in most previous estimates for the isovector GDH integral
even the resulting sign of these analyses were wrong. \\

Unlike for the proton, where we are able to present a stringent
verification of the \gdhsr at the level of 8~\% accuracy, for the time
being, we can only give estimates for the neutron and the isovector
cases. The further analysis of the \mami data on the deuteron
with more statistics and with the detailed treatment of nuclear effects
will further clarify the role of the lower resonances. Data especially
on polarized deuterium or helium targets are needed at energies above
3~GeV to verify the Regge parameterization.
\vspace{7mm}

Today, we have no indication for a failure of the \gdhsr in either
case, the proton or the neutron. Especially the relevance of the
high energy part above the second resonance has been underestimated in
the past for the \gdhsr on the neutron but also for the proton.

\newpage
\section{Future activities
related to the \gdhsr}
Currently, the analysis of the data of the \gdhcol on the deuteron
taken at \mami is being worked on. The results will provide new
insight into the resonance structure with respect to isospin but also
will help to study nuclear effects with the analyzing power of
spin. Also, it should be possible to obtain  results with higher precision
for the neutron resonance contributions to the \gdhsr. This will move
the verification of the \gdhsr on the neutron from the domain of an
estimate to a more reliable result. 

In order to extend the experimental possibilities of \mami, a fourth
microtron acceleration stage is presently under construction, which
will increase the electron beam energy to 1.5~GeV. 
A fourth microtron acceleration stage is presently under construction
at \mami. This will increase the electron beam energy to 1.5~GeV and
extend the experimental possibilities.
As a detector to be
used with the tagged photon beam a combination of the Crystal Ball
detector~\cite{Badala:1991yh} and \textsc{Taps}~\cite{Marin:1998zf} in
forward direction will be used. This will allow to study partial
channels of the third resonance and part of the forth resonance. Also
the improved systematics with respect to the detection of neutral
final states will help with measurements of scattering off the
deuteron.\\

Several other experiments are planed that can confirm the
findings of the \gdhcol. These experiments use the laser backscattering technique to
obtain polarized photons instead of bremsstrahlung produced by
polarized electrons. The principal layout of the detection systems are
similar to that developed for the \gdhex at \elsa
(see Sec.~\ref{sec:detectors}).
\begin{itemize}
\item The \textsc{Legs} facility at BNL uses a polarized HD-target~\cite{Wei:2004ei}
to cover photon energies up to 470~MeV. Very first results have been
presented for the tip of the $\Delta$-resonance~\cite{Sandorfi:2002zx}.
\item The \textsc{Graal} facility also intends to use the HD-target
technique and will cover photon energies up to
550~MeV~\cite{Renard:2000jc}.
\item At \textsc{SPring-8} dynamically polarized PE-foils will be used
as a target and the energy coverage is 1.8~--~2.8~GeV~\cite{Iwata:2000jb}. 
\end{itemize}
Beyond the energy coverage of the \gdhcol there are
experiments planed to extend the measurements to higher energies:
\begin{itemize}
\item At \textsc{JLab} a measurement at
photon energies from 2.5 through 6~GeV is
proposed~\cite{Sober:2002ra}. A frozen spin target similar to that of
the \gdhcol is under development. 
\item \textsc{Slac} has an approved proposal to measure total cross
section asymmetries in the energy regime from 4 to 40~GeV. 
\end{itemize}
Experiments at higher energies will help to verify the Regge
parameterization and reduce the systematic error of the extrapolation
for the GDH integral.

\newpage
\section{Conclusion}
The \gdhsr exclusively relies on fundamental assumptions. These
assumptions represent the building blocks of modern physics. We have
pointed out that also the ``No-subtraction'' hypothesis, one of the
much questioned steps of its derivation, is of fundamental nature. The 
validity of the No-subtraction hypothesis is a consequence of
unitarity when restrictions to low orders of electromagnetic coupling are
avoided. Today no challenge or possible modification of the
\gdhsr discussed in the literature appears to be of substantial
relevance. 

The \gdhsr is a member of a family of so-called
super-convergence sum rules. Another prominent member is the Bjorken
Sum Rule. The statement of the Bjorken Sum Rule at $Q^2\to\infty$
represents the counter piece to the \gdhsr at $Q^2=0$. While
$Q^2\to\infty$ is experimentally only indirectly accessible by means
of a QCD evolution the \gdhsr is measurable directly at the real
photon point using bremsstrahlung photons.
The connection of the two sum rules by means of the generalized GDH
Sum allows to study the transition from partonic degrees of freedom to
the domain of hadronic interactions. It turns out that this transition
in terms of the GDH Sum in the range $0~\text{GeV}^2\le Q^2 \le
1~\text{GeV}^2$ is quite dramatic and exceeds the variations found at
higher $Q^2$.

The \gdhex was performed at two accelerators, \elsa and \mami, to
cover a very wide energy range from pion threshold to the onset of the
Regge regime. Extensive polarimetry of the beam was done to keep this
source of systematic errors low. Tagging systems with sub-percent energy
resolution were used to prepare the photon beams. A polarized solid
state target integrated in a horizontal cryostat was used minimizing
the impact on solid angle coverage. Two detectors designed for the two
energy regions at \elsa and \mami with high solid angle coverage
have determined the polarized total photoabsorption cross sections.

The results from the \gdhcol on the polarized photoabsorption show
that apart from the lowest lying $\Delta$-resonance the structure of
the nucleon's response is largely not understood. The data
verify the Gerasimov-Drell-Hearn Sum Rule
for the proton at the level of 8~\% including systematic errors of the
extrapolation to unmeasured energy regions. The \gdhsr for the neutron
and the isovector case also appear to be valid. Here the high energy
part above 1~GeV in photon energy plays an important role that was
underestimated in previous attempts that found discrepancies to the sum
rule prediction. 

More precise results on the deuteron and the neutron
can be expected from the \gdhex at \mami when the data already taken are analyzed.
Future experiments on total photoabsorption at
photon energies above 3~GeV will reduce the systematic error of the
extrapolation to the high-energy domain. 

\newpage
\noindent{\bf\large Acknowledgements}\\~\\
I am indebted to Gisela Anton for her continuous and steadfast support
and advise and her encouragement throughout the whole GDH project and
many years of challenges.

Also, I am grateful to Berthold Schoch for
initially suggesting the investigation of the GDH Sum Rule and for
his support for the GDH-Collaboration in general and for me
in pursuing the experiment in particular.

I wish to thank my colleagues of the GDH-Collaboration who all
contributed to the great success of the experiment; in particular,
Hans-Jürgen Arends, Jürgen Ahrens, Andrea Bock, Wolter von
Drachenfels, Hartmut Dutz, Frank Frommberger, Peter Grabmayr, Paolo
Pedroni, Jochen Krimmer, Werner Meyer, Thilo Michel, Jakob Naumann,
Markus Sauer, Thorsten Speckner, Robert Van de Vyver and Günter
Zeitler.  

It is a pleasure to thank Horst Rollnik for numerous very valuable and
illuminative discussions and for sharing his profound knowledge
of theoretical physics.
For several instructive conversations on aspects of the derivation of the
GDH sum rule, I am grateful to Robert L. Jaffe.
Several lively and challenging discussions with Ralf Panförder and
Steven Bass are gratefully acknowledged.

\end{document}